\documentclass[a4paper,oneside,11pt]{article}

\newcommand{\href}[1]{}

\usepackage[colorlinks=true,pdfstartview=FitV,linkcolor=black,citecolor=black,urlcolor=blue,plainpages=false]{hyperref}
\usepackage{appendix}

\usepackage[T1]{fontenc}
\usepackage[latin1]{inputenc}

\usepackage{amsmath}
\usepackage{amsthm}
\usepackage{amsfonts}
\usepackage{dsfont}
\usepackage{mathrsfs}
\usepackage{fullpage}
\usepackage{enumerate}

\usepackage[]{authblk}
\usepackage{times}
\usepackage{epsf}
\usepackage{psfrag}
\usepackage{epsfig}
\usepackage{amssymb}
\usepackage{subfigure}
\usepackage{comment} 

\newcommand{\Z}{{\mathds Z}}

\newcommand{\Enc}{{\mathsf{Enc}} \,}
\newcommand{\ep}{\epsilon}

\newcommand{\ze}{\zeta}
\newcommand{\C}{{\mathcal C}}

\newcommand{\T}{{\mathcal T}}
\newcommand{\Tau}{{\mathbb T}}
\newcommand{\E}{{\mathcal E}}
\newcommand{\M}{{\mathcal M}}
\newcommand{\W}{{\mathcal W}}

\newcommand{\Ls}{{\mathcal L}}
\newcommand{\N}{{\mathcal N}}

\newcommand{\R}{{\mathbb R}}

\newcommand{\MC}[3]{#1\leftrightarrow #2 \leftrightarrow #3}

\newcommand{\defi}{\triangleq}

\newcommand{\error}{\text{\rm error}}

\newcommand{\ol}{\overline}
\newcommand{\ul}{\underline}

\newcommand{\Remark}{\noindent \emph{Remark:}\;}

\newcommand{\st}{:}

\newcounter{constcount}
\newcounter{numcount}
\newcommand{\eqnum}{\stackrel{(\roman{numcount})}{=}\stepcounter{numcount}}

\newcommand{\leqnum}{\stackrel{(\roman{numcount})}{\leq\;}\stepcounter{numcount}}

\newcommand{\geqnum}{\stackrel{(\roman{numcount})}{\geq\;}\stepcounter{numcount}}

\newcommand{\cnt}{$(\roman{numcount})$ \stepcounter{numcount}}

\newcommand{\rescnt}{\setcounter{numcount}{1}}

\newcommand{\nueff}{\tilde \nu}

\newcounter{thmcnt}
  \let\Oldsection\section
  
\renewcommand{\section}{\stepcounter{thmcnt}\Oldsection}

\newenvironment{lemmarep}[1]{\noindent {\bf Lemma #1.}\begin{it}}{\end{it}}

\newenvironment{theoremrep}[1]{\noindent {\bf Theorem #1.}\begin{it}}{\end{it}}

\newtheorem{theorem}{Theorem}

\newtheorem{lemma}{Lemma}
\newtheorem{definition}{Definition}

\newtheorem{cor}{Corollary}

\renewcommand{\th}{^{\rm th}}

\newcounter{examplecounter}
\newcommand{\aln}[1]{\begin{align*}#1\end{align*}}

\newcommand{\al}[1]{\begin{align}#1\end{align}}

\def\Item$#1${\item $\displaystyle#1$
   \hfill\refstepcounter{equation}(\theequation)}

\begin{document}

\title{Diamond Networks with Bursty Traffic:\\Bounds on the Minimum Energy-Per-Bit}

\author{Ilan Shomorony$^{\dagger}$, Ra\'ul Etkin$^\ast$, Farzad Parvaresh$^\ast$ and A. Salman Avestimehr$^{\dagger}$\\
$^\dagger$Cornell University, Ithaca, NY\\
$^\ast$Hewlett-Packard Laboratories, Palo Alto, CA}

\maketitle

\begin{abstract}
When data traffic in a wireless network is bursty, small amounts of data sporadically become available for transmission, at times that are unknown at the receivers,
and an extra amount of energy must be spent at the transmitters to overcome 
this lack of synchronization between the network nodes.
In practice, pre-defined header sequences are used with the purpose of synchronizing the different network nodes.
However, in networks where relays must be used for communication, the overhead required for synchronizing the entire network may be very significant.
%

In this work, we study the fundamental limits of energy-efficient communication in an asynchronous diamond network with two relays.
We formalize the notion of relay synchronization by saying that a relay is synchronized if the conditional entropy of the arrival time of the source message given the received signals at the relay is small.
We show that the minimum energy-per-bit for bursty traffic in diamond networks is achieved with a coding scheme where each relay is either synchronized or not used at all.
A consequence of this result is the derivation of a lower bound to the \emph{minimum energy-per-bit} for bursty communication in diamond networks. 
This bound allows us to show that schemes that perform the tasks of synchronization and communication separately (i.e., with synchronization signals preceding the communication block) can achieve 
the minimum energy-per-bit to within a constant fraction that ranges from $2$ in the synchronous case to $1$ in the highly asynchronous regime. 
\end{abstract}


\section{Introduction}

Most theoretical studies of wireless networks assume that transmitters and receivers are synchronized, in the sense that the receiver knows when data transmission is about to start. 
This is in general justified by the fact that, if large amounts of data are to be transmitted, then the time and energy required for synchronization are negligible when compared to what is required for communication itself.
Several applications, such as Wi-Fi, fall into this category and, in their context, optimizing the time and energy required for establishing the connection is of small practical importance.

However, in certain applications such as wireless sensor networks and bursty data communication in cellular networks, small amounts of time-sensitive data are sporadically available for transmission, at times that are unknown to the receivers. 
In such scenarios, the receiver is constantly listening to the output of a noisy channel in an attempt to identify a message. 
An extra amount of energy is then spent at the transmitter to make sure that the message is not missed and the noise is not mistaken for the message. 
In the sporadic data model, this extra energy represents a significant part of the total energy spent and becomes a relevant quantity. 
There is a large body of work treating synchronization from a practical perspective with the goal of minimizing overheads and synchronization errors. 
However, these studies lack a fundamental characterization of the energy and bandwidth costs of synchronization.

Early work on the fundamental limits of \emph{asynchronous} communication involved characterizing the data  rates that can be achieved when the receiver does not know the beginning of the communication block  \cite{strongasynch}. 
Later, in \cite{asynchcap}, a similar model was considered, but the performance metric was instead the energy (or, in general, the cost) per bit required for reliable asynchronous communication.
The characterization of the minimum energy-per-bit is important from a practical point of view, especially since it is often the case that the sensors in a wireless sensor network are battery-operated.
Thus, in the case of short and sporadic transmissions, i.e., bursty traffic, when synchronization costs may in fact dominate the communication costs, the characterization of the minimum energy-per-bit is very relevant.

In this work, we follow the asynchronism model from \cite{asynchcap}.
However, we focus on the AWGN channel model, rather than on discrete channels.
We assume that $B$ bits of data become available at the source node at a random arrival time $\nu_B$, and must be communicated to a destination with a maximum delay $d_B$\footnote{We index the random arrival time $\nu$ and the delay constraint $d$ by $B$ since we will consider an asymptotic regime in $B$, as described in Section \ref{setupsec}.}. 
The arrival time $\nu_B$ is assumed to be unknown to all network nodes, and unknown to the source before the arrival time itself.
However, $\nu_B$ is known to be drawn from $\{1,...,A_B\}$, where $A_B$ quantifies the asynchronism level. 
Under this setting, and assuming that $\nu_B$ is drawn uniformly at random from $\{1,...,A_B\}$, it was shown in \cite{asynchcap} that the \emph{asynchronous} minimum energy-per-bit of a point-to-point AWGN channel is given by
\al{
\left( 1 + \frac{\log A_B}{B} \right) e_b^{\rm sync}, \label{asyncmeb}
}
where $e_b^{\rm sync} = 2 N_0 \ln 2$ is the minimum energy-per-bit for an AWGN channel with noise power $N_0$ in the synchronous setting.
Our first result is to show that the asynchronous minimum energy-per-bit in (\ref{asyncmeb}) can be achieved through a scheme where the tasks of synchronization and communication are performed separately.
In such a scheme, which we refer to as a \emph{separation-based} coding scheme, as soon as the message arrives (at time $\nu_B$) the source uses a synchronization signal in order to inform the destination that the message is about to be transmitted.
If this synchronization procedure succeeds, communication can then take place as if we were in the synchronous setting.
We focus on such \emph{separation-based} schemes due to their ease of design and practical implementation.


We then move on to the main topic of the paper: asynchronous communication in multi-hop networks.
This is motivated by the fact that multi-hop communication with relays increases network range and throughput and reduces power consumption.
The fundamental question we focus on is: ``how should relays facilitate the communication between source and destination when they do not know the beginning of the transmission block?''.
On the one hand, one could devise a scheme where relays are constantly assuming that communication is taking place.
However, this approach would intuitively waste energy outside the actual communication block.
On the other hand, we could consider a separation-based scheme which first synchronizes all relays and the destination, and then proceeds to communicate over a synchronous network.
However, this may also be potentially wasteful, since the relays are not required to decode the message, so they do not need to know the beginning of the transmission block precisely.
In essence, our goal is to understand whether intermediate relays should be synchronized and whether separation-based coding schemes perform well.

\begin{figure}[ht] 
     \centering
       \includegraphics[height=35mm]{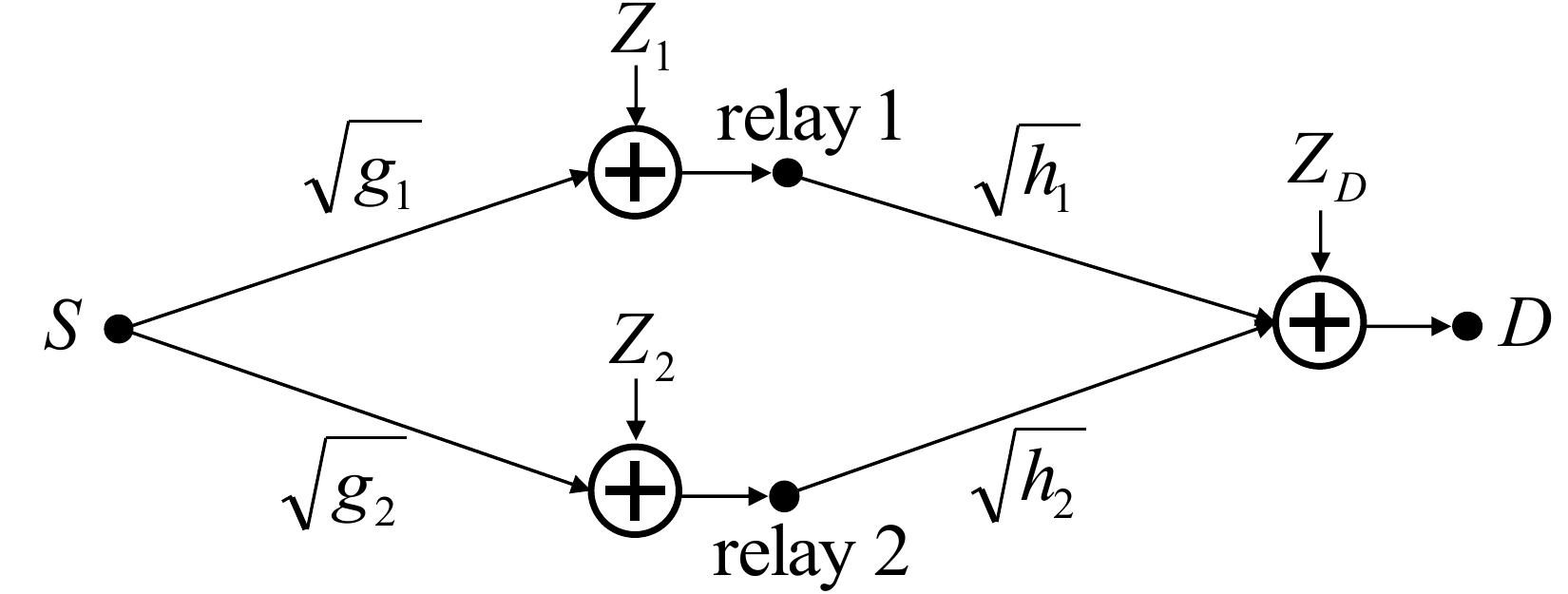} \caption{Two-relay diamond network. \label{netfig}}
\end{figure}

We study this problem in the context of the two-relay diamond network shown in Figure \ref{netfig}.
We say that a coding scheme synchronizes relay $i$ if, intuitively, the signals received by relay $i$ during times 
$1,2,...,A_B,A_B+1,...,A_B+d_B$, represented by $Y_i^{A_B+d_B}$, reveal a significant amount of information about $\nu_B$; or, more precisely, if $H(\nu_B \, | \, Y_i^{A_B+d_B})/B \to 0$ as $B \to \infty$. 
Under this notion of relay synchronization, we show that it is optimal from an energy-per-bit point of view to consider coding schemes 
that synchronize any relay that is used (i.e., that does not stay silent).
This result allows us to show that, depending on the specific values of $g_1, g_2, h_1$ and $h_2$, it is optimal from the energy-per-bit point of view to either only use relay $1$, only use relay $2$, or use both relay $1$ and relay  $2$.
This result is in contrast with the intuition provided by the synchronous case, in which, the capacity (and also the minimum energy-per-bit) is always improved if we utilize as many relays as are available.
Finally, we utilize the fact that relays must be synchronized to derive a lower bound to the minimum energy-per-bit for the asynchronous two-relay diamond network. 
We then verify that the energy-per-bit achieved by a separation-based scheme is within a constant factor of this lower bound.
This factor is $2$ in the synchronous case, but it drops towards $1$ as the asynchronism-per-bit $(\log A_B)/B$ increases.
We conclude that, in high-asynchronism regimes, where synchronization costs are high, separation-based schemes perform close to optimally.

The paper is organized as follows.
In section \ref{relworksec}, we summarize some of the previous work on asynchronous communication.
In section \ref{setupsec}, we describe our network model and formally define the notion of the asynchronous minimum energy-per-bit that we use.
In section \ref{simpleboundssection}, we provide some preliminary results.
First, we describe the known results on the minimum energy-per-bit of point-to-point AWGN channels.
Then we show how similar ideas can be used to derive upper and lower bounds for the minimum energy-per-bit for the asynchronous diamond network.
However, the ratio between these upper and lower bounds is unbounded, and the remainder of the paper is devoted to improving the lower bound (i.e., the converse direction).
In section \ref{mainresultssection}, we state our two main results.
The first main result, Theorem \ref{mainthm}, essentially states that it is optimal to consider coding schemes where any relay that is used (i.e., does not stay silent) must be synchronized.
We then state and prove our second main result, Theorem \ref{lboundthm}, which bounds the asynchronous minimum energy-per-bit of the two-relay diamond network.
The upper and lower bound are then verified to be within a constant fraction of each other.
The proof of Theorem \ref{mainthm} is left to section \ref{syncsection}.
We then conclude the paper in section \ref{conclusion}.


\section{Related Work} \label{relworksec}

The modeling of bursty data traffic used in this work builds up on the asynchronism models introduced in \cite{strongasynch} and \cite{asynchcap}.
In \cite{strongasynch}, asynchronism is modeled by having the message transmission block start at a randomly chosen time within a prescribed window.
The receiver knows the transmission window, but not the location of the transmission block.
The authors consider an asymptotic regime in which the size of the window grows exponentially with the number of bits to be transmitted, and they define the communication rate as the ratio between the number of transmitted bits and the average time elapsed between the beginning of the transmission block and the time when the decoder makes a decision.
Under this model, in \cite{strongasynch}, several aspects of the tradeoff between achievable communication rates and the asynchronism exponent were characterized.
Later on, in a follow-up work \cite{ChandarNovel}, the authors drew connections between this asynchronism model and the detection and isolation model introduced in \cite{NikiforovDetection}.

The asynchronism model considered in \cite{asynchcap} is very similar to the model from \cite{strongasynch}.
In \cite{asynchcap}, however, the performance metric is the data rate per unit cost, rather than just the data rate.
The authors also allow for the random variable associated with the beginning of the communication block to have more general probability distributions (not just the uniform distribution).
Their goal is to characterize the maximum achievable rate per unit cost, or the capacity per unit cost, which is the inverse of the minimum cost per bit.
For a discrete channel $p(y|x)$ with input alphabet $\mathcal X$ and output alphabet $\mathcal Y$, and an arbitrary cost function $k : \mathcal X \to [0,\infty]$, they show that the capacity per unit cost is given by
\al{ \label{result1}
{\bf C}(\bar \beta) = \max_{X} \min \left\{ \frac{I(X;Y)}{E[k(X)]}, \frac{I(X;Y)+D(Y\|Y_\star)}{E[k(X)](1+\bar \beta)}\right\},
}
where $Y_\star$ is the random variable corresponding to the output of the channel outside of the transmission block (i.e., when the transmitter is idle), and $\bar \beta$ is a parameter that characterizes how the uncertainty of the beginning of the transmission block grows with the number of bits to be sent.
In particular, for an AWGN channel with noise variance $N_0$, and quadratic cost function $k(x) = x^2$, and assuming that the beginning of the transmission block is drawn uniformly from $\{1,2,...,2^{\beta B}\}$, for $\beta > 0$, where $B$ is the number of bits to be transmitted, this expression reduces to
\al{ \label{result2}
{\bf C}(\beta) = \frac{1}{1+\beta} \frac{\log e}{2 N_0}.
}
Notice that, if we define the length of the window to be $A_B = 2^{\beta B}$, (\ref{result2}) implies that, for an AWGN channel, the asynchronous minimum energy per bit is given by
\aln{
(1+\beta) 2 \ln 2 N_0  = \left( 1 + \frac{\log A_B}{B}\right) e_b^{\rm sync},
}
where $e_b^{\rm sync} = 2 \ln 2 N_0$ is the usual (synchronous) minimum energy-per-bit of an AWGN channel.
In addition, the authors of \cite{asynchcap} also characterize the basic trade-off between the capacity per unit cost and the exponent of the delay within which the decoder must make a decision.

In \cite{WangAsynchronousExponents}, the same point-to-point asynchronous model from \cite{strongasynch} is considered, but the authors study the miss and false alarm error exponents. 
As a consequence, they are able to characterize the suboptimality of tranining-based schemes.

In \cite{PolyanskiyAsynchronous}, a strengthened version of the asynchronism model from \cite{strongasynch} is proposed, in which the decoder needs to estimate both the message and the location of the codeword exactly.
It is shown that the asynchronous capacity region remains unchanged under this formulation.
In addition, the finite blocklength regime is investigated.

\section{Problem Setup} \label{setupsec}

We consider the diamond network, shown in Figure \ref{netfig}.
We assume a discrete-time model where, at time $t$, each transmitter node $u \in \{S,1,2\}$ transmits a real-valued signal $X_u[t]$, each relay $i$, for $i \in \{1,2\}$, receives
$
Y_i[t] = \sqrt{g_i} X_S[t] + Z_i[t],
$
and the destination $D$ receives
$
Y_D[t] = \sqrt{h_1} X_1[t] + \sqrt{h_2} X_2[t] + Z_D[t],
$
where $Z_1[t], Z_2[t]$ and $Z_D[t]$ are independent i.i.d. $\N(0,N_0)$ noise terms.

Our bursty traffic model follows the asynchronous communication model introduced in \cite{asynchcap}.
%
%
%
The source receives a $B$-bits message $m$ at some random time $\nu \in [1:A]$, where, for $a > b$, we define $[a:b] \defi \{a,a+1,...,b\}$.
The source then needs to communicate this message to the destination with a delay of at most $d$ time-steps. 

In order to formally define reliable communication, we consider the asymptotic regime of $B \to \infty$.
Thus, we consider a sequence of arrival distributions $\{\nu_B\}_{B=1}^\infty$, where $\nu_B$ is uniform on $[1: A_B]$ and $A_B = 2^{\beta B}$, for $B=1,2,...$ and some $\beta > 0$.
Notice that, as $B \to \infty$, $B/A_B \to 0$, thus capturing the idea of short and sporadic messages.
Once the $B$ bits arrive at the source at time $\nu_B$, they must be communicated to the destination within a delay $d_B$.
Notice that, in order for the problem to be meaningful, $d_B$ should be small in comparison to $A_B$. 
Otherwise, it would be possible to devise a strategy where the source only starts its trasmission at pre-defined time-steps separated by $d_B$ time-steps, and the traffic would not be actually bursty.
Thus, since $A_B$ is exponential in $B$, we will require the delay $d_B$ to be subexponential in $B$.

An \emph{asynchronous code} $\C$ for the symmetric diamond network is designed to communicate a specific number of bits $B$ with a delay of $d_B$, assuming an arrival distribution $\nu_B$.
This code is comprised of 
\begin{itemize}
\item an encoding function for the source $f : [1:2^B] \times [1:A_B] \to \R^{d_B+1}$, which defines the source transmit signals for $[\nu_B:\nu_B+d_B]$, given the $B$ message bits and their arrival time $\nu_B$;
\item relaying functions $\rho_{i,t} : \R^{t-1} \to \R$, which define relay $i$'s transmit signal at time $t$ given its received signals in times $1,...,t-1$, for $t=1,...,A_B+d_B$;
\item a sequential decoder $(\tau,\hat m)$, which, at time $t$, decides to either decode the message (in which case it sets $\tau = t$ and outputs a decoded message $\hat m$) or to wait (in which case $\tau > t$). 
\end{itemize}
We then have the following definition.


\begin{definition} \label{achievedef}
Energy-per-bit $e_b$ is achievable if we can find a sequence of codes $\{\C_k\}_{k=1}^{\infty}$ and a sequence $\{B_k\}_{k=1}^{\infty}$, with $B_k \to \infty$ as $k \to \infty$, where code $\C_k$ can transmit $B_k$ bits with a maximum delay of $d_{B_k}$, assuming the input distribution is $\nu_{B_k}$, and we have
\begin{enumerate}[1. ]
\item $\lim_{k \to \infty} \Pr\left(\error(\C_{k})\right) = 0$
\vspace{1mm}
\item $\lim_{k \to \infty} \displaystyle \frac{ \log d_{B_k}}{B_k} = 0$
\vspace{1mm}
\item $\liminf_{k \to \infty} \displaystyle \frac{ E[\E_{\C_k}] }{B_k} \leq e_b,$
\end{enumerate}
where $\E_{\C_k} \defi \sum_{t=1}^{A_{k}+d_{B_k}} \left(X_S[t]^2 + X_1[t]^2 + X_2[t]^2\right)$ is the total energy used by code $\C_k$, 
$A_k \defi 2^{\beta B_k}$, 
and $\error(\C_k)$ is the event $\{ m \ne \hat m\} \cup \{\tau > \nu_{B_k} + d_{B_k}\}$ for code $\C_k$.
The asynchronous minimum energy-per-bit is the infimum over all achievable energy-per-bit values.
\end{definition}

Constraint 2 is what characterizes the data as time-sensitive, thus requiring the communication to be in fact bursty.
Notice that our definition of achievable energy-per-bit is similar to the ones in \cite{asynchcap} (with delay exponent $\delta = 0$), and in \cite{ElGamalEnergy} (by setting $\beta = 0$, i.e., in the synchronous case).

\section{Preliminary Results} \label{simpleboundssection}

In this section we first describe some known results for point-to-point AWGN channels.
Then we extend them in a simple way to the two-relay diamond network, and show that this approach yields upper and lower bounds on the asynchronous minimum energy-per-bit which can be arbitrarily far from each other.

\subsection{Point-to-point AWGN Channel}

In the synchronous case, the minimum energy-per-bit of a point-to-point AWGN channel is a special case of the inverse of the
capacity per unit cost, studied in \cite{verducapacityperunitcost}, where the cost is the average power.
Consider a simple AWGN point-to-point channel, 
where the channel gain between transmitter and receiver is $\sqrt h$.
Let $e_b^{\rm \, sync}$ be the minimum energy-per-bit of this channel in the synchronous setting and $e_b^{\rm \, async}$ be the minimum energy-per-bit of this channel in the asynchronous setting.
The following lemma can be obtained from the results in \cite{verducapacityperunitcost}.

\begin{lemma} \label{synclemma}
If $C(P)$ is the capacity of the synchronous AWGN channel with power constraint $P$, then 
\vspace{-2mm}
\aln{
e_b^{\rm \, sync}   = \inf_{P > 0} \frac{P}{C(P)} = \gamma / h,
}
where we define $\gamma = 2 N_0 \ln 2$.
\end{lemma}


The importance of Lemma \ref{synclemma} for us is that it guarantees that any energy-per-bit $e_b > e_b^{\rm sync}$ can be achieved with codes whose delay (i.e., the blocklength) is linear in the number of bits being sent.
To see this, consider any $e_b >  e_b^{\rm sync} = \inf_{P> 0} P/C(P)$.
We can find $P'>0$ such that $e_b > P'/C(P')$, and, for $\delta > 0$ sufficiently small, we will have $e_b \geq \frac{P'}{C(P')-\delta}$.
Now, for the rate $R = C(P') - \delta$,
we can find a sequence of (synchronous) 
codes $\{\C_k\}_{k=1}^\infty$, where $\C_k$ transmits $B_k = k R$ bits, with a blocklength equal to $k$, whose error probabilities go to $0$ as $k \to \infty$.
Therefore, the energy-per-bit of each code $\C_k$ satisfies 
\aln{
\frac{P' k}{B_k} = \frac{P'}{R} = \frac{P'}{C(P')-\delta} \leq e_b,
}
and the delay is $k = B_k/R$ is linear in $B_k$.


Next, we consider the same AWGN channel, but in the asynchronous setting.
We will make an additional assumption about the sequence of distributions of $\nu_B$.
Let $p_B(t) = \Pr[\nu_B = t]$ and let $p_B^{\max } = \max_t p_B(t)$.
We will require that $p_B^{\max } \to 0$ as $B \to \infty$.
Among the sequences of distributions satisfying this property, we have sequences of distributions whose probability mass functions have the same ``shape'' but stretched over the interval $[1:2^{\beta B}]$ for each $B$.
In particular, this is the case when $\nu_B$ is uniformly distributed on $\{1,...,2^{\beta B}\}$, which will be our focus when we consider the diamond network.
By using this restriction, we exclude the distributions for which the expression (6) in \cite{asynchcap} does not evaluate to the normalized entropy.
Under this assumption, we can state the following theorem, which is similar to the results in \cite{asynchcap}.
However, since we are not in the discrete alphabet setting which is the focus of \cite{asynchcap}, our achievability scheme is somewhat different, and it introduces the notion of a \emph{separation-based} scheme.
We present the achievability proof here, and the converse in Appendix \ref{apppttopt}.

\begin{theorem} \label{pttopt}
For an asynchronous AWGN channel, the minimum energy-per-bit is given by
\aln{
e_b^{\rm \, async} = (1+\bar H) e_b^{\rm \, sync}.
}
where $\bar H = \liminf_{k \to \infty} H(\nu_{B_k})/{B_k}$.
In particular, if each $\nu_{B_k}$ is uniformly distributed in $[1:2^{\beta B_k}]$, then $\bar H = \beta$.
\end{theorem}

\begin{proof}
\emph{Achievability}: 
We will show that the asynchronous energy-per-bit $(1+\bar H)e_b^{\rm \, sync}(1+\delta)^2$, for an arbitrarily small $\delta >0$, is achievable, which implies $e_b^{\rm \, async} \leq (1+\bar H) e_b^{\rm \, sync}$.
We will let $\{B_k\}$ be the subsequence of $1,2,...$ along which $\lim_{k \to \infty} H(\nu_{B_k})/B_k = \bar H$. 
We will then build a sequence of codes $\{\C_k\}_{k=1}^\infty$, where code $\C_k$ 
assumes arrival distribution $\nu_{B_k}$ and transmits $B_k$ bits.

Our scheme is based on having the transmitter send a large pulse as soon as the message arrives.
The receiver will use a threshold detector to detect the pulse.
Once the pulse is (correctly) detected, communication can proceed as in a synchronous channel.
For code $\C_k$, the total energy available for the pulse will be $H(\nu_{B_k}) e_b^{\rm \, sync}(1+\delta)^2 = H(\nu_{B_k}) \gamma (1+\delta)^2/h$. 
If the message arrives at time $t$ (which implies $p_{B_k}(t) > 0$), then the transmitter will first send a pulse of magnitude
\al{
(1+\delta)\sqrt{\frac{- \log{\left(p_{B_k}(t)\right)} \gamma}{h}} = (1+\delta)\sqrt{\frac{- 2 N_0\ln{\left(p_{B_k}(t)\right)}}{h}}.
}
Following the pulse, the transmitter sends a codeword from an optimal code designed to send $B_k$ bits with energy-per-bit $(1+\delta)^2 e_b^{\rm \, sync} = (1+\delta)^2 \gamma /h$ over the \emph{synchronous} version of the channel.
Therefore, the expected energy consumed by our code is given by
\aln{
E\left[\E_{\C_k}\right] & = \sum_{t=1}^{A_k} p_{B_k}(t) \left[ - (1+\delta)^2 \log{\left(p_{B_k}(t)\right)}\gamma / h+ B_k (1+\delta)^2 \gamma / h \right] \\
& = (1+\delta)^2 \frac{\gamma}h \left[ - \sum_{t=1}^{A_k} p_{B_k}(t) \log\left(p_{B_k}(t)\right) + B_k \right] \\
& = (1+\delta)^2 \frac{\gamma}h \left(H(\nu_{B_k}) + B_k\right),
}
where we used the convention that $0 \log 0 = 0$, in order to sum over all $t \in \{1,...,A_k\}$.
The energy-per-bit we will achieve will be $\lim_{k \to \infty} E\left[\E_{\C_k}\right] / B_k = (1+\delta)^2(1+\bar H) \gamma / h$, as we intended.
All we need to show is that the probability of error of our codes goes to $0$ as $k \to \infty$.
Since we are using an optimal code for the synchronous channel to actually communicate the bits, the probability of incorrect decoding, given that the pulse was detected goes to $0$ as $k \to \infty$. 
Moreover, from Lemma \ref{synclemma} we know that the blocklength required for these codes will be $B_k/R$ for some $R > 0$, which guarantees that the decoding delay $d_{B_k} = B_k/R + 1$ will satisfy $\lim_{k\to \infty}\log d_{B_k} /B_k = 0$, and the probability of late decoding also goes to $0$ as $k \to \infty$.
Thus, we only need to show that the probability of error in detecting the pulse goes to $0$.
For this to happen, we will set the detection threshold at the destination to be 
\aln{
(1+\delta/2)\sqrt{-2 N_0 \ln{\left(p_{B_k}(t)\right)}}
}
at time $t$.
We define the following two error events:
\begin{itemize}
\item $L_1 = \{ \text{Destination does not detect the pulse} \}$
\item $L_2 = \{ \text{Destination incorrectly detects a pulse before the pulse is sent} \}$
\end{itemize}
Clearly, the probability of error in detecting the pulse can be upper bounded by $\Pr(L_1 \cup L_2) \leq \Pr(L_1) + \Pr(L_2)$.
We will show that each of the terms goes to $0$ as $k \to \infty$.
For $L_1$, we have
\aln{
\Pr(L_1) & = \sum_{t=1}^{A_k} p_{B_k}(t) \Pr\left\{ (1+\delta)\sqrt{-2 N_0 \ln{\left(p_{B_k}(t)\right)}} + Z(t) < (1+\delta/2)\sqrt{- 2 N_0 \ln{\left(p_{B_k}(t)\right)}} \right\} \\
& = \sum_{t=1}^{A_k} p_{B_k}(t) \Pr\left\{ - Z(t) > \delta/2 \sqrt{-2 N_0 \ln{\left(p_{B_k}(t)\right)}} \right\} \\
& \leq \Pr\left\{ Z(t) > \delta/2 \sqrt{-2 N_0 \ln{\left(p_{B_k}^{\max }\right)}} \right\},
}
and, as $k \to \infty$, we have $p_{B_k}^{\max } \to 0$ and $- \log p_{B_k}^{\max } \to \infty$, implying that $\Pr(L_1) \to 0$.
For $L_2$, we have
\aln{
\Pr(L_2) & \leq \Pr\left\{ \exists \, t \in \{1,...,{A_k}\} \st Z(t) \geq (1+\delta/2)\sqrt{- 2 N_0 \ln{\left(p_{B_k}(t)\right)}} \right\} \\
& \leq \sum_{t=1}^{A_k} \Pr \left\{ Z(t) \geq (1+\delta/2)\sqrt{- 2 N_0 \ln{\left(p_{B_k}(t)\right)}} \right\} \\
& \leq \sum_{t=1}^{A_k} e^{(1+\delta/2)^2 \ln{\left(p_{B_k}(t)\right)}} = \sum_{t=1}^{A_k} p_{B_k}(t)^{(1+\delta/2)^2} \\
& = \sum_{t=1}^{A_k} p_{B_k}(t)^{1+\delta + \delta^2/4} \leq \sum_{t=1}^{A_k} p_{B_k}(t) (p_{B_k}^{\max })^{\delta + \delta^2/4} \\
& = (p_{B_k}^{\max })^{\delta + \delta^2/4},
}
which goes to $0$ as $k \to \infty$, since $p_{B_k}^{\max } \to 0$ as $k \to \infty$.
\end{proof}

\subsection{Two-relay Diamond Network}

We now start considering the two-relay diamond network shown in Figure \ref{netfig} in the asynchronous setting.
Unless otherwise noted, we will assume throughout the paper (wlog) that $g_2 \leq g_1$.
Moreover, we will focus in the case where $\nu_{B}$ is uniformly distributed on $[1:2^{\beta B}]$.


A simple achievable scheme for the two-relay diamond network in Figure \ref{netfig} is a separation-based scheme, similar to the one we used in the achievability of Theorem \ref{pttopt}.
Thus, we will have a \emph{synchronization phase}, where the source will send a pulse at time $\nu_{B_k}$ to synchronize the relays, and the relays will send a pulse at time $\nu_{B_k} + 1$ to synchronize the destination.
After this, provided that the pulses were correctly detected by all nodes, we are in a synchronous setting, and  we will have a \emph{communication phase}.
In this phase, any code for the synchronous two-relay diamond network can be used, as long as its delay is subexponential in the number of bits being sent.

To compute an achievable asynchronous energy-per-bit, we will use decode-and-forward for the communication phase.
Notice that several relaying schemes that outperform decode-and-forward are known \cite{ScheinThesis,ADTJ09,ParvareshEtkinDiamond}.
However, there is no closed-form expression for the energy-per-bit achieved by these schemes, making it difficult to compare their performance to the lower bound.
A careful calculation of the asynchronous energy-per-bit achieved by this separation-based scheme yields the following theorem, whose proof is in  Appendix \ref{appseparationachievable}.

\begin{theorem} \label{separationachievable}
The asynchronous minimum energy-per-bit for the network in Figure \ref{netfig} satisfies
\aln{
(1+\beta) \gamma\left(\frac{1}{g_2}+\frac{1}{h_1+h_2}\right) \geq e_b^{\min }.
}
\end{theorem}


In order to obtain lower bounds on the asynchronous minimum energy-per-bit, we will use a technique similar in flavor to cut-set bounds, but applied to minimum energy-per-bit.
The idea is to consider all four cuts in the network in Figure \ref{netfig}, and view it as a MIMO channel, thus being able to apply a lower bound to the asynchronous minimum energy-per-bit of a point-to-point channel, as in Theorem \ref{pttopt}.
This approach yields the following result.

\begin{theorem} \label{simplebounds}
The asynchronous minimum energy-per-bit for the network in Figure \ref{netfig} is lower bounded as $e_b^{\min } \geq {\rm LB}$, where ${\rm LB}$ is the optimal solution to
\al{
\textrm{Maximize}\quad   &  \gamma(1+\beta) \left[ y_1 + y_2 + y_3 + y_4 \right]      \nonumber \\
\textrm{subject to}\quad  & (g_1+g_2) y_1+ g_2 y_3+ g_1 y_4 \leq 1  \nonumber  \\
			\quad  & (h_1+h_2) y_2+h_1 y_3 \leq 1  \nonumber \\		
			\quad  & (h_1+h_2) y_2+h_2 y_4 \leq 1 \nonumber  \\
                     & y_1,y_2,y_3,y_4     \geq 0. \label{lpbound}
}
\end{theorem}

In order to prove this result, we will require the following two results, which bound the asynchronous minimum energy-per-bit of the MIMO channels obtained when we consider the different cuts.
Their proofs are in Appendices \ref{appptpt2} and \ref{appptpt3} respectively.

\begin{lemma} \label{ptpt2}
Consider the networks in Figures \ref{hop1} and \ref{hop2}, where the message arrival time $\nu_B$ is uniformly distributed in $[1:2^{\beta B}]$.
Then, the minimum asynchronous energy-per-bit $e_b^{\min }$ of these two networks is given respectively by
\aln{
e_b^{\min } = (1+\beta) \frac{\gamma}{g_1 + g_2}
\text{\; and \;}
e_b^{\min } = (1+\beta) \frac{\gamma}{h_1 + h_2}.
}
\end{lemma}

\begin{lemma} \label{ptpt3}
Consider the MIMO channel in Figure \ref{cutgen} in the asynchronous setting, where the message arrival time $\nu_B$ is uniformly distributed in $[1:2^{\beta B}]$.
Consider a sequence of codes $\{\C_k\}_{k=1}^\infty$ that achieves a finite energy-per-bit, and let $\E_{\C_k}^{(s_i)}$ be the energy spent by code $\C_k$ at the source transmitter $s_i$, for $i=1,2$.
Then, we must have
\aln{
a \liminf_{k\to\infty}\frac{E\left[\E_{\C_k}^{(s_1)}\right]}{B_k} + b \liminf_{k\to\infty}\frac{E\left[\E_{\C_k}^{(s_2)}\right]}{B_k} \geq \gamma(1+\beta).
}
\end{lemma}

\begin{proof}[Proof of Theorem \ref{simplebounds}]
We will use the networks in Figures \ref{hop1} and \ref{hop2} to bound the energy-per-bit used by the sources, and the energy-per-bit used by the relays respectively.
Notice that these networks correspond to two out of the four cuts in our diamond network.
\begin{figure*}[ht] 
     \centering
     \subfigure[]{
       \includegraphics[height=29mm]{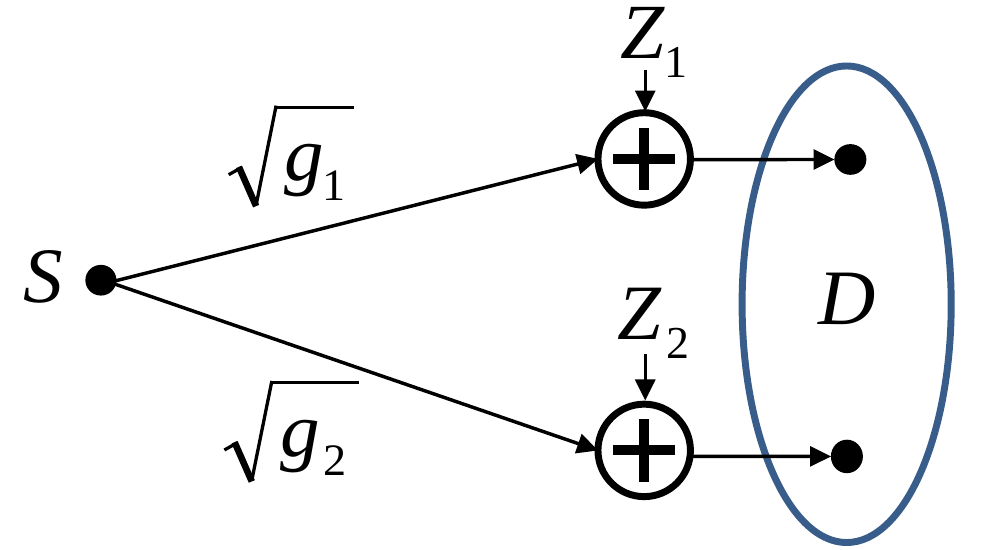} \label{hop1}} 
    \hspace{15mm}
    \subfigure[]{
       \includegraphics[height=25mm]{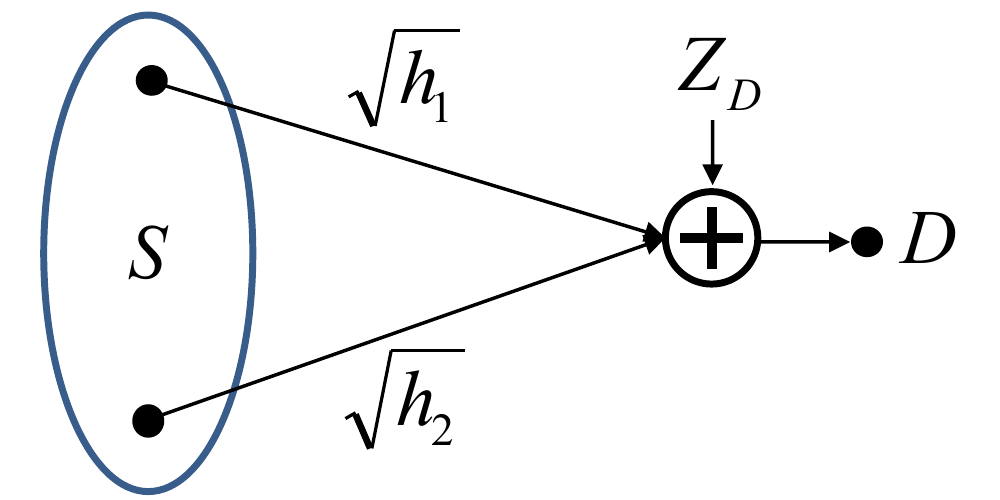} \label{hop2}}
    \hspace{0mm}
     \caption{MIMO Channels obtained from first and second hops of the diamond network in Figure \ref{netfig}.}
\end{figure*}

Now, suppose we have a sequence of codes $\{\C_k\}_{k=1}^\infty$ achieving a finite energy-per-bit $e_b$ on the diamond network in Figure \ref{netfig}.
Then we consider applying this sequence of codes to the network in Figure \ref{hop1}, where we assume the same asynchronism level.
In order to do this we let the source transmit as if it were in the network in Figure \ref{netfig}.
The destination, which has two receive antennas, which represent the relays from the original network, will first compute what the transmit signals of the relays would have been in the original network. 
Then it will simulate the second hop from the relays to the destination, and use the same sequential decoder used by the destination in the original network applied to the simulated received signal.
It is clear that the probability of error of this code applied to the network in Figure \ref{hop1} is identical to the probability of error of $\C_k$ on the network in Figure \ref{netfig}.
The main difference is that the energy from the relays is not consumed anymore, and the energy used by code $\C_k$ when applied to the network in Figure \ref{hop1}, is just the energy used by the source $\E_{\C_k}^{(s)}$.
This will allow us to bound the energy-per-bit used by the source.
From Lemma \ref{ptpt2}, we have
\al{
\liminf_{k \to \infty} \frac{E\left[\E_{\C_k}^{(s)}\right]}{B_k} \geq (1+\beta) \frac{\gamma}{g_1+g_2}. \label{senergy}
} 

Then we consider applying code $\C_k$ to the network in Figure \ref{hop2}.
This time, the source will simulate the transmit signals of the source in Figure \ref{netfig} and the received signals at the relays.
Then it can compute the transmit signals of the relays in Figure \ref{netfig} and use them one for each of its antennas.
If we let $\E_{\C_k}^{(r_1)}$ and $\E_{\C_k}^{(r_2)}$ be the energy used by the relays in code $\C_k$, Lemma \ref{ptpt2} tells us that
\al{
\liminf_{k \to \infty} \frac{E\left[\E_{\C_k}^{(r_1)} + \E_{\C_k}^{(r_2)}\right]}{B_k} \geq (1+\beta) \frac{\gamma}{h_1+h_2}. \label{renergy}
} 

Up to this point, we have only considered two out of the four possible cuts of the two-relay diamond network.
The other two cuts will yield MIMO channels that look like the network in Figure \ref{cutgen}, for $a = h_1$ and $b = g_2$, and $a = g_1$ and $b = h_2$.
\begin{figure}[ht] 
	\center
       \includegraphics[height=28mm]{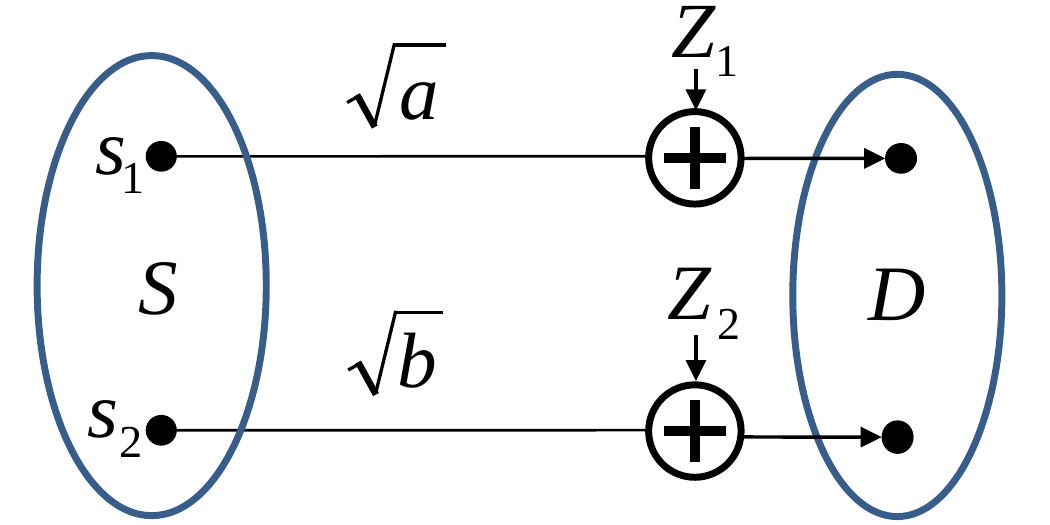} 
     \caption{MIMO channel obtained from remaining two cuts on the diamond network in Figure \ref{netfig}, by setting $a = h_1$ and $b = g_2$, or $a=g_1$ and $b=h_2$\label{cutgen}.}
\end{figure}
For a point-to-point channel such as the one in Figure \ref{cutgen}, it is not difficult to see that the asynchronous minimum energy-per-bit would be given by $\gamma (1+\beta) \min\left( \frac1a, \frac1b\right)$.
This is trivially achievable by just using one of the two source antennas ($s_1$ or $s_2$).
However, for the purposes of deriving a tighter lower bound, we will be interested in capturing a relationship between the energy that is spent in each of the two source antennas.
This relationship is stated in Lemma \ref{ptpt3}.

Recall that, in order to derive  (\ref{senergy}) and (\ref{renergy}) we used the fact that code $\C_k$ can be applied to the networks in Figures \ref{hop1} and \ref{hop2}.
Similarly, by applying code $\C_k$ to the network in Figure \ref{cutgen}, (with $a= h_1$ and $b = g_2$, and $a = g_1$ and $b = h_2$), we can use Lemma \ref{ptpt3} to obtain 
\al{
& g_2 \liminf_{k \to \infty} \frac{E\left[\E_{\C_k}^{(s)}\right]}{B_k} + h_1 \liminf_{k \to \infty} \frac{E\left[\E_{\C_k}^{(r_1)}\right]}{B_k} \geq \gamma (1+\beta),  \label{sr1energy} \\
& g_1 \liminf_{k \to \infty} \frac{E\left[\E_{\C_k}^{(s)} \right]}{B_k} + h_2 \liminf_{k \to \infty} \frac{E\left[\E_{\C_k}^{(r_2)}\right]}{B_k}  \geq \gamma (1+\beta). \label{sr2energy} 
}
Next, we notice that (\ref{senergy}), (\ref{renergy}), (\ref{sr1energy}) and (\ref{sr2energy}) imply that, for any $\delta > 0$, there exists a $k_0$ such that, for $k \geq k_0$,
\aln{
& ({g_1+g_2})\frac{E\left[\E_{\C_k}^{(s)}\right]}{B_k} \geq {\gamma}(1+\beta)  -\delta \\
& ({h_1+h_2})\frac{E\left[\E_{\C_k}^{(r_1)} + \E_{\C_k}^{(r_2)}\right]}{B_k} \geq {\gamma}(1+\beta)  -\delta \\
& g_2 \frac{E\left[\E_{\C_k}^{(s)}\right]}{B_k} + h_1 \frac{E\left[\E_{\C_k}^{(r_1)}\right]}{B_k} \geq \gamma (1+\beta) -\delta \\
& g_1 \frac{E\left[\E_{\C_k}^{(s)} \right]}{B_k} + h_2 \frac{E\left[\E_{\C_k}^{(r_2)}\right]}{B_k}  \geq \gamma (1+\beta) -\delta.
} 
Therefore, for any $k \geq k_0$, a lower bound to
\aln{
\frac{E\left[\E_{\C_k}\right]}{B_k} = \frac{E\left[\E_{\C_k}^{(s)}\right]}{B_k} + \frac{E\left[\E_{\C_k}^{(r_1)}\right]}{B_k} + \frac{E\left[\E_{\C_k}^{(r_2)}\right]}{B_k} 
} 
is the optimal value of the linear program
\al{
\textrm{Minimize}\quad   &  x_s + x_{r_1} + x_{r_2}      \nonumber \\
\textrm{subject to}\quad  & (g_1+g_2) x_s \geq \gamma (1+\beta) - \delta  \nonumber  \\
			\quad  & (h_1+h_2) (x_{r_1}+x_{r_2}) \geq \gamma (1+\beta) -\delta  \nonumber \\		
			\quad  & g_2 x_s +h_1 x_{r_1} \geq \gamma (1+\beta) -\delta \nonumber  \\
			\quad  & g_1 x_s +h_2 x_{r_2} \geq \gamma (1+\beta) -\delta \nonumber \\
			\quad  & x_s,x_{r_1},x_{r_2} \geq 0. \label{primallp}
}
This implies that, for any $\delta > 0$, the above linear program is also a lower bound to 
\al{
\liminf_{k\to \infty} \frac{E\left[\E_{\C_k}\right]}{B_k}, \label{minenergyexp0}
} 
and, after letting $\delta \to 0$, (\ref{primallp}) is still a lower bound to (\ref{minenergyexp0}).
Finally, by taking the dual of (\ref{primallp}) with $\delta = 0$, we conclude that (\ref{lpbound}) is a lower bound to the asynchronous minimum energy-per-bit of the diamond network in Figure \ref{netfig}.
The advantage of using the dual linear program in (\ref{lpbound}) rather than its primal is that any feasible solution $(y_1,y_2,y_3,y_4)$ yields a lower bound to the asynchronous minimum energy-per-bit of the diamond network.
\end{proof}

In order to explicitly compute a gap between the upper and lower bounds, we may consider a worse lower bound, obtained from the feasible solution to (\ref{lpbound}) $y_1 =(g_1+g_2)^{-1},y_2= (h_1+h_2)^{-1},y_3 = 0,y_4 =0$.
This tells us that 
\al{  \label{simplebound0}
\text{\rm LB} \geq \gamma (1+\beta) \left( \frac1{g_1+g_2} + \frac1{h_1+h_2} \right).
}
The gap between our upper bound and this lower bound is given by
\al{ \label{gap1}
(1+\beta) \gamma\left(\frac{1}{g_2}-\frac{1}{g_1+g_2}\right).
}
This result shows that our separation-based scheme performs well in cases where the channel gains of the first hop are much stronger than the channel gains of the second hop (since the gap in (\ref{gap1}) is small in comparison to the lower bound in (\ref{simplebound0})).
However, it is important to realize that our gap depends on $\beta$, suggesting that the separation-based scheme may be arbitrarily bad in high-asynchronism regimes (i.e., when $\beta$ is large).
Notice that, even if we consider the optimal solution to (\ref{lpbound}), our lower bound is still a multiple of $(1+\beta)$ and the gap to the upper bound from Theorem \ref{simplebounds} will still depend on $\beta$.





\section{Main Results} \label{mainresultssection}

Our first main result (Theorem \ref{mainthm}) is that a relay can only be helpful in a coding scheme (from the energy-per-bit point of view) if it is \emph{synchronized}.
From this, we can derive our second main result (Theorem \ref{lboundthm}), which consists of a lower bound for the asynchronous minimum energy per bit of the diamond network (tighter than the one in Theorem \ref{simplebounds}), whose ratio to the upper bound in Theorem \ref{separationachievable} is bounded by $2$, and decreases to $1$ as $\beta$ increases.
The proof of Theorem \ref{mainthm} is very technical, and is deferred to section \ref{syncsection}, while the proof of Theorem \ref{lboundthm} is presented in this section.

In this work, we will define synchronization as follows.
\begin{definition} \label{syncdef}
Relay $i$ is synchronized in the sequence of codes $\{\C_k\}_{k=1}^\infty$ if
\aln{
\lim_{k \to \infty} \frac{H(\nu_{B_k} | Y_i^{\tilde A_k})}{B_k} = 0,
}
where $Y_i^{\tilde A_k}$ is the vector of received signals of relay $i$ from time $1$ to time $\tilde A_k \defi A_k + d_{B_k}$, 
when using code $\C_k$.
\end{definition}
Our first main result states that it is optimal (in terms of minimum energy-per-bit) to consider only schemes where we either use both relays and synchronize them, or we just use relay $1$ and synchronize it.
We rule out the case where only relay $2$ is synchronized because, since $g_2 \leq g_1$, we have the Markov chain $\MC{\nu_{B_k}}{Y_1}{Y_2}$.
Thus, we must have $H(\nu_{B_k} | Y_1^{\tilde A_k}) \leq H(\nu_{B_k} | Y_2^{\tilde A_k} )$, which implies that if relay $2$ is synchronized, so is relay $1$.


%

\begin{theorem} \label{mainthm}
Suppose we have a sequence of codes $\{\C_k\}_{k=1}^\infty$ achieving a finite energy-per-bit $e_b$ on the asynchronous diamond network in Figure \ref{netfig}.
Then we can achieve arbitrarily close to the energy-per-bit $e_b$ with a sequence of codes $\{\C_k'\}$ for which one of the following is true:
\begin{enumerate}
\item Relay $i$, for $i=1,2$, can create a list $\Lambda_k^{(r_i)} \subset [1:A_k]$ based on their received signals, such that $\nu_{B_k} \in \Lambda_k^{(r_i)}$ with vanishing error probability and list size $|\Lambda_k^{(r_i)}|$ subexponential in $B_k$
\item Relay 1 can create a list $\Lambda_k^{(r_1)} \subset [1:A_k]$ based on its received signals, such that $\nu_{B_k} \in \Lambda_k^{(r_1)}$ with vanishing error probability and list size $|\Lambda_k^{(r_1)}|$ subexponential in $B_k$ and relay $2$ is inactive (i.e., does not transmit any signal)
\end{enumerate}
\end{theorem}
Theorem \ref{mainthm} states that we can assume wlog that any sequence of codes $\{\C_k\}_{k=1}^\infty$ achieving a finite energy-per-bit $e_b$ will allow any relay that is used (i.e., any relay that does not stay silent) to create a list $\Lambda_k^{(r_i)} \subset [1:A_k]$ that has size $|\Lambda_k^{(r_i)}|$ that is subexponential in $B_k$ and contains $\nu_{B_k}$ with vanishing error probability.
Therefore, if relay $i$ is used in the sequence of codes $\{\C_k\}_{k=1}^\infty$, and if we let $\mathds{1}\{\nu_{B_k}\in \Lambda_k^{(r_i)}\}$ be an indicator function for the event $\nu_{B_k}\in \Lambda_k^{(r_i)}$, then we must have
\aln{
\frac{H(\nu_{B_k}|Y_i^{\tilde A_k})}{B_k} & \leq \frac{H(\nu_{B_k}|\Lambda_k^{(r_i)})}{B_k} \leq \frac{H(\nu_{B_k}, \mathds{1}\{\nu_{B_k}\in \Lambda_k^{(r_i)}\}|\Lambda_k^{(r_i)})}{B_k} \nonumber \\
& \leq \frac{1 + H(\nu_{B_k}|\Lambda_k^{(r_i)}, \mathds{1}\{\nu_{B_k}\in \Lambda_k^{(r_i)}\})}{B_k} \nonumber \\
& \leq \frac{1+H(\nu_{B_k}|\Lambda_k^{(r_i)}, \nu_{B_k} \in \Lambda_k^{(r_i)})+H(\nu_{B_k}|\Lambda_k^{(r_i)}, \nu_{B_k} \notin \Lambda_k^{(r_i)})\Pr(\nu_{B_k}\notin \Lambda_k^{(r_i)})}{B_k} \nonumber \\
& \leq \frac{1+ \log|\Lambda_k^{(r_i)}|+\beta B_k \Pr(\nu_{B_k}\notin \Lambda_k^{(r_i)})}{B_k}, \nonumber 
}
which goes to $0$ as $k \to \infty$, because $|\Lambda_k^{(r_i)}|$ is subexponential in $B_k$, and $\Pr(\nu_{B_k} \notin \Lambda_k^{(r_i)}) \to 0$ as $k \to \infty$.
Thus, we have just shown the following.

\begin{cor} \label{synccor}
It is possible to achieve the minimum energy-per-bit of the asynchronous diamond network in Figure \ref{netfig} with codes where each relay is either synchronized or remains silent.
\end{cor}


%

In the remainder of this section, we show how Theorem \ref{mainthm} can be used to improve our lower bound, and, in section \ref{syncsection}, we prove Theorem \ref{mainthm}.
We will need some facts related to the capacity of a two-user degraded broadcast channel.
Let $C(P)$ be the capacity region of a degraded broadcast channel $\MC{X}{Y_1}{Y_2}$.
We know that this capacity region consists of all pairs $(R_1,R_2)$ such that
\aln{
& R_1 \leq I(X;Y_1 | U) \nonumber \\
& R_2 \leq I(U;Y_2), 
}
for some distribution $p(u,x)$ such that $E\left[\|X\|^2\right] \leq P$, where $R_2$ corresponds to the common rate, and $R_1$ to the private rate to the stronger user.
However, we will be interested in the multi-letter characterization of the same region; i.e., all pairs $(R_1,R_2)$ such that 
\al{ \label{mletter}
& R_1 \leq \tfrac1n I(X^n;Y_1^n | U) \nonumber \\
& R_2 \leq \tfrac1n I(U;Y_2^n), 
}
for some $n$ and some distribution $p(u,x^n)$  such that $E\left[\|X^n\|^2\right] \leq nP$.
An important quantity for us will be the $(1 : \gamma)$-capacity of this broadcast channel, which we define as
\aln{
C_{1:\gamma}(P) = \max \{R : (R,\gamma R) \in C(P)\},
}
for some $\gamma > 0$.
Using the multi-letter description of the capacity (\ref{mletter}), it is easy to see that we have
\al{ \label{cgamma}
C_{1:\gamma}(P) = \sup_{n, p(u,x^n) : E\|X^n\|^2 \leq nP} \min\left[ \frac{I(X^n;Y_1^n|U)}{n}, \frac{I(U;Y_2^n)}{\gamma n} \right].
}
Now we can state our new lower bound. 

\begin{theorem} \label{lboundthm}
The asynchronous minimum energy-per-bit for the network in Figure \ref{netfig} is lower bounded as
\aln{
e_b^{\min } \geq \min\left\{\text{\rm LB2},\gamma(1+\beta)\left( \frac1{g_1} + \frac{1}{h_1}\right)\right\}, 
}
where \text{\rm LB2} is the optimal solution to 
\al{
\textrm{\rm Maximize}\quad   &  \gamma\left[y_1\left(1+\beta+\beta \frac{g_1}{g_2}\right)+(1+\beta)( y_2 + y_3 + y_4 )\right]      \nonumber \\
\textrm{\rm subject to}\quad  & (g_1+g_2) y_1+ g_2 y_3+ g_1 y_4 \leq 1  \nonumber  \\
			\quad  & (h_1+h_2) y_2+h_1 y_3 \leq 1  \nonumber \\		
			\quad  & (h_1+h_2) y_2+h_2 y_4 \leq 1 \nonumber  \\
                     & y_1,y_2,y_3,y_4     \geq 0. \label{lpbound2}
}
\end{theorem}

\begin{proof}

First we assume that $\{\C_k\}_{k=1}^\infty$ falls into case (a) of Theorem \ref{mainthm}, and both relays are synchronized.
In this case, we will consider using code $\C_k$ on the degraded broadcast channel in Figure \ref{degbc}, in the synchronous setting.
\begin{figure}[ht] 
     \centering
       \includegraphics[height=42mm]{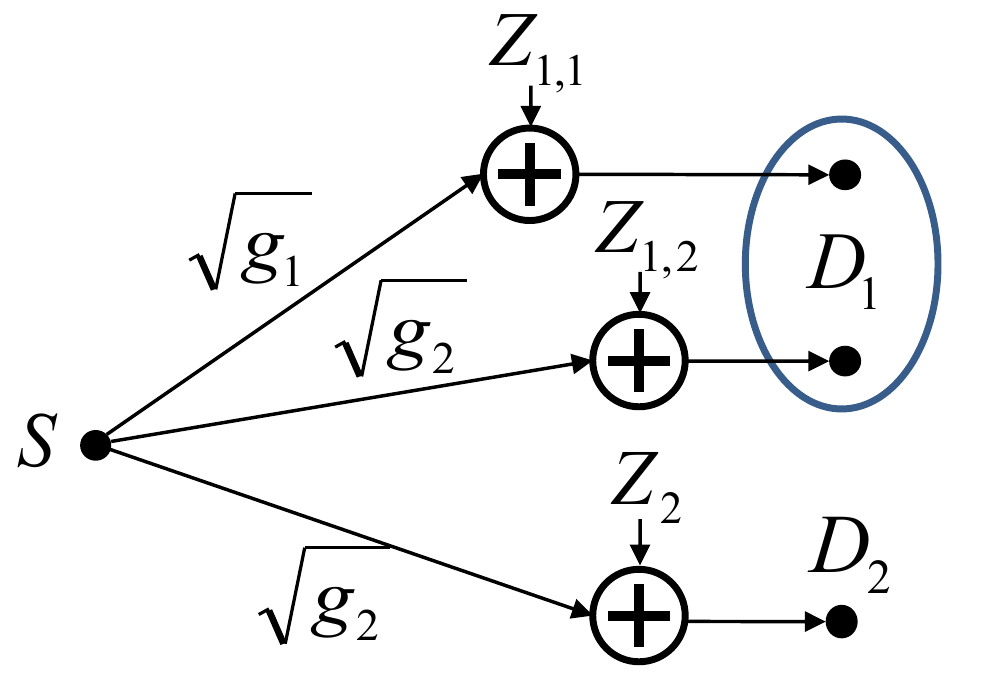} \caption{Degraded broadcast channel used in the proof of Theorem \ref{lboundthm}. \label{degbc}}
\end{figure}
Notice that, for this broadcast channel, $Y_2 = \sqrt{g_2} X + Z_2$ is a scalar, while $Y_1 = \left[\begin{smallmatrix} Y_{1,1}\\ Y_{1,2} \end{smallmatrix}\right] = \left[\begin{smallmatrix} \sqrt{g_1}X +Z_{1,1}\\ \sqrt{g_2}X + Z_{1,2} \end{smallmatrix}\right]$ is a vector.
When we consider using code $\C_k$ on this channel in the synchronous setting, we will have the source choosing an arrival time $\nu_{B_k}$ uniformly at random from $[1:2^{\beta B_k}]$ and transmitting the message as if we were in the asynchronous setting.
Notice that destination $D_1$ can simulate what the relays would have done in the diamond network, thus being able to simulate what the destination from the diamond network would have received.
This guarantees that, with probability $1-\ep_k$, where $\ep_k \to 0$, destination $D_1$ can decode $m$ correctly and output a list $[\tau-d_{B_k}:\tau]$ containing $\nu_{B_k}$.

We will let $X^{\tilde A_k}$ be the the random vector corresponding to the transmit signals of the source when using code $\C_k$ on the broadcast channel, and $Y_1^{\tilde A_k}$ and $Y_2^{\tilde A_k}$ be the corresponding outputs at $D_1$ and $D_2$ respectively.
Since we are assuming that relay $2$ is synchronized in the diamond network, destination $D_2$ will be synchronized here, which implies that
\al{ \label{syncproof}
\lim_{k \to \infty} \frac{H\left(\nu_{B_k} \left| Y_2^{\tilde A_k}\right.\right)}{B_k} = 0.
}
Now, using (\ref{cgamma}) with $U = \nu_{B_k}$, we obtain
\al{ \rescnt
C_{1:\beta}\left(E\left[\E_{\C_k}^{(s)}\right]/\tilde A_k\right) & \geq \frac{1}{\tilde A_k} \min\left[ I(X^{\tilde A_k};Y_1^{\tilde A_k}|\nu_{B_k}), I(\nu_{B_k};Y_2^{\tilde A_k})/\beta \right] \nonumber \\
& = \frac{1}{\tilde A_k} \min\left[ B_k - H(X^{\tilde A_k}|Y_1^{\tilde A_k},\nu_{B_k}), B_k - H\left(\nu_{B_k} \left| Y_2^{\tilde A_k}\right.\right)/\beta \right] \nonumber \\
& \geqnum \frac{1}{\tilde A_k} \min\left[  B_k - \left( H(\ep_k) + \ep_k B_k \right),B_k - H\left(\nu_{B_k} \left| Y_2^{\tilde A_k}\right.\right)/\beta \right], \label{capeval}
} \rescnt
where \cnt follows from Fano's inequality.
Next, we notice that the capacity $C(P)$ of the Gaussian degraded broadcast channel is known in closed-form, and in the case of Figure \ref{degbc}, it is comprised of all non-negative pairs $(R_1,R_2)$ satisfying
\aln{
& R_1 \leq \frac12 \log\left(1+{\alpha (g_1+g_2) P/N_0}\right) \\
& R_2 \leq \frac12 \log\left(1+\frac{(1-\alpha) g_2 P}{N_0+\alpha g_2 P}\right),
}
for some $\alpha \in [0,1]$.
It is then not difficult to see that the $(1:\beta)$-capacity of our broadcast channel can be expressed as
\aln{
C_{1:\beta}(P) = \max_{0 \leq \alpha \leq 1} \min \left[ \frac12 \log\left(1+\alpha (g_1+g_2) P/ N_0\right),  \frac{1}{2\beta}\log\left(1+\frac{(1-\alpha) g_2 P}{N_0+ \alpha g_2 P}\right)\right].
}
Then we have
\al{ 
\sup_{P>0} \frac{C_{1:\beta}(P)}{P} & \leq \max_{0 \leq \alpha \leq 1} \min \left[ \sup_{P > 0} \frac{\log\left(1+{\alpha (g_1+g_2) P/N_0}\right)}{2 P}, \sup_{P > 0} \frac{\log\left(1+\frac{(1-\alpha) g_2 P}{N_0+ \alpha g_2 P}\right)}{2 \beta P} \right] \nonumber \\
& = \max_{0 \leq \alpha \leq 1} \min \left[ \frac{\alpha (g_1+g_2)}{\gamma}, \frac{(1-\alpha) g_2}{\beta \gamma} \right] \nonumber \\
& = \frac{g_2 ( g_1 + g_2 )}{\gamma \left[ \beta(g_1 + g_2) + g_2\right]}. \label{supeval}
}
Now, by combining (\ref{capeval}) and (\ref{supeval}), we conclude that
\aln{
\frac{g_2 ( g_1 + g_2 )}{\gamma \left[ \beta(g_1 + g_2) + g_2\right]} & \geq \sup_{P>0} \frac{C_{1:\beta}(P)}{P} \geq \frac{C_{1:\beta}\left(E\left[\E_{\C_k}^{(s)}\right]/\tilde A_k\right)}{E\left[\E_{\C_k}^{(s)}\right]/\tilde A_k} \nonumber \\
& \geq \frac{\frac{1}{\tilde A_k} \min\left[ B_k - \left( H(\ep_k) + \ep_k B_k \right),  B_k - H\left(\nu_{B_k} \left| Y_2^{\tilde A_k}\right.\right)/\beta\right]}{E\left[\E_{\C_k}^{(s)}\right]/\tilde A_k} \nonumber \\
& = \frac{\min\left[ 1 - \left( H(\ep_k)/B_k + \ep_k \right),  1 - H\left(\nu_{B_k}\left| Y_2^{\tilde A_k}\right.\right)/(\beta B_k)\right]}{E\left[\E_{\C_k}^{(s)}\right]/B_k}.
}
Finally, by taking the $\limsup$ when $k \to \infty$ and using (\ref{syncproof}), we obtain
\al{
& \frac{g_2 ( g_1 + g_2 )}{\gamma \left[ \beta(g_1 + g_2) + g_2\right]} \geq \limsup_{k \to \infty}\frac{B_k}{E\left[\E_{\C_k}^{(s)}\right]} \nonumber \\
& \Rightarrow \liminf_{k\to\infty} {E\left[\E_{\C_k}^{(s)}\right]}/{B_k} \geq \frac{\gamma \left[ \beta(g_1 + g_2) + g_2\right]}{g_2 ( g_1 + g_2 )} = \gamma \left(\frac{\beta}{g_2} + \frac{1}{g_1+g_2} \right). \label{senergy2}
}
For the other three cuts in the network, we use the same 
analysis that we used in the proof of Theorem \ref{simplebounds} to obtain (\ref{renergy}), (\ref{sr1energy}) and (\ref{sr2energy}).
Then, by following very similar steps to those in the proof of Theorem \ref{simplebounds}, we conclude that a lower bound to
\aln{
\liminf_{k \to \infty} \frac{E \left[\E_{\C_k}\right]}{B_k}
}
is given by the optimal solution to the linear program
\al{
\textrm{Minimize}\quad   &  x_s + x_{r_1} + x_{r_2}      \nonumber \\
\textrm{subject to}\quad  & (g_1+g_2) x_s \geq \gamma \left[1+\beta\left(\frac{g_1+g_2}{g_2}\right)\right]  \nonumber  \\
			\quad  & (h_1+h_2) (x_{r_1}+x_{r_2}) \geq \gamma (1+\beta) \nonumber \\		
			\quad  & g_2 x_s +h_1 x_{r_1} \geq \gamma (1+\beta) \nonumber  \\
			\quad  & g_1 x_s +h_2 x_{r_2} \geq \gamma (1+\beta) \nonumber \\
			\quad  & x_s,x_{r_1},x_{r_2} \geq 0. \label{primallp2}
}
Then, by taking the dual of (\ref{primallp2}) we obtain (\ref{lpbound2}), which concludes the proof in the case where both relays are synchronized.

If the sequence of codes falls into case (b) of Theorem \ref{mainthm}, we may assume that only relay $1$ is synchronized an relay $2$ is silent. 
Then the analysis is much simpler.
We essentially have two concatenated point-to-point asynchronous AWGN channels, in which case the asynchronous minimum energy-per-bit is exactly given by
\al{ \label{2silentbound}
e_b^{\min } = \gamma (1+\beta)\left( \frac{1}{g_1} + \frac{1}{h_1} \right),
}
and the theorem follows.
\end{proof}

\section{Implications of the Main Results} \label{implicationssection}

The result in Theorem \ref{lboundthm} allows us to characterize the asynchronous minimum energy-per-bit of the diamond network to within a constant ratio, which ranges from $2$ in the synchronous case to $1$ in the highly asynchronous case.
First notice that, if we just use relay $1$ (the stronger relay), we can achieve energy-per-bit $\gamma (1+\beta)\left( \frac{1}{g_1} + \frac{1}{h_1} \right)$.
Therefore, in cases where
\al{ \label{onerelayoptimal}
\gamma (1+\beta)\left( \frac{1}{g_1} + \frac{1}{h_1} \right) \leq \text{LB2}, 
}
the optimal strategy for the two-relay diamond network is to just use relay $1$.
In these cases, there is no gap between upper and lower bound.
In cases where
\al{ \label{notonerelayoptimal}
\gamma (1+\beta)\left( \frac{1}{g_1} + \frac{1}{h_1} \right) > \text{LB2}, 
}
it is clear that $\text{LB2}$ is a lower bound on the asynchronous minimum energy-per-bit for \emph{any} sequence of codes (independent of which relays are synchronized).
Therefore, if we compare it to the simple upper bound from Theorem \ref{separationachievable}, and using the feasible solution to (\ref{lpbound2}) $y_1 =(g_1+g_2)^{-1},y_2= (h_1+h_2)^{-1},y_3 = 0,y_4 =0$, the gap will satisfy
\aln{ 
& (1+\beta) \gamma\left(\frac{1}{g_2}+\frac{1}{h_1+h_2}\right)-\text{LB2} \nonumber \\
& \quad \quad \leq 
(1+\beta) \gamma\left(\frac{1}{g_2}+\frac{1}{h_1+h_2}\right)-\gamma\left(\frac{\beta}{g_2} + \frac{1}{g_1+g_2} +\frac{1+\beta}{h_1+h_2}\right) \nonumber \\
& \quad \quad = \gamma \left( \frac{1}{g_2} - \frac{1}{g_1+g_2} \right).
}
Notice that this gap does not depend on $\beta$ anymore.
Therefore, we conclude that the separation-based scheme, although suboptimal, has a performance that does not become worse for large $\beta$; it in fact becomes relatively better.
\begin{figure*}[ht] 
     \centering
    \subfigure[$\beta = 0.8$]{
       \includegraphics[height=55mm]{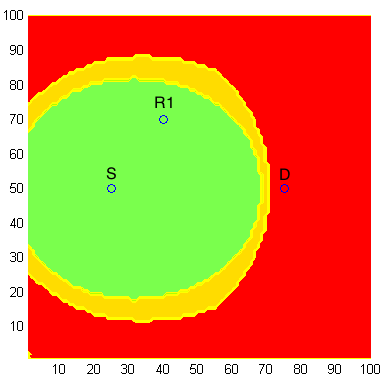}}
    \hspace{7mm}
    \subfigure[$\beta = 1$]{
       \includegraphics[height=55mm]{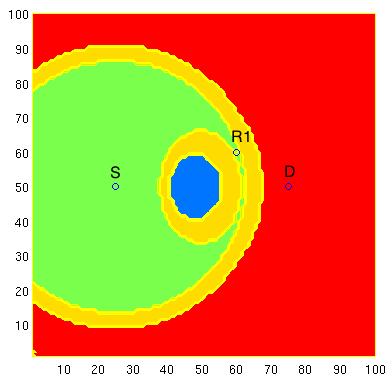}}
    \subfigure[$\beta = 0.2$]{
       \includegraphics[height=55mm]{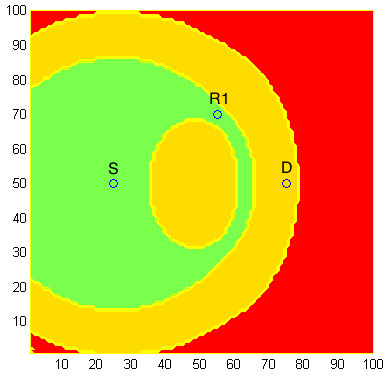}} 
    \hspace{7mm}
    \subfigure[$\beta = 0.5$]{
       \includegraphics[height=55mm]{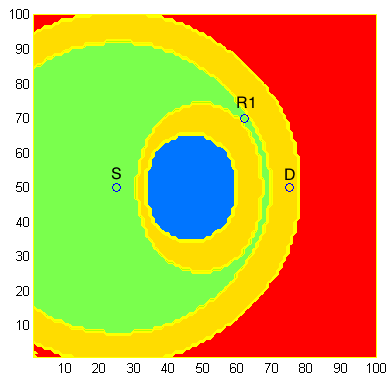}}
     \caption{Optimal relay selection for a fixed position of source, destination and relay 1, and a varying position of relay 2. We show in green the positions of relay $2$ for which it would be optimal to use both relays, in red the region where it would be optimal to use only relay $1$, in blue the region where it would be optimal to use only relay $2$, and in yellow the region for which our result does not provide an answer. The channel gains are assumed to be inversely proportional to the cube of the distance.\label{optrelays1}}
\end{figure*}
An important outcome of these results is that, for some values of $g_1,g_2,h_1$ and $h_2$, we can decide whether it is optimal to use one or both relays.
As we observed before, in the cases where (\ref{onerelayoptimal}) holds, it is optimal to only use relay $1$.
The intuition is that the cost of using relay $2$ is high, since it must be synchronized in order to be useful, and using it does not improve the achievable energy-per-bit.
On the other hand, in cases where our upper bound is below the lower bound when using only relay $1$, i.e., when
\al{ \label{twoerelaysoptimal}
 \frac{1}{g_1} + \frac{1}{h_1} \geq \frac{1}{g_2}+\frac{1}{h_1+h_2},
}
we know that the optimal strategy involves using both relays, and paying the price for synchronizing them.
The plots in Figure \ref{optrelays1} illustrate these results.
For a given value of $\beta$ and for fixed positions of the source, relay $1$ and the destination, we show in green the positions of relay $2$ for which it would be optimal to use both relays, in red the region where it would be optimal to use only relay $1$, in blue the region where it would be optimal to use only relay $2$, and in yellow the region for which our result does not provide an answer.
To create these plots, we assumed that the channel gains are proportional to the cube of the inverse of the distance.

Next, we consider the ratio $\text{upper-bound}/\text{lower-bound}$.
The upper-bound that we use is simply the minimum between a decode-and-forward scheme using only relay 1 and a decode-and-forward scheme using both relays, i.e.,
\al{ \label{achievedenergy}
\min\left\{\gamma(1+\beta)\left( \frac1{g_1} + \frac{1}{h_1} \right), \gamma(1+\beta)\left( \frac1{g_2} + \frac{1}{h_1+h_2}\right)\right\}.
}
As shown in \cite{asyncdiamondisit}, in the special case where the first hop of the diamond network is symmetric, i.e., when $g_1 = g_2 = g$, the upper and lower bounds are within a constant factor of each other.
To see this, notice that, if $g_1 = g_2 = g$, the upper-bound always reduces to $\gamma (1+\beta) (1/g + 1/(h_1+h_2))$.
Moreover, by computing the bound (\ref{lpbound2}) with $y_1 =(2g)^{-1},y_2= (h_1+h_2)^{-1},y_3 = 0,y_4 =0$, we obtain the lower bound of $\gamma (\beta/g + 1/(2g) + (1+\beta)/(h_1+h_2))$.
We then have
\aln{
\frac{\gamma (1+\beta)\left(\frac1g+\frac1{h_1+h_2}\right)}{\gamma \left(\frac{\beta}{g}+\frac1{2g}+\frac{1+\beta}{h_1+h_2}\right)} & \leq \frac{(1+\beta)\frac1g}{\frac{\beta}{g}+\frac{1}{2g}} = \frac{1+\beta}{\frac12 +\beta}.
}
Therefore, if $g_1 = g_2$, separation-based schemes achieve to within a factor of $(1+\beta)/(\frac12+\beta)$ from the minimum energy-per-bit.
This ratio equals $2$ when $\beta = 0$ (i.e., in the synchronous case) but it decreases towards $1$ as $\beta$ increases.

\begin{figure}[ht] 
     \centering
	\includegraphics[height=75mm]{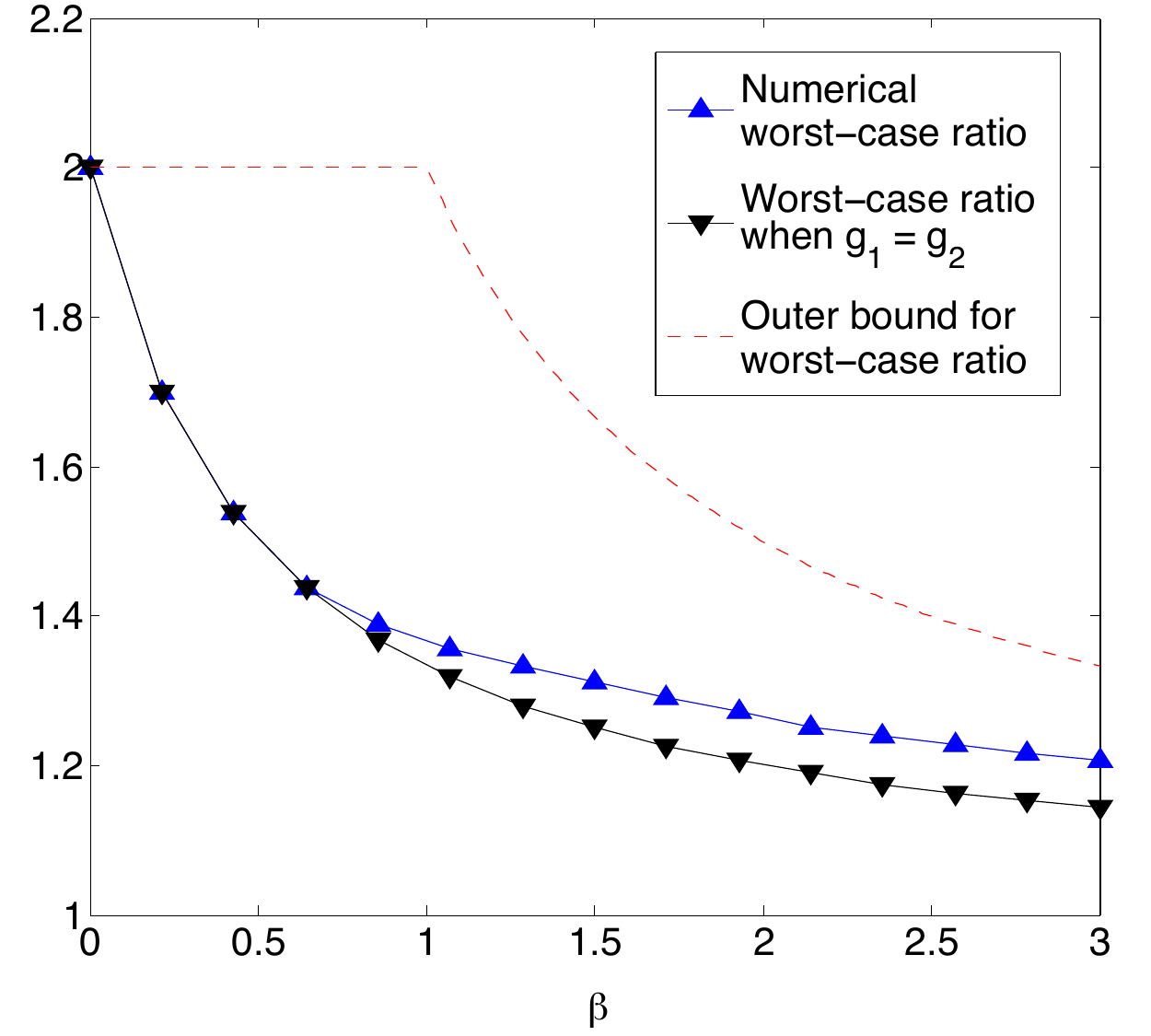}
          \caption{Worst-case upper bound to lower bound ratio.\label{ratiocurve}}
\end{figure}

In the general case, however, finding a good analytical bound on the worst-case ratio between the upper and lower bounds is not as easy.
As we noticed before, if (\ref{onerelayoptimal}) holds, then the gap between upper and lower bound is zero, and the ratio is one.
Therefore, we may assume that, for the worst-case ratio, (\ref{notonerelayoptimal}) holds, and by plugging $y_1 =(g_1+g_2)^{-1},y_2= (h_1+h_2)^{-1},y_3 = 0,y_4 =0$ into (\ref{lpbound2}), we have the lower bound
\aln{
\gamma\left(\frac{\beta}{g_2} + \frac{1}{g_1+g_2} +\frac{1+\beta}{h_1+h_2}\right). }
Then, an upper bound to the worst-case ratio is
\aln{
\frac{\gamma (1+\beta)\left(\frac1{g_2}+\frac1{h_1+h_2}\right)}{\gamma \left(\frac{\beta}{g_2}+\frac1{g_1+g_2}+\frac{1+\beta}{h_1+h_2}\right)} & \leq \frac{\frac{1+\beta}{g_2}}{\frac{\beta}{g_2}+\frac{1}{g_1+g_2}} \leq \frac{\frac{1+\beta}{g_2}}{\frac{\beta}{g_2}} = \frac{1+\beta}{\beta}.
}
This clearly shows that, as $\beta \to \infty$, the worst-case ratio tends to $1$.
However, this bound tends to infinity when $\beta \to 0$.
%
%
To verify that this is not the case for the worst-case ratio, we consider two regimes.
\begin{enumerate}[(i)]
\item $h_1 \leq g_2$: By considering only the second term in (\ref{achievedenergy}), and the lower bound provided by plugging \aln{y_1=0, y_2= \frac{1}{2(h_1+h_2)},y_3 = \frac12 \min \left(\frac1{h_1},\frac1{g_2}\right) = \frac{1}{2g_2},y_4 =0} into (\ref{lpbound2}), we can upper bound the worst-case ratio as
\aln{
\frac{\gamma (1+\beta)\left(\frac1{g_2}+\frac1{h_1+h_2}\right)}{\gamma (1+\beta) \left( \frac{1}{2g_2}+\frac{1}{2(h_1+h_2)}\right)} = 2.
}
\item $h_1 > g_2$: By considering only the first term in (\ref{achievedenergy}), and the lower bound provided by plugging \aln{y_1=\frac{h_1-g_2}{h_1(g_1+g_2)}, y_2= 0,y_3 = \min \left(\frac1{h_1},\frac1{g_2}\right) = \frac{1}{h_1},y_4 =0} into (\ref{lpbound2}), we can upper bound the worst-case ratio as
\aln{
& \frac{\gamma (1+\beta) \left(\frac1{g_1}+\frac1{h_1}\right)}{\gamma  \left[ \left(1+\beta +\beta \frac{g_2}{g_1}\right)\frac{h_1-g_2}{h_1(g_1+g_2)} + (1+\beta)\frac{1}{h_1}\right]}  \leq \frac{\frac1{g_1}+\frac1{h_1}}{ \frac{h_1-g_2}{h_1(g_1+g_2)} + \frac{1}{h_1} }  = \frac{ \frac1{g_1}+\frac1{h_1} }{\frac{h_1+g_1}{h_1(g_1+g_2)} } \\ 
& \quad \quad \quad \quad = \frac{\left(1+\frac{g_2}{g_1}\right) h_1 + g_1 + g_2}{h_1+g_1} = \frac{\left(1+\frac{g_2}{g_1}\right) (h_1 + g_1)}{h_1+g_1} = 1 + \frac{g_2}{g_1} \leq 2.
}
\end{enumerate}
We conclude that, in the worst case, the upper bound in (\ref{achievedenergy}) and the lower bound from Theorem \ref{lboundthm} are within a factor of $2$ of each other, and that this factor goes to $1$ as $\beta \to \infty$.
Since this bound is very crude, we considered finding the approximate worst-case ratio between the upper bound in (\ref{achievedenergy}) and the lower bound from Theorem \ref{lboundthm} numerically.
First we notice that for any choice of $g_1, g_2, h_1, h_2$, if we normalize all the channel gains by $\max(g_1,g_2,h_1,h_2)$ we obtain the same ratio between upper bound and lower bound.
Therefore, we may restrict our search for the worst-case ratio to the case where all channel gains lie in $[0,1]$.
Thus, we considered the ratio between upper bound and lower bound for $g_1,g_2,h_1,h_2 \in \{1/30,2/30,...,30/30\}$, and found the worst-case for several values of $\beta$.
We obtained the plot in Figure \ref{ratiocurve}.
%
This plot confirms that the worst-case ratio is uniformly upper bounded by $2$, and decreases to $1$ as $\beta$ increases.
The ratio decreases to $1$ faster than $1+1/\beta$, but not as fast as $(1+\beta)/(1/2+\beta)$ (the case where $g_1 = g_2$).

\section{Proof of Theorem \ref{mainthm}} \label{syncsection}

Our main objective in this section is to prove Theorem \ref{mainthm}.
The main idea is to show that, in order to achieve the minimum energy-per-bit on the network in Figure \ref{netfig}, 
there is no point in using a relay if it is not \emph{synchronized}, where the notion of a synchronized relay is formalized in Definition \ref{syncdef}.
%

Recall that, in our definition of what it means for a sequence of codes to achieve an energy-per-bit $e_b$ (Definition \ref{achievedef}), we require that code $\C_k$, which operates on a channel with arrival distribution $\nu_{B_k}$, transmits $B_k$ bits.
In this section, it will be useful to use an equivalent definition of achievable asynchronous energy-per-bit $e_b$.
Under the assumption that the distribution $\nu_{B_k}$ is uniform over $[1:2^{\beta B_k}]$, we obtain the following result, whose proof is in Appendix \ref{appflemma}.

\begin{lemma} \label{flemma}
Suppose we have a sequence of codes $\{\C_k\}_{k=1}^\infty$, where code $\C_k$ operates on a channel with uniform arrival distribution on $[1:2^{\beta B_k}]$ but only transmits $B_k - f(B_k)$ bits, with $f(\cdot) \geq 0$ and
\al{
\lim_{k \to \infty} \frac{f(B_k)}{B_k} = 0. \label{limf}
}
Suppose that, in addition, this sequence of codes satisfies the following:
\begin{itemize}
\item $\lim_{k \to \infty} \Pr\left(\error(\C_{k})\right) = 0$
\item $\lim_{k \to \infty} \frac{ \log d_{B_k}}{B_k-f(B_k)} = 0$
\item $\liminf_{k \to \infty} \frac{ E[\E_{\C_k}] }{B_k-f(B_k)} \leq e_b$
\end{itemize}
Then, for any $\eta > 0$, this sequence can be used to construct a new sequence of codes $\{\C_{k}'\}_{k=1}^{\infty}$, where code $\C_k'$ operates on a channel with uniform arrivals on $[1:2^{\beta B_k'}]$ and transmits $B_k'$ bits, satisfying
\begin{itemize}
\item $\lim_{k \to \infty} \Pr\left(\error(\C_{k}')\right) = 0$
\item $\lim_{k \to \infty} \frac{ \log d_{B_k}'}{B_k'} = 0$
\item $\liminf_{k \to \infty} \frac{ E[\E_{\C_k'}] }{B_k'} \leq (1+\eta) e_b,$
\end{itemize}
i.e., $\{\C_k'\}$ achieves an energy-per-bit $(1+\eta)e_b$ according to the original definition.
\end{lemma}

%
%
%

This Lemma allows us to regard the three conditions satisfied by the sequence of codes $\{\C_k\}_{k=1}^\infty$ in the statement of Lemma \ref{flemma} as an equivalent definition of what it means for a sequence of codes to achieve energy-per-bit $e_b$.

\begin{figure}[ht] 
     \centering
       \includegraphics[height=120mm]{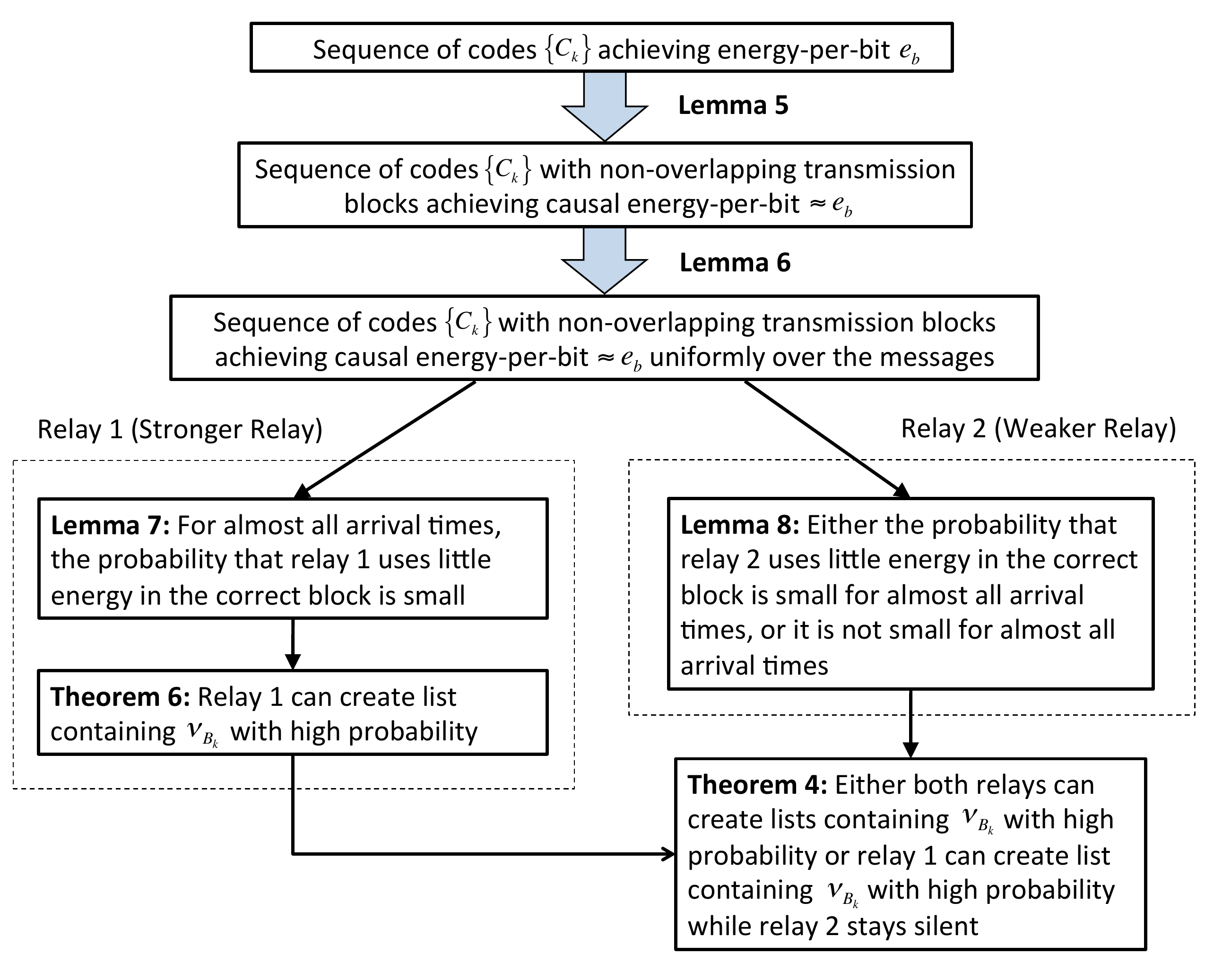} \caption{Main steps in the proof of Theorem \ref{mainthm}, and informal descriptions of main lemmas and theorems. \label{diagramfig}}
\end{figure}

In order to prove Theorem \ref{mainthm}, we will start with a sequence of asynchronous codes $\{\C_k\}$ achieving energy-per-bit $e_b$, and we will make several modifications to it, until we obtain another sequence of codes with the properties stated in the statement of the theorem.
These steps and the lemmas and theorems that construct the proof are summarized in the diagram in Figure \ref{diagramfig}.

Our first goal is to convert any given scheme into another scheme where the transmissions by the source are restricted to start at a few special ``transmission times'', and last at most $\ell$ time steps.
Then, if each pair of consecutive transmission times are separated by more than $\ell$ time steps, at a given time $t$, if the source is transmitting, there is only one possible starting time for the transmission block.
Intuitively, this will facilitate the relays' task.
We formally define this notion as follows.

\begin{definition} \label{nol}
An asynchronous code $\C_k$ is said to have non-overlaping transmission blocks if there is a set of times $\{t_1,t_2,...,t_q\}$ with $t_1 < t_2 < ... < t_q$ and a transmission block length $\ell_k$,
satisfying the following properties:
\begin{itemize}
\item $t_{i+1} - t_{i} \geq \ell_k+1$, for $i=1,...,q-1$ 
\item The interval $[1:A_k]$ is divided into $q$ subintervals $I_i = \left[\frac{(i-1)A_k}q+1:\frac{iA_k}q\right]$, $i=1,...,q$ (disregarding edge effects), such that $t_i > \frac{iA_k}q$, and if the message arrives at time $\nu_{B_k} \in I_i$, then the source keeps it buffered until the start of the block $[t_i:t_i + \ell_k -1]$, where the transmissions occur.
Moreover, given that the message arrived in $I_i$, the signals transmitted by the source during $[t_{i}:t_{i}+\ell_k-1]$ are only a function of the message value, and not the actual arrival time.
\item At a time $t \in [t_{i}:t_{i}+\ell_k-1]$, the relays only need the signals received in $[t_{i}:t]$ to compute their relaying functions, and the destination only needs the signals received in $[t_{i}:t]$ to apply its detecting/decoding functions.
If $t \notin [t_{i}:t_{i}+\ell_k-1]$ for $i=1,2,...,q$, then the relays stay silent, and the destination does not apply any detection/decoding function.
\end{itemize}
\end{definition}


We will in general refer to the set of transmission times of code $\C_k$ as $S_k$.
Notice that a non-overlapping transmission blocks scheme effectively induces a new message arrival distribution $\nueff_{B_k}$, where $\Pr(\nueff_{B_k} = t) = 1/|S_k|$ if $t \in S_k$ and $\Pr(\nueff_{B_k} = t) = 0$ otherwise.
Then we have the following key result, whose proof is in Appendix \ref{appnollem}.

\begin{lemma} \label{nollem}
Suppose we have a sequence of codes $\{\C_k\}_{k=1}^\infty$ achieving a finite energy-per-bit $e_b$ on the asynchronous diamond network in Figure \ref{netfig}.
Then we can build another sequence of codes $\{\C_k'\}$ with delay constraint $d_{B_k}'$ subexponential in $B_k' = B_k$, with non-overlapping transmission blocks of length $\ell_k$, for which 
\aln{
\liminf_{k \to \infty} \frac{E\left[\E_{\C_k'}[1:\nueff_{B_k}+\ell_k-1]\right]}{B_k} \leq (1+\eta)e_b,
}
for any arbitrarily small $\eta > 0$, and whose probability of error goes to $0$ as $k \to \infty$.
\end{lemma}
 
\noindent \emph{Remark 1:} The statement of Lemma \ref{nollem} does not require the sequence of codes $\{\C_k'\}$ to in fact achieve any energy-per-bit, according to Definition \ref{achievedef}.

\vspace{2mm}

\noindent \emph{Remark 2:} Notice that the energy spent in the code $\C_k'$ after time $\nueff_{B_k} + \ell_k-1$ is only spent by the relays. 
Therefore, if somehow the relays were able to decode $\nueff_{B_k}$ at time $\nueff_{B_k}+\ell_k - 1$, they could stop transmitting, and Theorem \ref{nollem} would imply that we have a sequence of codes with non-overlapping transmission blocks achieving arbitrarily close to energy-per-bit $e_b$.

We will now move toward our main result for the $2$-relay diamond network.
Since we will be frequently dealing with the sequence of codes constructed via Lemma \ref{nollem}, the following definition will be useful.

\begin{definition} A sequence of codes $\{\C_k\}_{k=1}^\infty$ with non-overlapping transmission blocks achieves a \emph{causal} energy-per-bit $e_b$, if it satisfies properties 1 and 2 in Definition \ref{achievedef}, and
\aln{
\liminf_{k \to \infty} \frac{\left[\E_{\C_k}[1:\nueff_{B_k}+\ell_k-1]\right]}{B_k} \leq e_b,
}
where $\ell_k$ is the transmission block length. \label{causaldef}
\end{definition}

\Remark
Consider a sequence of codes with non-overlapping transmission blocks that achieves a causal energy-per-bit $e_b$ with delay $d_{B_k}$ (which must be, according to Definition \ref{achievedef}, subexponential in $B_k$). 
Notice that a message that arrives in the first half of $I_i$, for any $i$, cannot be decoded with a delay smaller than $|I_i|/2 = A_k/(2|S_k|)$. 
Since the message arrives in the first half of some $I_i$ with probability $1/2$, and the error probability goes to $0$, Definition \ref{causaldef} implicitly requires that $d_{B_k} \geq A_k/(2|S_k|)$, for k sufficiently large. 
Therefore, since the delay $d_{B_k}$ must be subexponential in $B_k$, we must also have $A_k/|S_k|$ subexponential in $B_k$.
This fact will be used in subsequent proofs.

\vspace{2mm}

Lemma \ref{nollem} states that we can take any sequence of codes $\{\C_k\}_{k=1}^\infty$ achieving energy-per-bit $e_b$ and use it to build another sequence of codes $\{\C_k'\}$ which achieves a causal energy-per-bit arbitrarily close to $e_b$ and has non-overlapping transmission blocks.
Our first goal will be to show that any such sequence of codes can be converted into yet another sequence of codes, achieving the same causal energy-per-bit, where either both relays decode $\nueff_{B_k}$ exactly, or relay $1$ decodes $\nueff_{B_k}$ exactly and relay $2$ is not used at all.
This way, the relays that are actually used for communication can decode $\nueff_{B_k}$ and stop transmitting at time $\nueff_{B_k}+\ell_k-1$.
This will allow us to convert our sequence of codes that achieve causal energy-per-bit $e_b$, to a sequence of codes that in fact achieves energy-per-bit $e_b$.
In addition, this new sequence will have the property that any relay that is used must be synchronized.

In the process of proving Theorem \ref{mainthm}, an important step will be to restrict the set of messages that can be sent by a given code to only those with some special properties.
Because of that, it will be interesting that the energy spent by the code does not vary too much depending on the message that is sent.
This will allow us to restrict a code to only sending a certain subset of the messages without having the average energy-per-bit change much.

Consider a code $\C_k$ with non-overlapping transmission blocks.
Now, suppose that for each transmission time $t_i$ we have an injective mapping $\phi_{k,i} : \{1,...,M_k\} \to \{1,...,2^{B_k}\}$.
We will let $\phi_k$ represent the ensemble of all these mappings, i.e., $\phi_k = \{\phi_{k,1},\phi_{k,2},...,\phi_{k,|S_k|}\}$.
Then we have the following definition.

\begin{definition}
The restriction of code $\C_k$ according to $\phi_k$, denoted by $\C_k^{\phi}$, is a new code with message set $\{1,...,M_k\}$, which, given that message $m \in \{1,...,M_k\}$ (effectively) arrives at time $\nueff_{B_k} = t_i$, transmits the message $\phi_{k,i}(m)$ using code $\C_k$.
The destination applies the same decoder of code $\C_k$, and then uses $\phi_{k,i}^{-1}$ to decode $m$ (declaring an error if $\phi_{k,i}^{-1}$ does not map to any element in $\{1,...,M_k\}$).
\end{definition}

We will be interested in codes $\C_k$ for which any restriction yields a good code, as defined next.


\begin{definition}
A sequence of codes $\{\C_k\}_{k=1}^\infty$ with non-overlapping transmission blocks that achieves a causal energy-per-bit $e_b$ 
is said to achieve a causal energy-per-bit $e_b$ uniformly over the messages if 
for any sequence of message restrictions $\{\phi_k\}$ we have
\al{ \label{uniformeq}
\liminf_{k \to\infty} \frac{E\left[ \E_{\C_k^{\phi}}[1:\nueff_{B_k}+\ell_k-1]  \right]}{\log B_k} \leq e_b.
}
\end{definition}


Then we have the following Lemma, whose proof is in Appendix \ref{appuniflemma}.  

\begin{lemma} \label{uniflemma}
Suppose we have a sequence of codes $\{\C_k\}_{k=1}^\infty$ with non-overlapping transmission blocks achieving a causal energy-per-bit $e_b$ on the asynchronous diamond network in Figure \ref{netfig}.
Then we can have a sequence of codes $\{\C_k'\}$ that have non-overlapping transmission blocks, achieving a causal energy-per-bit $(1+\eta)e_b$ uniformly over the messages, for any $\eta > 0$.
\end{lemma}

In order to show our main result, i.e., that any relay that is used may be assumed to be synchronized, we first focus on relay $1$, which is the relay that has a stronger channel from the source. 

\begin{theorem} \label{listthm}
For any sequence of codes $\{\C_k\}_{k=1}^\infty$ for the asynchronous diamond network with non-overlapping transmission blocks  achieving a causal energy-per-bit $e_b$ uniformly over the messages, relay $1$ can create a list $\Lambda_k \subset S_k$ which contains $\nueff_{B_k}$ with vanishing error probability, and has a size $|\Lambda_k|$ that is subexponential in $B_k$.
Moreover, each $t$ in the list is added to it no later than at time $t + \ell_k - 1$.
\end{theorem}

\begin{proof}
Let $\E_{\C_k}^{(r_i)}[W(\nu_{B_k})]$ be the energy used by relay $i$ in the transmission block $W(\nu_{B_k}) = [\nu_{B_k}:\nu_{B_k}+\ell_k-1]$, when using code $\C_k$.
Consider a sequence of non-negative numbers $\{\ep_k\}$ for which $\ep_k \to 0$.
Let the set $\T(\alpha,\C_k,\ep_k)$ of arrival times be defined as
\al{
\T(\alpha,\C_k,\ep_k) & = \left\{ t \in S_k \st \Pr \left(\left. \frac{\E_{\C_k}^{(r_1)}[W(\nueff_{B_k})]}{B_k} \leq \alpha \text{ and } \frac{\E_{\C_k}^{(r_2)}[W(\nueff_{B_k})]}{B_k} \leq \alpha \, \right| \nueff_{B_k} = t \right) \leq \ep_k    \right\}. \label{one}
}
The proof of the following Lemma is in Appendix \ref{appalphaeplemma}.
\begin{lemma} \label{alphaeplemma}
There exists an $\alpha > 0$ and a non-negative sequence $\{\ep_k\}$ with $\ep_k \to 0$, such that
\al{
\limsup_{k \to \infty}  \Pr \left( \nueff_{B_k} \in \T(\alpha,\C_k,\ep_k)  \right) = 1. \label{limsup}
}
\end{lemma}
Notice that, for the $\alpha>0$ and the sequence $\{\ep_k\}$ provided by Lemma \ref{alphaeplemma}, we can actually replace the $\limsup$ in (\ref{limsup}) with a limit, 
since one can consider the subsequence of $\{\C_k\}_{k=1}^\infty$ for which the limit exists and is the $\limsup$ of the original sequence.
Therefore, we will assume that (\ref{limsup}) holds with $\limsup$ replaced by $\lim$.

We will show that, if (\ref{limsup}) holds, relay $1$ can implement a list detector $\Lambda_k$ for $\nu_{B_k}$ with probability of error going to $0$, and whose list size is subexponential in $B_k$.
But first we describe a scheme in which each relay $i$, $i=1,2$, implements a list decoder $\Lambda_{k,i}$ for $\nu_{B_k}$  based on its transmit signals $X_i(t)$, $t=1,...,A_k+d_{B_k}-1$, and we show that the probability that \emph{both} decoders make an error at the same time goes to $0$.
The list decoder $\Lambda_{k,i}$ 
for $\nueff_{B_k}$ selects the first $N_k \defi \frac{E\left[{\E_{\C_k}}[1:\nueff_{B_k}+\ell_k-1]\right]}{\alpha}$ transmission blocks where the energy consumed by relay $i$ is at least $\alpha B_k$, 
and lists the corresponding transmission times.
Let $\E_{\C_k}^{(r_i)}{[1:\nueff_{B_k}+\ell_k-1]}$ be the total energy consumed by relay $i$ up to time $\nueff_{B_k}+\ell_k-1$.
Notice that, if 
\al{
& \E_{\C_k}^{(r_i)}{[1:\nueff_{B_k}+\ell_k-1]} < B_k E\left[{\E_{\C_k}}[1:\nueff_{B_k}+\ell_k-1]\right] \text{ and} \label{exp1f} \\
& \E_{\C_k}^{(r_i)}{[W(\nueff_{B_k})]} > B_k \alpha, \label{exp2f}} 
then $\frac{\E_{\C_k}^{(r_i)}{[W(\nueff_{B_k})]}}{ \E_{\C_k}^{(r_i)}{[1:\nueff_{B_k}+\ell_k-1]}} > \frac{\alpha}{E\left[{\E_{\C_k}}[1:\nueff_{B_k}+\ell_k-1]\right]} = N_k^{-1}$.
Moreover, there can be at most $N_k$ transmission blocks $W(t_j)$ in $[1:\nueff_{B_k}+\ell_k-1]$ satisfying $\frac{\E_{\C_k}^{(r_i)}{[W(t_j)]}}{ \E_{\C_k}^{(r_i)}{[1:\nueff_{B_k}+\ell_k-1]}} > N_k^{-1}$, which implies that, if (\ref{exp1f}) and (\ref{exp2f}) are satisfied, the list decoder 
$\Lambda_k$ will be correct, i.e., 
$\nueff_{B_k} \in \Lambda_k$.
The probability of error at both decoders is then given by 
\al{ \rescnt
\Pr & (\text{error at $\Lambda_{k,1}$  and $\Lambda_{k,2}$}) = \Pr(\nueff_{B_k} \notin \Lambda_{k,1} \text{ and } \nueff_{B_k} \notin \Lambda_{k,2}) \nonumber \\
& \leq \Pr\left[ \left( \E_{\C_k}^{(r_1)}[W(\nueff_{B_k})] \leq \alpha B_k \cup  \E_{\C_k}^{(r_1)}{[1:\nueff_{B_k}+\ell_k-1]} \geq B_k E\left[{\E_{\C_k}}[1:\nueff_{B_k}+\ell_k-1]\right] \right) \right. \nonumber \\ 
& \quad \quad \quad \left. \cap \left( \E_{\C_k}^{(r_2)}[W(\nueff_{B_k})] \leq \alpha B_k \cup  \E_{\C_k}^{(r_2)}{[1:\nueff_{B_k}+\ell_k-1]} \geq B_k E\left[{\E_{\C_k}}[1:\nueff_{B_k}+\ell_k-1]\right]\right) \right] \nonumber \\
& \leqnum \Pr\left[ \left( \E_{\C_k}^{(r_1)}[W(\nueff_{B_k})] \leq \alpha B_k \cap \E_{\C_k}^{(r_2)}[W(\nueff_{B_k})] \leq \alpha B_k \right) \right. \nonumber \\
& \quad \quad \quad \cup \left( \E_{\C_k}^{(r_1)}{[1:\nueff_{B_k}+\ell_k-1]} \geq B_k E\left[{\E_{\C_k}}[1:\nueff_{B_k}+\ell_k-1]\right] \right) \nonumber \\
& \quad \quad \quad \left. \cup \left( \E_{\C_k}^{(r_2)}{[1:\nueff_{B_k}+\ell_k-1]} \geq B_k E\left[{\E_{\C_k}}[1:\nueff_{B_k}+\ell_k-1]\right] \right) \right] \nonumber \\ 
& \leq \Pr\left( \E_{\C_k}^{(r_1)}[W(\nueff_{B_k})] \leq \alpha B_k \cap \E_{\C_k}^{(r_2)}[W(\nueff_{B_k})] \leq \alpha B_k \right) \nonumber \\
& \quad \quad \quad + 2 \Pr\left( \E_{\C_k}{[1:\nueff_{B_k}+\ell_k-1]} \geq B_k E\left[{\E_{\C_k}}[1:\nueff_{B_k}+\ell_k-1]\right] \right) \nonumber \\
& \leqnum \Pr\left(\left.  \E_{\C_k}^{(r_1)}[W(\nueff_{B_k})] \leq \alpha B_k \cap \E_{\C_k}^{(r_2)}[W(\nueff_{B_k})] \leq \alpha B_k \, \right| \nueff_{B_k} \in \T(\alpha,\C_k,\ep_k) \right) \nonumber \\
& \quad \quad \quad + \Pr\left(\nueff_{B_k} \notin \T(\alpha,\C_k,\ep_k)\right)  + 2/B_k \nonumber \\
& \leqnum 2/B_k + \ep_k + \Pr\left(\nueff_{B_k} \notin \T(\alpha,\C_k,\ep_k)\right), \label{errorboundbeg} 
} \rescnt
where \cnt follows by noticint that if we have four events $A_1,A_2,B_1$ and $B_2$, then $(A_1 \cup B_1) \cap (A_2 \cup B_2)$ implies $(A_1 \cap A_2) \cup B_1 \cup B_2$,
\cnt follows from Markov's inequality and \cnt follows from the fact that 
\aln{
\Pr & \left(\left.  \E_{\C_k}^{(r_1)}[W(\nueff_{B_k})] \leq \alpha B_k \cap \E_{\C_k}^{(r_2)}[W(\nueff_{B_k})] \leq \alpha B_k \, \right| \nueff_{B_k} \in \T(\alpha,\C_k,\ep_k) \right) \\
 &= \sum_{t \in \T(\alpha,\C_k,\ep_k)} \Pr\left(\nueff_{B_k} = t \left| \nueff_{B_k} \in \T(\alpha,\C_k,\ep_k) \right. \right) \noindent \\
& \quad \quad \quad \quad\quad \quad \Pr \left(\left.  \E_{\C_k}^{(r_1)}[W(\nueff_{B_k})]/B_k \leq \alpha \text{ and } \E_{\C_k}^{(r_2)}[W(\nueff_{B_k})]/B_k \leq \alpha \, \right| \nueff_{B_k} = t \right)  \leq \ep_k.
}
Therefore, 
since $\Pr\left(\nueff_{B_k} \notin \T(\alpha,\C_k,\ep_k)\right) \to 0$, we have that $\Pr(\text{error at $\Lambda_{k,1}$  and $\Lambda_{k,2}$}) \to 0$.


Next, we want to use a similar argument to show that relay $1$ can implement by itself a list detector with a list size linear in $B_k$ and probability of error going to $0$ as $k \to \infty$.
In order to do that, notice that, since the channel gain from source to relay $1$, $g_1$, is stronger than the channel gain from source to relay $2$, $g_2$, relay $1$ can ``virtually'' simulate the received signal of relay $2$, and then simulate the output of the relaying functions of relay $2$, thus being able to implement the list decoder based on the transmit signals of relay $2$ as well.
To simulate the received signal at relay $2$, relay $1$ multiplies its received signal by $\sqrt{g_2/g_1}$ and then adds a Gaussian noise with  variance $1-g_2/g_1$ to it.
It is easy to see that the resulting signal has the same marginal statistics as the signal received at relay $2$.

Assume that $\hat X_2(t)$ is the signal that would be transmitted from relay $2$ at time $t$ according to relay $1$'s simulation.
Relay $1$ can use the list decoder $\Lambda_k = \Lambda_{k,1} \cup \hat \Lambda_{k,2}$, where $\hat \Lambda_{k,2}$ is the list decoder based on $\hat X_2(t)$, $t=1,...,A_k+d_{B_k}-1$.
Notice that $\hat X_2$ has the same distribution as $X_2$, but the joint distributions of $(X_1,X_2)$ and $(X_1,\hat X_2)$ are different, which is why the previous argument does not work to show that the error probability of this list decoder goes to $0$.
In particular, if we let $\hat \E_{\C_k}^{(r_2)}[W(\nueff_{B_k})]$ be the energy used in the simulated signal $\hat X_2$ in the window $W(\nueff_{B_k}) = [\nueff_{B_k},\nueff_{B_k}+d_{B_k}-1]$, we \emph{cannot} say that if $t \in \T(\alpha,\C_k,\ep_k)$ then
\aln{
\Pr \left(\left. \frac{\E_{\C_k}^{(r_1)}[W(\nueff_{B_k})]}{B_k} \leq \alpha \text{ and } \frac{\hat \E_{\C_k}^{(r_2)}[W(\nueff_{B_k})]}{B_k} \leq \alpha \, \right| \nueff_{B_k} = t \right) \leq \ep_k.
}
To solve this problem, we start by noticing that we can write, for $t \in \T(\alpha,\C_k,\ep_k)$,
\al{ \rescnt
 \ep_k & \geq \Pr \left(\left. \frac{\E_{\C_k}^{(r_1)}[W(\nueff_{B_k})]}{B_k} \leq \alpha \text{ and } \frac{\E_{\C_k}^{(r_2)}[W(\nu_{B_k})]}{B_k} \leq \alpha \, \right| \nueff_{B_k} = t \right) \nonumber \\
& \eqnum  \sum_{m=1}^{2^{B_k}} 2^{-{B_k}} \Pr \left(\left. \frac{\E_{\C_k}^{(r_1)}[W(\nueff_{B_k})]}{B_k} \leq \alpha \text{ and } \frac{\E_{\C_k}^{(r_2)}[W(\nueff_{B_k})]}{B_k} \leq \alpha \, \right| \nueff_{B_k} = t, m \text{ is sent} \right) \nonumber \\
& \eqnum \sum_{m=1}^{2^{B_k}} 2^{-{B_k}} \Pr \left(\left. \frac{\E_{\C_k}^{(r_1)}[W(\nueff_{B_k})]}{B_k} \leq \alpha \, \right| \nueff_{B_k} = t, m \text{ is sent} \right)  \\
& \quad \quad \quad \times \Pr \left(\left. \frac{\E_{\C_k}^{(r_2)}[W(\nueff_{B_k})]}{B_k} \leq \alpha \, \right| \nueff_{B_k} = t, m \text{ is sent} \right) \nonumber \\
& \geq \sum_{m=1}^{2^{B_k}} 2^{-{B_k}} \min_{j\in\{1,2\}} \left[ \Pr \left(\left. \frac{\E_{\C_k}^{(r_j)}[W(\nueff_{B_k})]}{B_k} \leq \alpha \, \right| \nueff_{B_k} = t, m \text{ is sent} \right) \right]^2, \label{bound1a}
} \rescnt
where \cnt follows from the independence of $\nueff_{B_k}$ and $m$, and \cnt follows from the fact that, given $\nueff_{B_k}$ and $m$, $X_1$ and $X_2$ are independent.
Now we notice that
\al{ \rescnt
\Pr & \left(\left. \frac{\E_{\C_k}^{(r_1)}[W(\nueff_{B_k})]}{B_k} \leq \alpha \text{ and } \frac{\hat \E_{\C_k}^{(r_2)}[W(\nueff_{B_k})]}{B_k} \leq \alpha \, \right| \nueff_{B_k} = t \right) \nonumber \\
& = \sum_{m=1}^{2^{B_k}} 2^{-{B_k}} \Pr \left(\left. \frac{\E_{\C_k}^{(r_1)}[W(\nueff_{B_k})]}{B_k} \leq \alpha \text{ and } \frac{\hat \E_{\C_k}^{(r_2)}[W(\nueff_{B_k})]}{B_k} \leq \alpha \, \right| \nueff_{B_k} = t, m \text{ is sent} \right) \nonumber \\
& \leq \sum_{m=1}^{2^{B_k}} 2^{-{B_k}} \min_{j\in\{1,2\}} \left[ \Pr \left(\left. \frac{\E_{\C_k}^{(r_j)}[W(\nueff_{B_k})]}{B_k} \leq \alpha \, \right| \nueff_{B_k} = t, m \text{ is sent} \right) \right]. \label{bound2a}
}
From the Cauchy-Schwarz inequality, for any numbers $a_1,...,a_M$, we have
\aln{
\frac1M \sum_{i=1}^M a_i \leq \sqrt{\frac1M \sum_{i=1}^M a_i^2}.
}
Thus, we can combine (\ref{bound1a}) and (\ref{bound2a}) to obtain
\al{ \rescnt
\Pr & \left(\left. \frac{\E_{\C_k}^{(r_1)}[W(\nueff_{B_k})]}{B_k} \leq \alpha \text{ and } \frac{\hat \E_{\C_k}^{(r_2)}[W(\nueff_{B_k})]}{B_k} \leq \alpha \, \right| \nueff_{B_k} = t \right)  \nonumber \\ 
& \leq \sqrt{\sum_{m=1}^{2^{B_k}} 2^{-{B_k}} \min_{j\in\{1,2\}} \left[ \Pr \left(\left. \frac{\E_{\C_k}^{(r_j)}[W(\nueff_{B_k})]}{B_k} \leq \alpha \, \right| \nueff_{B_k} = t, m \text{ is sent} \right) \right]^2} \leq \sqrt{\ep_k}. \label{bound3}
}

Now, using (\ref{bound3}), it is possible to repeat the same steps we used in (\ref{errorboundbeg}) to obtain 
\aln{
\Pr( \nueff_{B_k} \notin \Lambda_{k,1} \cup \hat \Lambda_{k,2}) \leq 2/{B_k} + \sqrt{\ep_k} + \Pr\left(\nueff_{B_k} \notin \T(\alpha,\C_k,\ep_k)\right),
}
and we conclude that $\Pr( \nueff_{B_k} \notin \Lambda_{k,1} \cup \hat \Lambda_{k,2}) \to 0$.
This implies that the list decoder implemented by relay $1$ alone, which has list size $2 d_{B_k} E[\E_{\C_k}]/ \alpha$ (which is subexponential in $B_k$), contains $\nu_{B_k}$ with vanishing error probability. 
%
%
%
\end{proof}

Notice that the result above implies that, in the case where $g_1 = g_2$, both relays can implement the list decoders for $\nueff_{B_k}$, and each of them will have vanishing error probability.

In Theorem \ref{listthm},
we learned that in any scheme that achieves a finite causal energy-per-bit, relay $1$ can approximately decode $\nu_{B_k}$ with vanishing error probability.
Next we address relay $2$, the weaker relay.
Similar to what we did in Theorem \ref{listthm}, we will define the set of arrival times
\al{
\T_2(\alpha,\C_k,\ep_k) & = \left\{ t_i \in S_k \st \Pr \left(\left. \frac{\E_{\C_k}^{(r_2)}[W(\nueff_{B_k})]}{B_k} \leq \alpha \, \right| \nueff_{B_k} = t_i \right) \leq \ep_k    \right\}, \label{two}
}
where $S_k$ is the set of transmission times of code $\C_k$ and $W(\nueff_{B_k}) = [\nueff_{B_k} : \nueff_{B_k} + \ell_k -1 ]$ is the transmission block associated to $\nueff_{B_k}$. 
As in Theorem \ref{listthm}, where we used Lemma \ref{alphaeplemma} to characterize the asymptotic behavior of $\Pr[\nueff_{B_k} \in \T(\alpha,\C_k,\ep_k)]$, we have the following result.


\begin{lemma} \label{lim10lem}
Suppose we have a sequence of codes $\{\C_k\}_{k=1}^\infty$ achieving a finite energy-per-bit $e_b$ on the asynchronous diamond network in Figure \ref{netfig}.
Consider any $\alpha > 0$ and any non-negative sequence $\{\ep_k\}$, with $\ep_k \to 0$.
Then, for any $\eta > 0$, we can have  a sequence of codes $\{\C_k'\}$ achieving a causal energy-per-bit $(1+\eta)e_b$ uniformly over the messages that have non-overlapping transmission blocks, 
and for which one of the following is true:
\begin{enumerate}
\item $\limsup_{k \to \infty} \Pr\left(\nueff_{B_k} \in \T_2(\alpha,\C_k,\ep_k)\right) = 1$,
\item $\liminf_{k \to \infty} \Pr\left(\nueff_{B_k} \in \T_2(\alpha,\C_k,\ep_k)\right) = 0$,
\end{enumerate}
where $\T_2(\alpha,\C_k,\ep_k)$ is defined in (\ref{two}).
\end{lemma}

Lemma \ref{lim10lem} will be the basis of the proof of Theorem \ref{mainthm}.
Intuitively, if a sequence of codes satisfies (a) 
then relay $2$ should be able to approximately decode the arrival time $\nu_{B_k}$.
Otherwise, if (b) is satisfied,
then we can find yet another sequence of codes which does not use relay $2$ and achieves the same energy-per-bit.
We can now prove Theorem \ref{mainthm}, which we restate here.

\vspace{3mm}
\begin{theoremrep}{\ref{mainthm}}
Suppose we have a sequence of codes $\{\C_k\}_{k=1}^\infty$ achieving a finite energy-per-bit $e_b$ on the asynchronous diamond network in Figure \ref{netfig}.
Then we can achieve arbitrarily close to the energy-per-bit $e_b$ with a sequence of codes $\{\C_k'\}$ for which one of the following is true:
\begin{enumerate}
\item Relay $i$, for $i=1,2$, can create a list $\Lambda_k^{(r_i)} \subset [1:A_k]$ based on their received signals, such that $\nu_{B_k} \in \Lambda_k^{(r_i)}$ with vanishing error probability and list size $|\Lambda_k^{(r_i)}|$ subexponential in $B_k$
\item Relay 1 can create a list $\Lambda_k^{(r_1)} \subset [1:A_k]$ based on its received signals, such that $\nu_{B_k} \in \Lambda_k^{(r_1)}$ with vanishing error probability and list size $|\Lambda_k^{(r_1)}|$ subexponential in $B_k$ and relay $2$ is inactive (i.e., does not transmit any signal)
\end{enumerate}
\end{theoremrep}

\begin{proof}
Fix some $\alpha > 0$ and some non-negative sequence $\{\ep_k\}$ with $\ep_k \to 0$.
We start by using Lemma \ref{lim10lem} in order to assume that our original sequence of codes $\{\C_k\}_{k=1}^\infty$ has non-overlapping transmission blocks of length $\ell_k$, achieves a causal energy-per-bit $e_b$ uniformly over the messages, and satisfies either
\al{&\lim_{k \to \infty} \Pr\left(\nueff_{B_k} \in \T_2(\alpha,\C_k,\ep_k)\right) = 1\text{, or} \label{lim1} \\
&\lim_{k \to \infty} \Pr\left(\nueff_{B_k} \in \T_2(\alpha,\C_k,\ep_k)\right) = 0. \label{lim2}}
Notice that the $\limsup$ and $\liminf$ in the statement of Lemma \ref{lim10lem} can be replaced by limits by simply restricting $\{C_k'\}$ to the corresoponding subsequences.
Also notice that, if the set of transmission times for the code $\C_k$ is given by $S_k$, our delay for $\{\C_k\}_{k=1}^\infty$ is at most $2\frac{A_k}{|S_k|} + \ell_k$, which must be subexponential in $B_k$.

We consider case (\ref{lim1}) first.
We follow very similar steps to those used when we created the list decoder for the relays in Theorem \ref{listthm}.
Relay $2$ will use its transmit signals to implement a list decoder $\Lambda_k$ 
for $\nueff_{B_k}$ with probability of error going to $0$, whose list size is subexponential in $B_k$.
Since each effective arrival $\nueff_{B_k}$ corresponds to exactly $\frac{A_k}{|S_k|}$ actual arrival times $\nu_{B_k}$, we see that the list decoder for $\nueff_{B_k}$ can then be converted to a list decoder for $\nu_{B_k}$ with a list $\frac{A_k}{|S_k|}$ times longer.
Since $\frac{A_k}{|S_k|}$ is subexponential in $B_k$, so is the size of the list for the resulting list decoder for the actual arrival time $\nu_{B_k}$.

The list decoder $\Lambda_k$ 
for $\nueff_{B_k}$ selects the first $N_k \defi \frac{E\left[{\E_{\C_k}}[1:\nueff_{B_k}+\ell_k-1]\right]}{\alpha}$ transmission blocks where the energy consumed by relay $2$ is at least $\alpha B_k$, 
and lists the corresponding transmission times.
Let $\E_{\C_k}^{(r_2)}{[1:\nueff_{B_k}+\ell_k-1]}$ be the total energy consumed by relay $2$ up to time $\nueff_{B_k}+\ell_k-1$.
Notice that, if 
\al{
& \E_{\C_k}^{(r_2)}{[1:\nueff_{B_k}+\ell_k-1]} < B_k E\left[{\E_{\C_k}}[1:\nueff_{B_k}+\ell_k-1]\right] \text{ and} \label{exp1} \\
& \E_{\C_k}^{(r_2)}{[W(\nueff_{B_k})]} > B_k \alpha, \label{exp2}} 
then $\frac{\E_{\C_k}^{(r_2)}{[W(\nueff_{B_k})]}}{ \E_{\C_k}^{(r_2)}{[1:\nueff_{B_k}+\ell_k-1]}} > \frac{\alpha}{E\left[{\E_{\C_k}}[1:\nueff_{B_k}+\ell_k-1]\right]} = N_k^{-1}$.
Moreover, there can be at most $N_k$ transmission blocks $W(t_i)$ in $[1:\nueff_{B_k}+\ell_k-1]$ satisfying $\frac{\E_{\C_k}^{(r_2)}{[W(t_i)]}}{ \E_{\C_k}^{(r_2)}{[1:\nueff_{B_k}+\ell_k-1]}} > N_k^{-1}$, which implies that, if (\ref{exp1}) and (\ref{exp2}) are satisfied, the list decoder 
$\Lambda_k$ will be correct, i.e., 
$\nueff_{B_k} \in \Lambda_k$. 
The probability of error of the list detector is thus given by
\al{ \rescnt
\Pr & (\nueff_{B_k} \notin \Lambda_k) \leq \Pr\left( \E_{\C_k}^{(r_2)}{[1:\nueff_{B_k}+\ell_k-1]} \geq B_k E\left[{\E_{\C_k}}[1:\nueff_{B_k}+\ell_k-1]\right] \cup \E_{\C_k}^{(r_2)}{[W(\nueff_{B_k})]}  \leq B_k \alpha\right)  \nonumber \\
& \leq \Pr\left( \E_{\C_k}^{(r_2)}{[W(\nueff_{B_k})]}  \leq B_k \alpha \right) + \Pr\left( \E_{\C_k}^{(r_2)}{[1:\nueff_{B_k}+\ell_k-1]} \geq B_k E\left[{\E_{\C_k}}[1:\nueff_{B_k}+\ell_k-1]\right] \right)  \nonumber \\
& \leq \Pr\left( \E_{\C_k}^{(r_2)}{[W(\nueff_{B_k})]}  \leq B_k \alpha \right) + \Pr\left( \E_{\C_k}{[1:\nueff_{B_k}+\ell_k-1]} \geq B_k E\left[{\E_{\C_k}}[1:\nueff_{B_k}+\ell_k-1]\right] \right)  \nonumber \\
& \leqnum \Pr\left(\left.  \E_{\C_k}^{(r_2)}{[W(\nueff_{B_k})]}  \leq B_k \alpha  \, \right| \nueff_{B_k} \in \T_2(\alpha,\C_k,\ep_k) \right) + \Pr\left(\nueff_{B_k} \notin \T_2(\alpha,\C_k,\ep_k)\right)  + 1/B_k \nonumber \\
& \leqnum 1/B_k + \ep_k + \Pr\left(\nueff_{B_k} \notin \T_2(\alpha,\C_k,\ep_k)\right), \label{errorbound} 
} \rescnt
where \cnt follows from Markov's inequality and \cnt follows from the fact that 
\aln{
\Pr & \left(\left.  {\E_{\C_k}^{(r_2)}[W(\nueff_{B_k})]} \leq \alpha {B_k}\, \right| \nueff_{B_k} \in \T_2(\alpha,\C_k,\ep_k) \right) \\
 &= \sum_{t \in \T_2(\alpha,\C_k,\ep_k)} \Pr\left(\nueff_{B_k} = t \left| \nueff_{B_k} \in \T_2(\alpha,\C_k,\ep_k) \right. \right) \Pr \left(\left. {\E_{\C_k}^{(r_2)}[W(\nueff_{B_k})]} \leq \alpha{B_k} \, \right| \nueff_{B_k} = t \right)  \leq \ep_k.
}
Since we are in case (\ref{lim1}), we have $\Pr\left(\nueff_{B_k} \notin \T_2(\alpha,\C_k,\ep_k)\right) \to 0$, and therefore $\Pr(\nueff_{B_k} \notin \Lambda_k) \to 0$.
Moreover, since $g_1 \geq g_2$, we know that, by adding some extra Gaussian noise to its received signal, relay 1 can simulate the received signals at relay $2$, and compute what the output of relay 2 would have been at each time $t$.
Thus, relay 1 can create a list decoder for $\nueff_{B_k}$ based on the simulated output of relay 2, and since it will be statistically equal to the actual list decoder from relay 2, its error probability will also tend to $0$ as $k \to \infty$.

Now, we need to take care of the fact that our codes only achieve causal energy-per-bit $e_b$.
To fix this, we will use the fact that both relays are approximately decoding the effective arrival time $\nueff_{B_k}$ (they have a list of subexponential size in $B_k$ containing $\nueff_{B_k}$ with high probability), to improve the coding scheme such that both relays can decode $\nueff_{B_k}$ exactly.
In order to do that, we will have the source transmitting a pulse after the transmission block. 
Define $U_k \defi 2 B_k e_b/\alpha$.
Notice that the fact that $\{\C_k\}_{k=1}^\infty$ achieves a causal energy-per-bit $e_b$ implies that $\liminf_{k \to \infty} \frac{N_k}{B_k} \leq \frac{e_b}{\alpha}$.
This, in turn, implies that for a subsequence of codes $\{\C_{k_j}\}$, $N_{k_j} \leq \frac{2 B_{k_j} e_b}{\alpha} = U_{k_j}$.
Therefore, we will restrict ourselves to this subsequence and drop the notation $k_j$ for simplicity.
We will have the source transmitting a pulse of magnitude 
\aln{
2\sqrt{\frac{4 \ln U_k}{g_2}},
}
at time $\nueff_{B_k} + \ell_k$.
Notice that this time was previously not used by the scheme due to the non-overlapping transmission blocks requirement that $t_{i+1} \geq t_i + \ell_k + 1$.
The relays, after adding a transmission time $t_i$ to the list, use a threshold detector at time $t_i + \ell_k$ with threshold $\sqrt{4 \ln U_k}$.
If a pulse is found at $t_i + \ell_k$, the relay declares $\hat \nueff_{B_k} = t_i$, and stops transmitting after that point. 
This way, we will be converting our scheme that achieves a causal energy-per-bit $e_b$ to a scheme that actually achieves an energy-per-bit $e_b$.
However, a further modification needs to be made, before we can bound the energy used by this code.
We let $L_k$ be the event that both relays correctly detect the pulse, thus decoding $\nueff_{B_k}$ correctly. 
We also let $\Lambda_k^{(r_i)}$ be the list decoder from relay $i$.
Then we have
\aln{
\Pr(\ol{L_k}) & = \Pr( \nueff_{B_k} \notin \Lambda_k^{(r_1)} \cup \nueff_{B_k} \notin \Lambda_k^{(r_2)}) + \Pr( \text{error in pulse detection} \cap \nueff_{B_k} \in \Lambda_k^{(r_1)} \cap \nueff_{B_k} \in \Lambda_k^{(r_2)} ) \nonumber \\
& \leq 2 \Pr( \nueff_{B_k} \notin \Lambda_k^{(r_2)}) + \Pr( \text{error in pulse detection} | \nueff_{B_k} \in \Lambda_k^{(r_1)} \cap \nueff_{B_k} \in \Lambda_k^{(r_2)}),
}
and, from (\ref{errorbound}), we know that the first term tends to $0$ as $k \to \infty$.
For the second term, we have
\al{
&\Pr( \text{error in pulse detection} | \nueff_{B_k} \in \Lambda_k^{(r_1)} \cap \nueff_{B_k} \in \Lambda_k^{(r_2)})  \nonumber \nonumber \\
& \quad \quad \leq \Pr  \left[\exists \, t \in \Lambda_k^{(r_i)}, i =1 \text{ or } i = 2,  t\ne \nueff_{B_k} \st Y_i(t+\ell_k) \geq \sqrt{4 \ln U_k} \right] \nonumber \\
& \quad \quad \quad \quad + \Pr\left[ Y_i(\nueff_{B_k}+\ell_k) < \sqrt{4 \ln U_k}, i \in \{1,2\} \right] \nonumber\\ 
& \quad \quad \leq 2 |\Lambda_k| \Pr \left[Z \geq \sqrt{4 \ln U_k} \right] + 2\Pr\left[ Z < - \sqrt{4 \ln U_k} \right] \nonumber \\ 
& \quad \quad \leq 2 U_k e^{-2 \ln U_k} + 2e^{-2 \ln U_k} = 2 ( U_k^{-1} + U_k^{-2}), \label{errorpulse}
}
and we conclude that $\Pr(\ol{L_k}) \to 0$ as $k \to \infty$.
Then we define $\gamma_k \defi \Pr(\ol{L_k})$, and we will have both relays stay silent in the last $\sqrt{\gamma} |S_k|$ transmission blocks, where $S_k$ is the set of transmission times.
It is easy to see that the probability of error of the resulting code still goes to $0$ as $k \to \infty$.
Since the relays stop transmitting after detecting a pulse at time $t_i + \ell_k$ for $t_i \in \Lambda_k$, we can now bound the energy used by the resulting code $\C_k'$ as
\al{ \rescnt
& E\left[\E_{\C_k'}\right] \leqnum  E\left[ \E_{\C_k}[1:\nueff_{B_k}+\ell_k-1]\right]+ E\left[ \E_{\C_k'}{[\nueff_{B_k}+\ell_k:t_{(1-\sqrt{\gamma_k})|S_k|-1}+\ell_k-1]}\right] \nonumber \\
& \eqnum  E\left[ \E_{\C_k}{[1:\nueff_{B_k}+\ell_k-1]}\right]+ \frac{16 \ln U_k}{g_2} + E\left[ \left. \E_{\C_k'}{[\nueff_{B_k}+\ell_k:t_{(1-\sqrt{\gamma_k})|S_k|-1}+\ell_k-1]}\right| \ol{L_k} \right] \Pr(\ol{L_k}) \nonumber \\
& \leqnum E\left[ \E_{\C_k}{[1:\nueff_{B_k}+\ell_k-1]}\right]+ \frac{16 \ln U_k}{g_2} + E\left[ \left.\E_{\C_k}{[1:\nueff_{B_k}+\ell_k-1]}\right|  \nueff_{B_k} \geq t_{(1-\sqrt{\gamma_k})|S_k|} \right] \gamma_k \nonumber \\
& \leq E\left[ \E_{\C_k}{[1:\nueff_{B_k}+\ell_k-1]}\right]+ \frac{16 \ln U_k}{g_2} + \gamma_k \frac{E\left[ \E_{\C_k}{[1:\nueff_{B_k}+\ell_k-1]}\right]}{\Pr(\nueff_{B_k} \geq t_{(1-\sqrt{\gamma_k})|S_k|})}  \nonumber \\
& = (1+\sqrt{\gamma_k}) E\left[ \E_{\C_k}{[1:\nueff_{B_k}+\ell_k-1]}\right]+ \frac{16 \ln U_k}{g_2}, 
\label{energypulse}
} \rescnt
where \cnt follows because, up to time $\nueff_{B_k}+\ell_k-1$, the energy used by code $\C_k'$ is the same as the energy used by $\C_k$ unless a pulse is incorrectly detected, in which case it is less;
\cnt follows because the energy spent from time $\nueff_{B_k}+\ell_k$ on is the energy used in the pulse and then either $0$ if the pulse is detected, or the energy that would be spent otherwise; \cnt follows from the fact that $\C_k$ has non-overlapping transmission blocks, and, if the pulse is missed, $\C_k'$ behaves as $\C_k$ would have behaved if the message had not arrived yet.
Now, since 
the sequence of codes $\{\C_k\}_{k=1}^\infty$ achieves a causal energy-per-bit $e_b$, it is easy to see that
\aln{
\liminf_{k \to \infty} \frac{E\left[\E_{\C_k'}\right]}{B_k} \leq e_b,
}
which means that $\C_k'$ achieves an energy-per-bit $e_b$ with both relays decoding the effective arrival time $\nueff_{B_k}$ exactly.
Therefore, this decoder for $\nueff_{B_k}$ can be converted into a list decoder for $\nu_{B_k}$ with a list of size $\frac{A_k}{|S_k|}$ which is subexponential in $B_k$.

The previous arguments imply that if, for some $\alpha > 0$ and some non-negative sequence $\{\ep_k\}$ with $\ep_k \to 0$, we have (\ref{lim1}), then the sequence of codes $\{\C_k\}_{k=1}^\infty$ can be converted into another sequence of codes achieving the same energy-per-bit where both relays can have a list decoder $\Lambda_k$ for $\nueff_{B_k}$ (where $|\Lambda_k|$ is subexponential in $B_k$) with vanishing error probability.
In this case, we fall into case (a).

Therefore, for case (b) we only need to consider sequences of codes $\{C_k\}$, such that for \emph{all} $\alpha>0$ and \emph{all} non-negative sequences $\{\ep_k\}$ with $\ep_k \to 0$, the sequence of codes built according to Lemma \ref{lim10lem} satisfies
(\ref{lim2}).
Thus, we assume that, for any $\alpha > 0$ and any $\{\ep_k\}$ with $\ep_k \to 0$, we have a sequence of codes $\{\C_k\}_{k=1}^\infty$ achieving the same energy-per-bit $e_b$ for which (\ref{lim2}) holds.

Our main modification will be to restrict the messages that our code $\C_k$ can send.
First we consider the set 
\aln{
\Ls_2(\alpha,\C_k,\ep_k) = \left\{ t \notin \T_2(\alpha,\C_k,\ep_k) \st \Pr \left(\error(\C_k) \left| \nueff_{B_k} = t \right.\right) \leq \sqrt{\Pr \left(\error(\C_k)\right)} \right\}.
}
To simplify the notation, we will refer to the sets $\T_2(\alpha,\C_k,\ep_k)$ and $\Ls_2(\alpha,\C_k,\ep_k)$ by simply $\T_2$ and $\Ls_2$.
We notice that
\aln{
\Pr\left(\error(\C_k)\right) & \geq \Pr\left(\error(\C_k)\left|  \nueff_{B_k}  \notin \T_2 \right.  \right) \Pr\left( \nueff_{B_k}  \notin \T_2 \right) \nonumber \\
& \geq \Pr\left(\error(\C_k)\left|  \nueff_{B_k}  \notin \T_2, \nueff_{B_k}  \notin \Ls_2 \right.  \right) \Pr\left( \nueff_{B_k}  \notin \Ls_2 | \nueff_{B_k}  \notin \T_2 \right) \Pr\left( \nueff_{B_k}  \notin \T_2 \right) \nonumber \\
& \geq   \sqrt{\Pr \left(\error(\C_k)\right)} \Pr\left( \nueff_{B_k}  \notin \Ls_2 | \nueff_{B_k}  \notin \T_2 \right) \Pr\left( \nueff_{B_k}  \notin \T_2 \right),
}
which implies that
\al{
\Pr\left(\nueff_{B_k} \notin \Ls_2 \right) & = \Pr\left(\nueff_{B_k} \notin \Ls_2 | \nueff_{B_k}  \notin \T_2  \right) \Pr(\nueff_{B_k}  \notin \T_2) + \Pr\left(\nueff_{B_k} \notin \Ls_2 | \nueff_{B_k}  \in \T_2  \right) \Pr(\nueff_{B_k}  \in \T_2) \nonumber \\
& \leq \sqrt{\Pr \left(\error(\C_k)\right)} +  \Pr(\nueff_{B_k}  \in \T_2). \label{l2bound}
}
From (\ref{lim2}), (\ref{l2bound}) and the fact that $\Pr\left( \error(\C_k) \right) \to 0$ as $k \to \infty$, we conclude that 
\al{
\lim_{k \to \infty}\Pr\left(\nueff_{B_k}\in \Ls_2(\alpha,\C_k,\ep_k)\right) = 1. \label{liml2}
}
We will now use the set $\Ls_2(\alpha,\C_k,\ep_k)$ to define the messages that can be sent by our code.
Since the sequence $\{\ep_k\}$ with $\ep_k \to 0$ can be chosen arbitrarily, we fix it to be $\ep_k = \max(1/B_k,\xi_k^{1/8})$, where we define $\xi_k \defi \Pr\left( \error(\C_k) \right)$.
If the set of messages for code $\C_k$ is $\M = \{1,2,...,{2^{B_k}}\}$, then we will let, for each $t \in \Ls_2(\alpha,\C_k,\ep_k)$,
\al{
\M_t = \left\{ m \in \M \left| \Pr \left(\left. \frac{\E_{\C_k}^{(r_2)}[W(\nueff_{B_k})]}{B_k} \leq \alpha \, \right| \nueff_{B_k} = t, \text{$m$ is sent} \right) > \frac{\ep_k}{2} \right. \right\}. \label{boundenergy}
}
In order to lower bound the size of $\M_t$, we notice that, for  $t \in \Ls_2(\alpha,\C_k,\ep_k) \subset S_k \setminus \T_2(\alpha,\C_k,\ep_k)$,
\aln{
\ep_k & < \Pr \left(\left. \E_{\C_k}^{(r_2)}[W(\nueff_{B_k})] \leq \alpha B_k \, \right| \nueff_{B_k} = t \right) \nonumber \\
& = \sum_{m=1}^{2^{B_k}} 2^{-B_k} \Pr \left(\left. \E_{\C_k}^{(r_2)}[W(\nueff_{B_k})] \leq \alpha B_k \, \right| \nueff_{B_k} = t, \text{$m$ is sent} \right) \nonumber \\
& = \sum_{m \in \M_t} 2^{-B_k} \Pr \left(\left. \E_{\C_k}^{(r_2)}[W(\nueff_{B_k})] \leq \alpha B_k \, \right| \nueff_{B_k} = t, \text{$m$ is sent} \right) \nonumber \\ 
& \quad \quad + \sum_{m \notin \M_t} 2^{-B_k} \Pr \left(\left. \E_{\C_k}^{(r_2)}[W(\nueff_{B_k})] \leq \alpha B_k \, \right| \nueff_{B_k} = t, \text{$m$ is sent} \right) \nonumber \\
& < |\M_t| 2^{-B_k} + \left(2^{B_k} - |\M_t| \right) 2^{-B_k} \frac{\ep_k}2 \leq |\M_t| 2^{-B_k}  + \frac{\ep_k}{2}. \nonumber
}
Therefore, we have
\al{ \label{mt}
|\M_t| \geq 2^{B_k} \frac{\ep_k}{2}.  
}
Next we notice that for any $t \in \Ls_2(\alpha,\C_k,\ep_k)$, we have
\al{
\xi_k^{1/2} & \geq \Pr \left(\error(\C_k) \left| \nueff_{B_k} = t \right.\right) \nonumber \\
& \geq  \Pr \left(\error(\C_k) \left| \E_{\C_k}^{(r_2)}[W(\nueff_{B_k})] \leq \alpha B_k, \nueff_{B_k} = t \right.\right) \Pr \left(\left. \E_{\C_k}^{(r_2)}[W(\nueff_{B_k})] \leq \alpha B_k \, \right| \nueff_{B_k} = t \right) \nonumber \\
& >  \Pr \left(\error(\C_k) \left| \E_{\C_k}^{(r_2)}[W(\nueff_{B_k})] \leq \alpha B_k, \nueff_{B_k} = t \right.\right) \ep_k, 
}
which implies that
\al{
\Pr \left(\error(\C_k) \left| \E_{\C_k}^{(r_2)}[W(\nueff_{B_k})] \leq \alpha B_k, \nueff_{B_k} = t \right.\right) 
\leq \frac{\xi_k^{1/2}}{\ep_k} \leq \xi_k^{3/8}.
}
We can now write
\al{
\xi_k^{3/8} & \geq \Pr \left(\error(\C_k) \left| \E_{\C_k}^{(r_2)}[W(\nueff_{B_k})] \leq \alpha B_k, \nueff_{B_k} = t \right.\right) \nonumber \\
& \geq \sum_{m \in \M_t} \Pr \left(\error(\C_k) \left| \E_{\C_k}^{(r_2)}[W(\nueff_{B_k})] \leq \alpha B_k, \nueff_{B_k} = t, \text{$m$ is sent} \right.\right)  \nonumber \\
& \quad \quad \times \Pr \left( \text{$m$ is sent}  \left| \E_{\C_k}^{(r_2)}[W(\nueff_{B_k})] \leq \alpha B_k, \nueff_{B_k} = t \right.\right). \label{step1}
}
Then we notice that, for any $m \in \M_t$, for $t \in \Ls_2(\alpha,\C_k,\ep_k)$, we have
\al{ 
& \Pr \left( \text{$m$ is sent}  \left| \E_{\C_k}^{(r_2)}[W(\nueff_{B_k})] \leq \alpha B_k, \nueff_{B_k} = t \right.\right) \nonumber \\
& \quad \quad = \frac{\Pr \left( \left. \E_{\C_k}^{(r_2)}[W(\nueff_{B_k})] \leq \alpha B_k   \right| \text{$m$ is sent }, \nueff_{B_k} = t \right) \Pr \left( \left. \text{$m$ is sent }  \right|  \nueff_{B_k} = t \right)}{\Pr \left( \left. \E_{\C_k}^{(r_2)}[W(\nueff_{B_k})] \leq \alpha B_k \right| \nueff_{B_k} = t \right)} \geq \frac{\ep_k}{2} 2^{-B_k}. \label{step2}
}
Now, if we define
\aln{
\M_t' = \left\{ m \in \M_t \st \Pr \left(\error(\C_k) \left| \E_{\C_k}^{(r_2)}[W(\nueff_{B_k})] \leq \alpha B_k, \nueff_{B_k} = t, \text{$m$ is sent}\right.\right) \leq 8 \frac{\xi_k^{1/4}}{\ep_k} \right\},
}
we can use (\ref{step1}) and (\ref{step2}) to obtain
\aln{
\xi_k^{3/8} & \geq \frac{\ep_k}{2} 2^{-B_k} \sum_{m \in \M_t} \Pr \left(\error(\C_k) \left| \E_{\C_k}^{(r_2)}[W(\nueff_{B_k})] \leq \alpha B_k, \nueff_{B_k} = t, \text{$m$ is sent} \right.\right) \nonumber \\
& \geq \frac{\ep_k}{2} 2^{-B_k} \sum_{m \in \M_t \setminus \M_t'} \Pr \left(\error(\C_k) \left| \E_{\C_k}^{(r_2)}[W(\nueff_{B_k})] \leq \alpha B_k, \nueff_{B_k} = t, \text{$m$ is sent} \right.\right) \nonumber \\
& \geq \frac{\ep_k}{2} 2^{-B_k} |\M_t \setminus \M_t'| \, 8 \frac{\xi_k^{1/4}}{\ep_k}
\geq 2^{-B_k} |\M_t \setminus \M_t'| \, 4 \xi_k^{1/4}, \nonumber 
}
and, thus,
\aln{
& |\M_t \setminus \M_t'| \leq \frac{\xi_k^{3/8}}{4 \xi_k^{1/4}}2^{B_k} \leq 2^{B_k} \frac{\ep_k}{4} \quad \stackrel{(\ref{mt})}{\Longrightarrow} \quad |\M_t'| \geq 2^{B_k} \frac{\ep_k}{4} \geq \frac{2^{B_k}}{4B_k}.
}
Moreover, for any $m \in \M_t'$, we have
\al{ \label{bounderror}
 \Pr \left(\error(\C_k) \left| \E_{\C_k}^{(r_2)}[W(\nueff_{B_k})] \leq \alpha B_k, \nueff_{B_k} = t, \text{$m$ is sent}\right.\right) \leq 8 \frac{\xi_k^{1/4}}{\ep_k} \leq 8 \xi_k^{1/8},
}
which goes to $0$ as $k \to \infty$.


For any effective arrival time $\nueff_{B_k} = t_i \in \Ls_2(\alpha,\C_k,\ep_k)$, 
we can fix a mapping $\phi_{k,i}$ from $\{1,...,\frac{2^{B_k}}{4B_k}\}$ onto a subset of $\M_t'$ with $\frac{2^{B_k}}{4{B_k}}$ messages.
We will choose the $\frac{2^{B_k}}{4{B_k}}$ messages $m$ with the smallest values of 
\aln{
\Pr \left(\error(\C_k) \left| \nueff_{B_k} = t, \text{$m$ is sent}\right.\right),
}
and we will call this subset $\M_t''$.
Notice that this choice implies that, for $m \in \M_t''$,
\al{ \label{step3}
\sum_{m \in \M_t''} \frac{4{B_k}}{2^{B_k}} \Pr \left(\error(\C_k) \left| \nueff_{B_k} = t, \text{$m$ is sent}\right.\right) & \leq \sum_{m \in \M_t'} |\M_t'|^{-1} \Pr \left(\error(\C_k) \left| \nueff_{B_k} = t, \text{$m$ is sent}\right.\right) \nonumber \\
& \leq \sum_{m \in \M} |\M_t'|^{-1} \Pr \left(\error(\C_k) \left| \nueff_{B_k} = t, \text{$m$ is sent}\right.\right) \nonumber \\
& = {2^{B_k}}|\M_t'|^{-1} \Pr \left(\error(\C_k) \left| \nueff_{B_k} = t \right.\right) \nonumber \\
& \leq {4}{\ep_k}^{-1}  \Pr \left(\error(\C_k) \left| \nueff_{B_k} = t \right.\right).
}
For an effective arrival time $\nueff_{B_k} = t_i \notin \Ls_2(\alpha,\C_k,\ep_k)$, 
we fix any injective mapping $\phi_i$ from $\{1,...,\frac{2^{B_k}}{4{B_k}}\}$ onto a subset of $\M$ with $\frac{2^{B_k}}{4{B_k}}$ messages.
We will build code $\C_k'$ with $B_k' = {B_k-\log 4B_k}$, as the restriction of code $\C_k$ according to $\phi_k$.
We can now upper bound the error probability of this new scheme as
\aln{ \rescnt
\Pr\left(\error(\C_k')\right) & = \sum_{i=1}^{|S_k|} |S_k|^{-1} 
\Pr \left(\error(\C_k') \left| \nueff_{B_k} = t_i \right.\right) \nonumber \\
& \leq \Pr(\nueff_{B_k} \notin \Ls_2) + \sum_{t_i\in \Ls_2} |S_k|^{-1} 
\Pr \left(\error(\C_k') \left| \nueff_{B_k} = t_i \right.\right) \nonumber \\
& = \Pr(\nueff_{B_k} \notin \Ls_2)  + \sum_{t_i\in \Ls_2} \sum_{m \in \M_{t_i}''} |S_k|^{-1} \frac{4B_k}{2^{B_k}} 
\Pr \left(\error(\C_k) \left| \nueff_{B_k} = t_i, \text{$m$ is sent}\right.\right) \nonumber \\
& \leqnum \Pr(\nueff_{B_k} \notin \Ls_2) + 4 \ep_k^{-1} \sum_{t_i\in \Ls_2} |S_k|^{-1} \Pr \left(\error(\C_k) \left| \nueff_{B_k} = t_i \right.\right) \nonumber \\
& \leq \Pr(\nueff_{B_k} \notin \Ls_2) + 4 \ep_k^{-1} \sum_{t_i\in S_k} |S_k|^{-1} \Pr \left(\error(\C_k) \left| \nueff_{B_k} = t_i \right.\right) \nonumber \\
& = \Pr(\nueff_{B_k} \notin \Ls_2) + 4 \ep_k^{-1}\Pr \left(\error(\C_k) \right) \nonumber \\
& \leq \Pr(\nueff_{B_k} \notin \Ls_2) + 4 \xi_k^{7/8},
} \rescnt
where \cnt follows from (\ref{step3}).
Thus, $\Pr\left(\error(\C_k')\right) \to 0$ as $k \to \infty$.
Moreover,  the restriction $\C_k'$ achieves a causal energy-per-bit
\aln{ \rescnt
\liminf_{k \to \infty} \frac{E\left[\E_{\C_k'}{[1:\nueff_{B_k}+\ell_k-1]}\right]}{B_k-\log 4B_k} = \liminf_{k \to \infty} \frac{E\left[\E_{\C_k'}{[1:\nueff_{B_k}+\ell_k-1]}\right]/ B_k}{1-\frac{\log 4B_k}{B_k}}
\leqnum 
e_b,
} \rescnt
where \cnt follows since the sequence of codes $\{\C_k\}_{k=1}^\infty$ was assumed to achieve causal energy-per-bit $e_b$ uniformly over the messages.
We will now consider using code $\C_k'$ in the network in Figure \ref{net2}.

\begin{figure}[ht] 
     \centering
       \includegraphics[height=30mm]{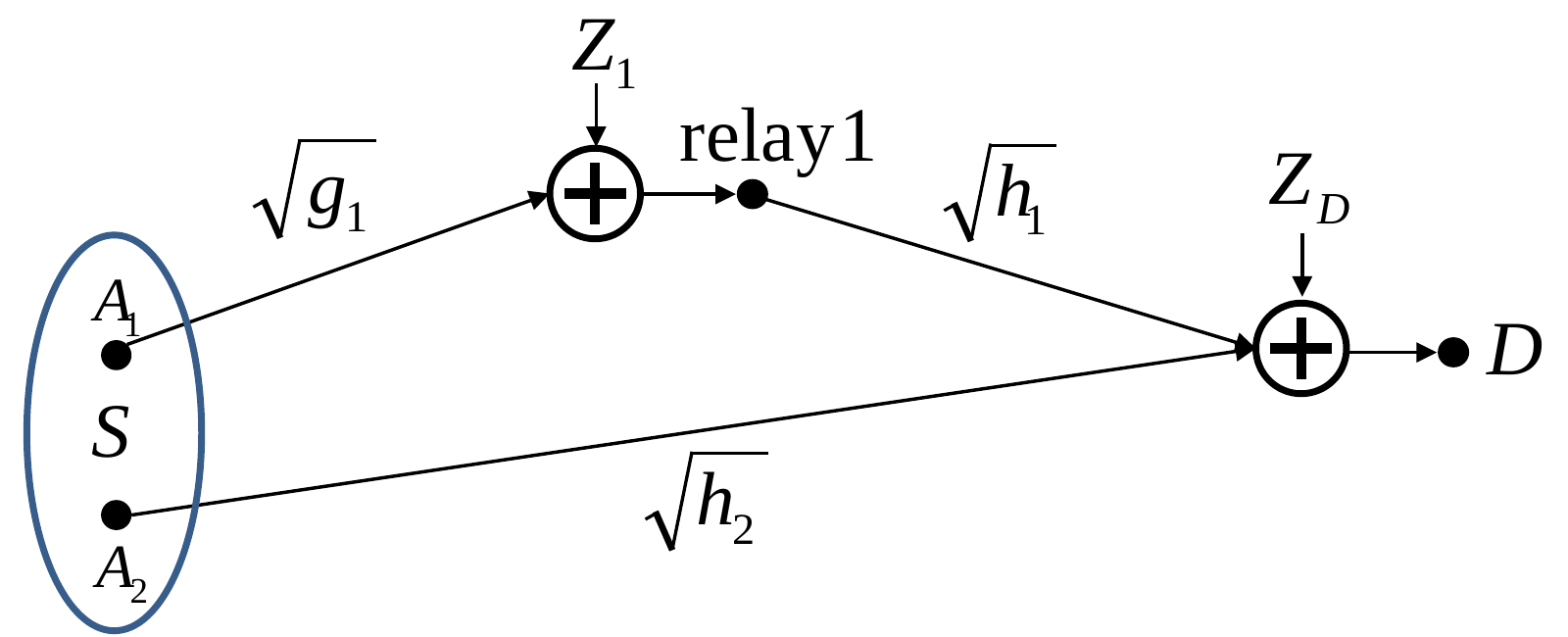} \caption{Contracted two-relay diamond network. \label{net2}}
\end{figure}

In this network, the source possesses two antennas, $A_1$ and $A_2$.
This network should be thought of as the previous diamond network when we allow the source and relay $2$ to cooperate.
Therefore, it is clear that any code used for the diamond network in Figure \ref{netfig} can be used in the network in Figure \ref{net2} by simply having the source simulate the received signals at relay $2$ and using its relaying functions to compute the transmit signals for $A_2$.
Clearly, the error probability and the expected energy when applying $\C_k$ to this new network are identical to those for the original network.

Working in the network in Figure \ref{net2}, the first modification we can make to the sequence of codes $\{\C_k'\}$ is to have them achieve energy-per-bit $e_b$, rather than causal energy-per-bit $e_b$.
This can be done since, according to Theorem \ref{listthm}, relay $1$ can have a list decoder for $\nueff_{B_k}$ with a list size that is subexponential in $B_k$.
Let $\Lambda_k$ be this list, and $U_k$ a subexponential function of $B_k$ satisfying $U_k \to \infty$ as $k \to \infty$ and $|\Lambda_k|/U_k \to 0$ as $k \to \infty$.
Similar to our previous argument, we can have the source send a pulse of magnitude $2\sqrt{\frac{4 \ln U_k}{g_1}}$ at time $\nueff_{B_k} + \ell_k$, and have relay $1$ use a threshold detector with threshold $\sqrt{4 \ln U_k}$ at time $t_i + \ell_k$, as long as $t_i$ was added to the list.
If the pulse is detected, relay $1$ can halt its transmissions.
Moreover, since relay $2$ and source are now together, relay $2$ can stop its transmissions at time $\nueff_{B_k} + \ell_k$ without having to detect any pulse.
We call the resulting code $\{\C_k''\}$.
Following (\ref{errorpulse}), it can be shown that the probability of error in decoding $\nueff_{B_k}$ at relay $1$ goes to $0$, and by following the argument in (\ref{energypulse}), it can be shown that our modified sequence of codes $\{\C_k''\}$ achieves an energy-per-bit $e_b$, and allows relay $1$ to decode $\nueff_{B_k}$ exactly with vanishing error probability.


The next modification we will make is to allow $A_2$ to stay silent up to time $\nueff_{B_k}$.
Notice that this does not happen when we use code $\C_k''$ in the network from Figure \ref{net2}, since the source is simulating what relay $2$ would be sending during each transmission block on the actual diamond network.
However, it is intuitive that the signals transmitted up to the time of the effective arrival time $\nueff_{B_k}$ are ``useless'' to the destination since they are independent of the actual message, and all they may be doing is preventing the destination from making a false alarm at the end of a transmission block prior to $\nueff_{B_k}$.
In order to fix that, we notice that, during transmission blocks prior to $\nueff_{B_k}$, all the source is doing is drawing noise sequences for the received signals at relay $2$, computing the corresponding outputs for relay $2$ and transmitting them on $A_2$.
Therefore, we will draw $|S_k|$ i.i.d. $\N(0,1)$ noise sequences of length $\ell_k$ prior to the communication session and share them among source and destination.
This way, instead of source transmitting what relay $2$ would have transmitted on a transmission block before $\nueff_{B_k}$ on antenna $A_2$, the destination can compute what relay $2$ would have transmitted during the transmission blocks and add that (multiplied by $\sqrt{h_2}$) to its received signal.
Notice that, during the transmission blocks prior to $\nueff_{B_k}$, the statistics of this modified received signal at the destination will be the same as in the code $\C_k''$, and $A_2$ will not be transmitting anything.
However, we need to be careful once we get to the actual transmission block $W(\nueff_{B_k})$.
Since the destination does not know $\nueff_{B_k}$, it will once again add to its received signals the contribution of what relay $2$ would have transmitted if it had received just noise.
To compensate for this, recall that the noise sequences used by the destination to compute the contribution of relay $2$ were shared with the source.
Therefore, the source can compute the output of the relay as well, and transmit its opposite on $A_2$, thus cancelling the addition done by the destination.
Notice that the result of this operation is that antenna $A_2$ will stay silent in all the transmission blocks, except for $W(\nueff_{B_k})$.
However, $A_2$ will utilize (possibly) more energy during $W(\nueff_{B_k})$ than it would have in code $\C_k''$, since it will have to add to its transmit signals a compensation sequence, i.e., a sequence of signals that will cancel the effect of the addition performed by the destination.
Later we will show that this extra energy is in fact negligible.

Now we need to describe the signal that the source will send on $A_2$ during $W(\nueff_{B_k})$.
If the effective arrival time is $\nueff_{B_k} \notin \Ls_2(\alpha,\C_k,\ep_k)$, then $A_2$ will remain silent.
If the effective arrival time is $\nueff_{B_k} = t_i$, for $i \geq (1-B_k^{-1})|S_k|$, $A_2$ will also remain silent.
This will likely cause an error, but notice that, from (\ref{liml2}), the probabilities of these two events goes to $0$.
Moreover, this modification can only decrease the energy used.
If the effective arrival time is $\nueff_{B_k} = t_i \in \Ls_2(\alpha,\C_k,\ep_k)$ and $i < (1-B_k^{-1})|S_k|$, then the source will repeatedly simulate what the output of relay $2$ would have been during $W(\nueff_{B_k})$ on the diamond network from Figure \ref{netfig} using code $\C_k$, until it finds one output signal sequence $X^{\ell_k}$, satisfying $\sum_{i=1}^{\ell_k} X_i^2 \leq \alpha B_k$.
Notice that, from (\ref{boundenergy}) and the fact that the non-overlapping transmission blocks imply that the distribution of the transmit signals of relay $2$ in $W(\nueff_{B_k})$ is independent of the previously received signals, it should be possible to find such output signal.
Moreover, notice that simulating the source repeatedly until finding such an output signal is equivalent to drawing an output signal from the distribution of $X[W(\nueff_{B_k})]$ of relay $2$ conditioned on the fact that $\nueff_{B_k} = t_i$ and $\E_{\C_k}^{(r_2)}[W(\nueff_{B_k})] \leq \alpha B_k$.
The source will thus find such an output sequence, and transmit it on $A_2$ together with the compensation sequence we mentioned previously.
Therefore, for any effective arrival time $\nueff_{B_k} = t_i \in \Ls_2(\alpha,\C_k,\ep_k)$ with $i < (1-B_k^{-1})|S_k|$, since the message we will be sending is from $\M_{t_i}''$, we have that the error probability will satisfy (\ref{bounderror}).
Now, if we let 
\aln{
\Ls_2'(\alpha,\C_k,\ep_k) = \left\{ t_i \in \Ls_2(\alpha,\C_k,\ep_k) \st i < (1-B^{-1})|S_k| \right\},
}
it is clear that we have 
\aln{
\Pr\left(\nueff_{B_k} \notin \Ls_2'(\alpha,\C_k,\ep_k)\right) \to 0, \text{as $k \to \infty$}.
}
Thus, the error probability of this new code, which we will call $\C_k'''$, satisfies
\aln{ \rescnt
\Pr\left(\error(\C_k''')\right) & = \sum_{i=1}^{|S_k|} |S_k|^{-1} 
\Pr \left(\error(\C_k''') \left| \nueff_{B_k} = t_i \right.\right) \nonumber \\
& = \Pr\left(\nueff_{B_k} \notin \Ls_2'\right) \nonumber \\
&\quad + \sum_{t_i \in \Ls_2'} \sum_{m \in \M_{t_i}''} |S_k|^{-1} \frac{4B_k}{2^{B_k}} 
\Pr \left(\error(\C_k) \left| \E_{\C_k}^{(r_2)}[W(\nueff_{B_k})] \leq \alpha B_k, \nueff_{B_k} = t_i, \text{$m$ is sent}\right.\right) \nonumber \\
& \leqnum \Pr\left(\nueff_{B_k} \notin \Ls_2'\right) + \sum_{t_i \in \Ls_2'} \sum_{m \in \M_{t_i}''} |S_k|^{-1} \, \frac{4{B_k}}{2^{B_k}} \,8 \, \xi_k^{1/8} \nonumber \\
& = \Pr\left(\nueff_{B_k} \notin \Ls_2'\right) + \sum_{i=1}^{|S_k|} |S_k|^{-1} \,8 \, \xi_k^{1/8} \nonumber \\
& = \Pr\left(\nueff_{B_k} \notin \Ls_2'\right) + 8 \xi_k^{1/8},
} \rescnt
where \cnt follows from (\ref{bounderror}).
We conclude that $\Pr\left(\error(\C_k''')\right) \to 0$, as $k \to \infty$.
We will let $\E_{\C_k'''}^{(A_1,r_1)}$ be the energy spent by code $\C_k'''$ at antenna $A_1$ and relay $1$.
Notice that antenna $A_1$ and relay $1$ perform exactly as the source and relay $1$ perform when code $\C_k''$ is used.
Therefore we have
\al{ \label{step5}
E\left[ \E_{\C_k'''}^{(A_1,r_1)} \right] = E\left[ \E_{\C_k''}^{(A_1,r_1)} \right] \leq E\left[ \E_{\C_k''} \right]
}
Now we need to compute the expected energy consumed by antenna $A_2$.
Let $V \in \R^{\ell_k}$ be the compensation sequence that is added to the transmit signal of $A_2$ during $W(\nueff_{B_k})$, and let $X\in \R^{\ell_k}$ be the actual transmit signal that the source draws, satisfying $\|X\|^2 \leq \alpha B$, to be transmitted on $A_2$.
Then if $\E_{\C_k'''}^{(A_2)}$ is the energy used by $A_2$, we have
\al{ \label{step6}
E\left[ \E_{\C_k'''}^{(A_2)} \right] & = E\left[ \E_{\C_k'''}^{(A_2)}[W(\nueff_{B_k})] \right] \nonumber \\
& = E\left[ \sum_{i=1}^{\ell_k} (X_i + V_i)^2  \right] \nonumber \\
& \leq  E\left[ 2\sum_{i=1}^{\ell_k} X_i^2 + 2 \sum_{i=1}^{\ell_k} V_i^2  \right] \nonumber \\
& = 2 E\left[ \|X\|^2\right] +  2 E\left[ \|V\|^2\right] \nonumber \\
& \leq 2 \alpha B_k +  2 E\left[ \|V\|^2\right].
}
In order to upper bound the value of $E\left[ \|V\|^2\right]$, we recall that, if $\nueff_{B_k}=t_i$, $V$ is a shared random sequence that was drawn as the output of relay $2$ during $W(t_i)$, assuming that only noise was received at relay $2$, i.e., assuming $\nueff_{B_k} >t_i$.
Moreover, if $\nueff_{B_k} = t_i$ with $i \geq (1-B_k^{-1})|S_k|$, since $A_2$ stays silent, we have $\|V\|^2 = 0$.
Thus we obtain
\al{
E\left[\|V\|^2\right] & = \sum_{i=1}^{(1-B_k^{-1})|S_k|-1} |S_k|^{-1} E\left[ \left. \|V\|^2\, \right| \nueff_{B_k} = t_i\right] \nonumber \\
& = |S_k|^{-1} E\left[ \left. \E_{\C_k''}^{(r_2)}[1:t_{(1-B_k^{-1})|S_k|-1} + \ell_k -1] \right| \nueff_{B_k} > t_{(1-B_k^{-1})|S_k|-1} \right] \nonumber \\
& \leq |S_k|^{-1} E\left[ \left. \E_{\C_k''} \right| \nueff_{B_k} \geq t_{(1-B_k^{-1})|S_k|} \right] \nonumber \\
& \leq |S_k|^{-1} \frac{E\left[ \E_{\C_k''}\right]}{ \Pr\left( \nueff_{B_k} \geq t_{(1-B_k^{-1})|S_k|} \right)} \nonumber \\
& = |S_k|^{-1} \frac{E\left[ \E_{\C_k''}\right]}{ B_k^{-1}} = |S_k|^{-1} B_k {E\left[ \E_{\C_k''}\right]} \label{step7}
}
Notice that $|S_k|$ is exponential in $B_k$ (since $\frac{A_k}{|S_k|}=\frac{2^{\beta B_k}}{|S_k|}$ is subexponential in $B_k$) and thus $|S_k|^{-1}{B_k} \to 0$ as $k \to \infty$.
Therefore, the energy-per-bit used on antenna $A_2$ satisfies
\al{ \rescnt \label{step8}
\liminf_{k \to \infty} \frac{E\left[\E_{\C_k'''}^{(A_2)}\right]}{B_k-\log 4B_k} & \leqnum 
\liminf_{k \to \infty} \frac{2 \alpha B_k+ 2|S_k|^{-1}B_k  E\left[\E_{\C_k''}\right]}{B_k-\log 4B_k} \nonumber \\
& = 
\liminf_{k \to \infty} \frac{2 \alpha}{1-\frac{\log 4B_k}{B_k}} + 2|S_k|^{-1}B_k \frac{1}{1-\frac{\log 4B_k}{B_k}} \frac{ E\left[\E_{\C_k''}\right]}{B_k} \nonumber \\
& \eqnum 2 \alpha,
} \rescnt
where \cnt follows from (\ref{step6}) and (\ref{step7}), and \cnt follows from the fact that $|S_k|^{-1}{B_k} \to 0$ as $k \to \infty$, and $\liminf_{k \to\infty} E\left[\E_{\C_k''}\right]/B_k \leq e_b$.
Clearly, we can find a subsequence of codes $\{\C_{k_j}'''\}_{j=1}^{\infty}$, for which the $\liminf$ in (\ref{step8}) holds as limit, and for which $E\left[\E_{\C_{k_j}'''}^{(A_2)}\right] \leq 3 \alpha (B_{k_j} -\log 4 B_{k_j}) \leq 3\alpha B_{k_j}$.
Thus, by only keeping the codes in $\{\C_{k_j}'''\}$, 
we conclude that the sequence of codes $\{\C_k'''\}$ can be used on the network in Figure \ref{net2} with the addtional constraint that any sequence of codes $\{\C_k\}_{k=1}^\infty$ satisfies
\al{ \label{extraconstraint}
E\left[ \E_{\C_k}^{(A_2)} \right] \leq 3 \alpha B_k.
}
Intuitively, if $\alpha$ is very small (recall that we could have fixed $\alpha > 0$ arbitrarily small), antenna $A_2$ should not be very useful for the scheme.
This idea is captured in the following Lemma, whose proof we present in the Appendix.

\begin{lemma} \label{finallemma}
Consider the network shown in Figure \ref{net2} in the asynchronous setting.
Suppose a sequence of codes $\{\C_k\}_{k=1}^\infty$ satisfies (\ref{extraconstraint}) and achieves a finite energy-per-bit.
Then we must have
\aln{
\liminf_{k \to\infty} \frac{E\left[\E_{\C_k}\right]}{B_k} \geq \gamma (1+\beta) \left(\frac{1}{g_1}+\frac{1}{h_1}\right) - f(\alpha),
}
where $f(\alpha)$ is a function satisfying $f(\alpha) \to 0$ as $\alpha \to 0$.
\end{lemma}

Then we notice that the sequence of codes $\{\C_k'''\}$, once restricted to the subsequence $\{\C_{k_j}'''\}$, achieves an energy-per-bit
\aln{ \rescnt
\liminf_{k \to \infty} \frac{E\left[\E_{\C_k'''}\right]}{B_k-\log 4B_k} & = 
\liminf_{k \to \infty} \frac{E\left[\E_{\C_k'''}^{(A_1,r_1)}\right]+E\left[\E_{\C_k'''}^{(A_2)}\right]}{B_k-\log 4B_k} \nonumber \\
& \leq
\liminf_{k \to \infty} \frac{E\left[\E_{\C_k'''}^{(A_1,r_1)}\right]}{B_k-\log 4B_k}+ \limsup_{k \to\infty} \frac{E\left[\E_{\C_k'''}^{(A_2)}\right]}{B_k-\log 4B_k} \nonumber \\
& \leqnum
\liminf_{k \to \infty} \frac{E\left[\E_{\C_k''}^{(A_1,r_1)}\right]}{B_k-\log 4B_k}+2 \alpha \nonumber \\
& \leqnum e_b + 3 \alpha,
} \rescnt
where \cnt follows from (\ref{step8}) (since the $\liminf$ can be replaced with the limit) and the fact that $A_1$ and $r_1$ perform identically in code $\C_k'''$ and $\C_k''$; and \cnt follows from the fact that code $\C_k''$ achieves an energy-per-bit $e_b$ on the network in Figure \ref{net2}.
Therefore, by applying Lemma \ref{finallemma}, we conclude that
\aln{
e_b \geq \gamma (1+\beta) \left(\frac{1}{g_1}+\frac{1}{h_1}\right) - f(\alpha) - 3\alpha.
}
Since this inequality should hold for any $\alpha > 0$, we may let $\alpha \to 0$ to obtain
\aln{
e_b \geq \gamma (1+\beta) \left(\frac{1}{g_1}+\frac{1}{h_1}\right).
}
Finally, we consider the energy-per-bit that can be achieved by using only relay $1$.
Once we remove relay $2$, the network can be viewed as two concatenated point-to-point channels, i.e., the relay first functions as a destination for the first hop, and then as a source for the second hop.
The minimum energy-per-bit for these channels, according to Theorem \ref{pttopt}, is respectively
\aln{
\frac{\gamma (1+\beta)}{g_1}  \text{\; and \;} \frac{\gamma (1+\beta)}{h_1}.
}
Moreover, since the communication delay in each hop can be made subexponential in $B_k$, the total delay (the sum of the two) will still be subexponential in $B_k$.
Therefore, we conclude that the scheme using only relay $1$ achieves the same or smaller energy-per-bit than the scheme using both relays, and we are in case (b).
This concludes the proof of Theorem \ref{mainthm}.
\end{proof}

\section{Concluding Remarks} \label{conclusion}

In this work we started studying the fundamental limits of energy-efficient communication in relay networks with bursty traffic.
For the diamond relay network, we showed that the minimum energy-per-bit can be achieved with codes where each relay is either synchronized or not used.
Intuitively, this result should not be surprising.
The idea is that a relay that is not synchronized will most likely waste energy outside of the actual communication block and harm the achieved energy-per-bit.
This result was then used to derive a lower bound for the asynchronous minimum energy-per-bit for the diamond network which allows us to prove that separation-based schemes are nearly optimal in high asynchronism regimes.

But the intuition that a relay that is not synchronized cannot be helpful from an energy-per-bit point of view extends beyond a simple diamond network.
Thus, it seems reasonable to expect that a result similar to Corollary \ref{synccor} holds for general wireless networks.
Such a result would have interesting consequences.
It would essentially imply that a separation-based scheme that synchronizes a certain optimal subset of the relays can achieve close to the asynchronous minimum energy-per-bit.
Moreover, it would raise the questions of how to find the optimal subset of relays to be synchronized and what the correct strategy for synchronization is.
In a large non-layered network, it is not even clear in what order the relays should be synchronized.

However, it should be noted that the techniques used to prove Theorem \ref{mainthm} cannot be easily extended to larger networks.
In particular, we notice that Lemma \ref{finallemma} essentially implies that if we have a code where relay 2 uses very little energy, then it is possible to come up with a new code where relay 2 is not used at all.
In order to prove that, we used the fact that the capacity (and, thus, the minimum energy-per-bit) of two concatenated point-to-point AWGN channels is known.
However, even if we just wanted to extend this result to an $N$-relay diamond network, we would no longer be able to prove such a result, since the capacity of the $(N-1)$-relay diamond network is not known.
Therefore, new techniques must be developed in order to prove that a relay that uses a negligible amount of energy-per-bit can in fact be turned off, without affecting the performance of the coding scheme.





\appendix \label{appsection}

\appendixpage
\addappheadtotoc

\renewcommand{\thesection}{\Roman{section}}

\section{Proof of Theorem \ref{pttopt}} \label{apppttopt}

\begin{theoremrep}{\ref{pttopt}}
For an asynchronous AWGN channel, the minimum energy-per-bit is given by
\aln{
e_b^{\rm \, async} = (1+\bar H) e_b^{\rm \, sync}.
}
where $\bar H = \liminf_{k \to \infty} H(\nu_{B_k})/{B_k}$.
\end{theoremrep}

\begin{proof}
\noindent \emph{Converse}: 
Consider an arbitrary sequence $\{\C_k\}_{k=1}^\infty$ of asynchronous codes, achieving a finite energy-per-bit $e_b$, and let the error probabilities be $\Pr\left(\error(\C_k)\right) = \ep_k$, where $\ep_k \to 0$.

We consider using code $\C_k$ in a \emph{synchronous} AWGN channel.
We let $X_k^{\tilde A_k}$ be a (discrete) random vector of length $\tilde A_k \defi A_k + d_{B_k} -1$ that has the distribution induced on the input of the channel by first choosing a message $M$ uniformly at random from $\{1,...,2^{B_k}\}$, then choosing a time index $T$ from $\{1,...,{\tilde A_k}\}$ according to the distribution of $\nu_{B_k}$, and then having the source and the destination operate according to the asynchronous code $\C_k$.
Let $Y_k^{\tilde A_k}$ be the corresponding received signal at the destination.
Notice that, from $Y_k^{\tilde A_k}$, it is possible to decode $M$ 
and output a set $\{t,t+1,...,t+d_{B_k}-1\}$ of consecutive time steps such that $T$ belongs to it with probability at least $1-\ep_k$.
Moreover, notice that there is a one-to-one correspondence between values of $(T,M)$ and values of $X_k^{\tilde A_k}$, and therefore $H(X_k^{\tilde A_k}) = H(T)+B_k = H(\nu_{B_k}) + B_k$.
If $C(P)$ is the capacity of the synchronous AWGN channel with average power constraint $P$, we have
\al{
C\left(E[\E_{\C_k}] /{\tilde A_k}\right) & = \sup_{n,f(x^n) \st \frac1n E\|X^n\|^2\leq E[\E_{\C_k}] /{\tilde A_k} } \frac1n I(X^n;Y^n) \geq \frac1{\tilde A_k} I(X_k^{\tilde A_k};Y_k^{\tilde A_k}) \nonumber \\ 
& \geqnum \frac1{\tilde A_k} I(X_k^{\tilde A_k};Y_k^{\tilde A_k}| T\bmod{d_{B_k}} )  = \frac1{\tilde A_k} \left[ H(X_k^{\tilde A_k}| T\bmod{d_{B_k}} ) - H(X_k^{\tilde A_k}|Y_k^{\tilde A_k}, T\bmod{d_{B_k}} )\right] \nonumber  \\
& = \frac1{\tilde A_k} \left[ H(X_k^{\tilde A_k}) - H( T\bmod{d_{B_k}} ) - H(X_k^{\tilde A_k}|Y_k^{\tilde A_k}, T\bmod{d_{B_k}} )\right] \nonumber \\
& \geq \frac1{\tilde A_k} \left[ H(\nu_{B_k})+B_k - \log d_{B_k} - H(X_k^{\tilde A_k}|Y_k^{\tilde A_k}, T\bmod{d_{B_k}} )\right] \nonumber \\
& \geqnum \frac1{\tilde A_k}\left\{ H(\nu_{B_k})+B_k - \log d_{B_k} - \left[H(\ep_k)+ \ep_k (1+\beta) B_k \right]\right\}, \label{fano} 
} \rescnt
where $f(x^n)$ refers to any distribution on $X^n$; $\cnt$ follows since $T$ is a function of $X_k^{\tilde A_k}$ and thus $\MC{Y_k^{\tilde A_k}}{X_k^{\tilde A_k}}{T\bmod d_{B_k}}$; and $\cnt$ follows from Fano's inequality, since from $Y_k^{\tilde A_k}$ and $T \bmod d_{B_k}$ we can decode $X_k^{\tilde A_k}$ with probability of error smaller than $\ep_k$.
Inequality (\ref{fano}) implies that
\aln{
\inf_{P > 0} \frac{P}{C(P)} & \leq \frac{E[\E_{\C_k}] /{\tilde A_k}}{C\left(E[\E_{\C_k}] /{\tilde A_k}\right)} \leq \frac{E[\E_{\C_k}] /{\tilde A_k}}{1/{\tilde A_k} \left[ H(\nu_{B_k})+B_k - \log d_{B_k} - H(\ep_k) -\ep_k (1+\beta) B_k \right]} \\
& = \frac{E[\E_{\C_k}] /B_k}{ H(\nu_{B_k})/B_k +1 - \frac{\log d_{B_k}}{B_k} - \frac{H(\ep_k)}{B_k} -  \ep_k (1+\beta) }.
}
Finally, 
using Lemma \ref{synclemma}, we conclude that
\aln{
\liminf_{k\to \infty} \frac{E[\E_{\C_k}]}{B_k} \geq e_b^{\rm sync} \liminf_{k \to \infty} \left[H(\nu_{B_k})/B_k +1 - \frac{\log d_{B_k}}{B_k} - \frac{H(\ep_k)}{B_k} -  \ep_k (1+\beta)\right] = e_b^{\rm sync} (1+\bar H)
}
and, thus, $(1+\bar H)e_b^{\rm \, sync} \leq e_b^{\rm \, async}$.
\end{proof}

\section{Proof of Theorem \ref{separationachievable}} \label{appseparationachievable}

\begin{theoremrep}{\ref{separationachievable}}
The asynchronous minimum energy-per-bit for the network in Figure \ref{netfig} satisfies
\aln{
(1+\beta) \gamma\left(\frac{1}{g_2}+\frac{1}{h_1+h_2}\right) \geq e_b^{\min }.
}
\end{theoremrep}

\begin{proof}
We construct a sequence of codes $\{\C_k\}_{k=1}^\infty$, where $\C_k$ transmits $B_k$ bits assuming arrival distribution $\nu_{B_k}$, for any sequence $\{B_k\}_{k=1}^\infty$, where $B_k \to \infty$ as $k \to \infty$.
We use a separation-based scheme scheme which achieves asynchronous energy-per-bit
\aln{(1+\delta)^2 \gamma (1+ \beta) \left(\frac{1}{g_2}+\frac{1}{h_1+h_2}\right)}
for an arbitrarily small $\delta > 0$. 
Similar to the scheme described in the achievability of Theorem \ref{pttopt}, the source sends a pulse as soon as the message arrives.
This pulse is detected by the relays, which send another pulse to the destination, taking advantage of beamforming. 
If relays and destination detect their pulses correctly, the network becomes a synchronous network, and we may employ decode-and-forward to communicate the $B_k$ bits.

More precisely, upon receiving the message, at time $\nu_{B_k}$, the source will transmit a pulse of magnitude
\aln{
(1+\delta)\sqrt{(\gamma \beta B_k)/g},
}
for $\delta > 0$.
This is analogous to the pulse used in the proof of Theorem \ref{pttopt} when $p_{B_k}(t) = 2^{-\beta B_k}$ for $t \in [1: A_k]$.
Notice that we use $g_2$ in the denominator, since $g_2 \leq g_1$, and we want to synchronize both relays.
Relay $i$ declares that the pulse is detected at time $t$ if $Y_i[t]$ is the first received signal larger than
$
(1+\delta/2)\sqrt{\gamma \beta B_k}.
$
By following the same steps in the achievability proof of Theorem \ref{pttopt}, it is not difficult to see that, for any $\delta > 0$, the probability of the relays not detecting the pulse tends to $0$ as $k \to \infty$.
Notice that we use $g_2$ in the denominator, since $g_2 \leq g_1$, and we want to synchronize both relays.

If the relays correctly detect the pulse, they can transmit pulses to the destination in the next time slot.
Since they can use beamforming to reduce the total energy required, at time $\nu_{B_k}+1$, each relay will send a pulse of magnitude
\aln{
(1+\delta)\sqrt{(\gamma \beta B_k)/(4 h)}.
}
Again by following the steps in the proof of Theorem \ref{pttopt}, it can be shown that if the destination declares that a pulse has been detected if the received signal value exceeds $(1+\delta/2)\sqrt{\gamma \beta B_k}$, the probability that the destination does not decode the pulse correctly also tends to $0$ as $k \to \infty$.
The total energy-per-bit consumed in the synchronization phase is \aln{(1+\delta)^2 \gamma \beta \left[1/g+1/{(2h)}\right].}

To compute the energy used in the communication phase, we notice that we can analyze the two hops separately, since we are employing decode-and-forward.
For the first hop, the energy-per-bit can be chosen arbitrarily close to the minimum energy-per-bit of a point-to-point channel with channel gain $g_2$.
Thus we choose the energy-per-bit used by the source to be $(1+\delta)^2 \gamma / g_2$.
Since $g_1 \geq g_2$, both relays are guaranteed to decode the message with high probability.
For the second hop, we again use beamforming to reduce the energy-per-bit that is consumed.
Thus, relay 1 and relay 2 will use the same codebook, but with different scaling coefficients.
More precisely, relay $i$ will use a codebook where the energy-per-bit of \emph{each codeword} is at most
\al{ \label{maxenergyperbit}
(1+\delta)^2 \gamma \frac{{h_i}}{(h_1+h_2)^2}. 
}
This can be done by using Gaussian random codebooks and replacing the codewords that exceed the energy-per-bit in (\ref{maxenergyperbit}) with zero codewords.
Since this constraint is satisfied by every codeword, even in the event that the pulse or the message from the source is not decoded by both relays, the energy-per-bit consumed in the communication phase will be $(1+\delta)^2 \gamma (1/g_2 + 1/(h_1+h_2))$, and the total energy-per-bit is 
\aln{(1+\delta)^2 \gamma (1+ \beta) \left(\frac{1}{g_2}+\frac{1}{h_1+h_2}\right).}
\end{proof}

\section{Proof of Lemma \ref{ptpt2}} \label{appptpt2}

\begin{lemmarep}{\ref{ptpt2}}
Consider the networks in Figures \ref{hop1} and \ref{hop2}, where the message arrival time $\nu_B$ is uniformly distributed in $[1:2^{\beta B}]$.
Then, the minimum asynchronous energy-per-bit $e_b^{\min }$ of these two networks is given respectively by
\aln{
e_b^{\min } = (1+\beta) \frac{\gamma}{g_1 + g_2}
\text{\; and \;}
e_b^{\min } = (1+\beta) \frac{\gamma}{h_1 + h_2}.
}
\end{lemmarep}

\begin{proof}\emph{Achievability.}
For the network in Figure \ref{hop1}, we consider having the destination do a pre-processing on the received signals. 
From $Y_1 = \sqrt{g_1} X + Z_1$ and $Y_2 = \sqrt{g_2} X + Z_2$, the destination will build an effective received signal 
\aln{
\tilde Y &= \sqrt{\frac{g_1}{g_1+g_2}}Y_1 + \sqrt{\frac{g_2}{g_1+g_2}}Y_2 = (\sqrt{g_1 + g_2}) X + \tilde Z,
}
where $\tilde Z = \sqrt{\frac{g_1}{g_1+g_2}}Z_1 + \sqrt{\frac{g_2}{g_1+g_2}}Z_2 \sim \N(0,1)$.
Therefore, we now effectively have a single-antenna point-to-point AWGN channel with channel gain $\sqrt{g_1+g_2}$, and Theorem \ref{pttopt} guarantees that the minimum energy-per-bit is at most $(1+\beta) \frac{\gamma}{g_1+g_2}$.
The same idea can be used to convert the channel in \ref{hop2} into a single-antenna point-to-point AWGN channel with channel gain $\sqrt{h_1+h_2}$, by using a pre-processing at the source.

\emph{Converse.} Notice that the same argument used in the converse of Theorem \ref{pttopt} can be used in order to guarantee that if a sequence of codes $\{\C_k\}_{k=1}^\infty$ achieves asynchronous energy-per-bit $e_b$, then we have
\aln{
\inf_{P>0} \frac{P}{C(P)} \geq \frac{e_b}{1+\beta},
}
where $C(P)$ is the capacity with average power constraint $P$ of either the network in Figure \ref{hop1} or the network in Figure \ref{hop2}.
It is not difficult to see that we have $C(P) = \frac12 \log\left( 1 + (g_1+g_2) P \right)$ in the former case and  $C(P) = \frac12 \log\left( 1 + (h_1+h_2) P \right)$ in the latter case, thus establishing the result.
\end{proof}

\section{Proof of Lemma \ref{ptpt3}} \label{appptpt3}

\begin{lemmarep}{\ref{ptpt3}}
Consider the MIMO channel in Figure \ref{cutgen} in the asynchronous setting, where the message arrival time $\nu_B$ is uniformly distributed in $[1:2^{\beta B}]$.
Consider a sequence of codes $\{\C_k\}_{k=1}^\infty$ that achieves a finite energy-per-bit, and let $\E_{\C_k}^{(s_i)}$ be the energy spent by code $\C_k$ at the source transmitter $s_i$, for $i=1,2$.
Then, we must have
\aln{
a \liminf_{k\to\infty}\frac{E\left[\E_{\C_k}^{(s_1)}\right]}{B_k} + b \liminf_{k\to\infty}\frac{E\left[\E_{\C_k}^{(s_2)}\right]}{B_k} \geq \gamma(1+\beta).
}
\end{lemmarep}

\begin{proof}
Consider any sequence of codes $\{\C_k\}_{k=1}^\infty$ for this channel achieving a finite energy-per-bit with delay $d_{B_k}$.
The expected energy used by code $\C_k$ can be written as
\aln{
E[\E_{\C_k}] & = E[\E_{\C_k}| \nu_{B_k} \leq A_k/2] \Pr(\nu_{B_k} \leq A_k/2) + E[\E_{\C_k}| \nu_{B_k} > A_k/2] \Pr(\nu_{B_k} > A_k/2) \\
& = \frac12 E[\E_{\C_k}| \nu_{B_k} \leq A_k/2] + \frac12 E[\E_{\C_k}| \nu_{B_k} > A_k/2],
}
and we also have
\aln{
a E\left[\E_{\C_k}^{(s_1)}\right] + b E\left[\E_{\C_k}^{(s_2)}\right] & = \frac12 \left( a E\left[ \left. \E_{\C_k}^{(s_1)} \right| \nu_{B_k} \leq A_k/2 \right] + b E\left[ \left. \E_{\C_k}^{(s_2)} \right| \nu_{B_k} \leq A_k/2\right] \right)  \\
& \quad \quad + \frac12 \left( a E\left[ \left. \E_{\C_k}^{(s_1)} \right| \nu_{B_k} > A_k/2 \right] + b E\left[ \left. \E_{\C_k}^{(s_2)} \right| \nu_{B_k} > A_k/2 \right] \right).
}
Thus we must have either $a E\left[ \left. \E_{\C_k}^{(s_1)} \right| \nu_{B_k} \leq A_k/2 \right] + b E\left[ \left. \E_{\C_k}^{(s_2)} \right| \nu_{B_k} \leq A_k/2 \right] \leq a E\left[\E_{\C_k}^{(s_1)}\right] + b E\left[\E_{\C_k}^{(s_2)}\right] $ or  $a E\left[ \left. \E_{\C_k}^{(s_1)} \right| \nu_{B_k} > A_k/2 \right] + b E\left[ \left. \E_{\C_k}^{(s_2)} \right| \nu_{B_k} > A_k/2 \right] \leq a E\left[\E_{\C_k}^{(s_1)}\right] + b E\left[\E_{\C_k}^{(s_2)}\right] $.
Suppose the former case without much loss of generality.
Then we will modify code $\C_k$ to obtain a code $\C_k'$ that only uses $s_1$ in the following way.
An arrival time $\nu_{B_k} \in \{2k-1,2k\}$ will correspond to an arrival time $\nu_{B_k} = k$ in the original scheme, for $k=1,...,A_k/2$.
If a message arrives at time $\nu_{B_k} \in \{2k-1,2k\}$ the sequence of $d_{B_k}$ transmit signals on antenna $s_1$ will be sent at times $2k+1,2(k+1)+1,2(k+2)+1,...,2(k+d_{B_k})+1$.
The sequence of $d_{B_k}$ transmit signals that should be sent over antenna $s_2$ according to code $\C_k$ will be sent on antenna $s_1$ multiplied by a factor $\sqrt{\frac{b}{a}}$ at times $2k+2,2(k+1)+2,2(k+2)+2,...,2(k+d_{B_k})+2$.
Now the destination can simply interpret the signals received at times $2k+1$ and $2k+2$, for $k=1,...,A_k/2$ as the signals received on its two antennas.
With this interpretation of the received signals, the destination can apply the same decoder from code $\C_k$.
The delay of the new code is at most $d_{B_k}' = 2d_{B_k}+2$, and its error probability satisfies
\aln{
\Pr\left(\error(\C_k')\right) = \Pr\left(\error(\C_k)| \nu_{B_k} \leq A_k/2 \right) \leq 2 \Pr\left(\error(\C_k)\right),
}
and also tends to $0$ as $k \to \infty$.
The energy used by code $\C_k'$ satisfies
\aln{
E\left[ \E_{\C_k'}\right] & = E\left[ \E_{\C_k}^{(s_1)}|\nu_{B_k} \leq A_k/2 \right] + \frac{b}{a}E\left[ \E_{\C_k}^{(s_2)}|\nu_{B_k} \leq A_k/2 \right].
}
Since  we now have a sequence of codes $\C_k'$ that achieves a finite energy-per-bit on a point-to-point channel with channel gain $a$, we must have, from Theorem \ref{pttopt},
\aln{
& \liminf_{k\to\infty} \left( \frac{E\left[ \E_{\C_k}^{(s_1)}|\nu_{B_k} \leq A_k/2 \right]}{B_k} + \frac{b}{a} \frac{E\left[ \E_{\C_k}^{(s_2)}|\nu_{B_k} \leq A_k/2 \right]}{B_k} \right) \geq \frac{\gamma(1+\beta)}{a} \\
& \Rightarrow \liminf_{k\to\infty} \left( \frac{a E\left[ \E_{\C_k}^{(s_1)}|\nu_{B_k} \leq A_k/2 \right]}{B_k} + \frac{ b E\left[ \E_{\C_k}^{(s_2)}|\nu_{B_k} \leq A_k/2 \right]}{B_k} \right) \geq {\gamma(1+\beta)} \\
& \Rightarrow a \liminf_{k\to\infty} \frac{E\left[ \E_{\C_k}^{(s_1)} \right]}{B_k} + b \liminf_{k\to\infty} \frac{E\left[ \E_{\C_k}^{(s_2)} \right]}{B_k} \geq {\gamma(1+\beta)}
}
\end{proof}

\section{Proof of Lemma \ref{flemma}} \label{appflemma}

\begin{lemmarep}{\ref{flemma}}
Suppose we have a sequence of codes $\{\C_k\}_{k=1}^\infty$, where code $\C_k$ operates on a channel with uniform arrival distribution on $[1:2^{\beta B_k}]$ but only transmits $B_k - f(B_k)$ bits, with $f(\cdot) \geq 0$ and
\al{
\lim_{k \to \infty} \frac{f(B_k)}{B_k} = 0. \label{limf}
}
Suppose that, in addition, this sequence of codes satisfies the following:
\begin{itemize}
\item $\lim_{k \to \infty} \Pr\left(\error(\C_{k})\right) = 0$
\item $\lim_{k \to \infty} \frac{ \log d_{B_k}}{B_k-f(B_k)} = 0$
\item $\liminf_{k \to \infty} \frac{ E[\E_{\C_k}] }{B_k-f(B_k)} \leq e_b$
\end{itemize}
Then, for any $\eta > 0$, this sequence can be used to construct a new sequence of codes $\{\C_{k}'\}_{k=1}^{\infty}$, where code $\C_k'$ operates on a channel with uniform arrivals on $[1:2^{\beta B_k'}]$ and transmits $B_k'$ bits, satisfying
\begin{itemize}
\item $\lim_{k \to \infty} \Pr\left(\error(\C_{k}')\right) = 0$
\item $\lim_{k \to \infty} \frac{ \log d_{B_k}'}{B_k'} = 0$
\item $\liminf_{k \to \infty} \frac{ E[\E_{\C_k}] }{B_k'} \leq (1+\eta) e_b,$
\end{itemize}
i.e., $\{\C_k'\}$ achieves an energy-per-bit $(1+\eta)e_b$ according to the original definition.
\end{lemmarep}

\begin{proof}
Notice that we may assume wlog that $f(B_k) \to \infty$ as $k \to \infty$.
Otherwise, we can simply modify each code $\C_k$ to transmit only $B_k - f(B_k) - \sqrt{B_k}$ bits (and the remaining $\sqrt{B_k}$ bits can be chosen uniformly at random by the source, so that they are not message bits).
Clearly, if we define $\hat f(B_k) = f(B_k) + \sqrt{B_k}$, $\hat f(B_k)$ still satisfies (\ref{limf}), and we have $\hat f(B_k) \geq \sqrt{B_k}$ for all $k$, and $\hat f(B_k) \to \infty$ as $k \to \infty$.

We will let $B_k' = B_k - f(B_k)$, for $k=1,2,...$.
It is clear from (\ref{limf}) that $B_k' \to \infty$ as $k \to \infty$.
Then, we will use code $\C_k$, which transmits $B_k - f(B_k)$ bits for an arrival distribution $\nu_{B_k}$, to create the code $\C_{k}'$ which transmits $B_k'$ bits for an arrival distribution $\nu_{B_k'}$.
Notice that the arrival time window assumed by code $\C_k$, $W_k = [1 : 2^{\beta B_k}]$, must be modified into a shorter arrival time window $w_k' = [1 : 2^{\beta B_k'}]$, and since $B_k - B_k' \to \infty$ as $k \to \infty$, $|w_k'|/|W_k| = 2^{-\beta (B_k - B_k')} \to 0$ as $k \to \infty$.
We consider 
partitioning $W_k$ into $M = \left\lceil |W_k|/|w_k'| \right\rceil$ subintervals $I_j$, $j=1,...,M$. 
The $j\th$ subinterval is $I_j = \left[(j-1)|w_k'| + 1 : j|w_k'|\right]$, for $j=1,...,M - 1$, and $I_M = \left[(M-1)|w_k'| + 1 : |W_k| \right]$.

Now we fix any $\eta > 0$. 
We will show that, for $k$ sufficiently large, we can find an interval $I_j$, with $j<M$, such that
\aln{
E\left[\left. \E_{\C_k} \right| \nu_{B_k} \in I_j \right] \leq (1+\eta) E\left[ \E_{\C_k} \right]
\text{ \; and \; } \Pr\left(\left. \error(\C_k)  \right| \nu_{B_k} \in I_j \right) \leq \frac{2(1+\eta)}{\eta} \Pr\left( \error(\C_k)   \right).
}
To see this we first define the set $J = \left\{ j < M : E\left[\left. \E_{\C_k} \right| \nu_{B_k} \in I_j \right] \leq (1+\eta) E\left[ \E_{\C_k} \right] \right\}$.
Then we have
\al{
E\left[ \E_{\C_k} \right] & \geq \sum_{j=1}^{M-1}\frac{|w_k'|}{|W_k|} E\left[\left. \E_{\C_k} \right| \nu_{B_k} \in I_j \right] \geq \sum_{j \notin J, j\ne M} \frac{|w_k'|}{|W_k|} E\left[\left. \E_{\C_k} \right| \nu_{B_k} \in I_j \right] \nonumber \\
& \geq \sum_{j \notin J, j\ne M} \frac{|w_k'|}{|W_k|} (1+\eta)  E\left[ \E_{\C_k} \right]
= \left| \{1,...,M-1\} \setminus J \right| \frac{|w_k'|}{|W_k|} (1+\eta)  E\left[ \E_{\C_k} \right] \nonumber \\
& \Longrightarrow \left| \{1,...,M-1\} \setminus J \right| \leq \frac{|W_k|}{(1+\eta) |w_k'|} 
\leq \frac{M}{(1+\eta)} \label{jstep1} 
}
Similarly, we define the set $J' = \left\{ j < M :  \Pr\left(\left. \error(\C_k)  \right| \nu_{B_k} \in I_j \right) \leq \frac{2(1+\eta)}{\eta} \Pr\left( \error(\C_k)   \right) \right\}$.
Then we obtain
\al{
\Pr & \left( \error(\C_k)   \right)  \geq \sum_{j=1}^{M-1}\frac{|w_k'|}{|W_k|} \Pr\left(\left. \error(\C_k)  \right| \nu_{B_k} \in I_j \right) \geq \sum_{j \notin J', j\ne M}\frac{|w_k'|}{|W_k|} \Pr\left(\left. \error(\C_k)  \right| \nu_{B_k} \in I_j \right) \nonumber \\
& \geq \sum_{j \notin J', j\ne M}\frac{|w_k'|}{|W_k|} \frac{2(1+\eta)}{\eta} \Pr\left( \error(\C_k) \right)
= \left| \{1,...,M-1\} \setminus J' \right| \frac{|w_k'|}{|W_k|} \frac{2(1+\eta)}{\eta} \Pr\left( \error(\C_k) \right) \nonumber \\
& \Longrightarrow \left| \{1,...,M-1\} \setminus J' \right| \leq \frac{|W_k| \eta}{2(1+\eta) |w_k'|} 
\leq \frac{M \eta}{2(1+\eta)} \label{jstep2}
}
From (\ref{jstep1}) and (\ref{jstep2}), we conclude that
\aln{
\left| J \cap J' \right| & \geq M - 1 - \left| \{1,...,M-1\} \setminus J \right| -  \left| \{1,...,M-1\} \setminus J' \right| \nonumber \\
& \geq M - 1 - \frac{M}{(1+\eta)} - \frac{M \eta}{2(1+\eta)} = M\left( \frac{\eta}{2(1+\eta)} \right) - 1.
}
Therefore, since $M \to \infty$ as $k \to \infty$, for $k$ sufficiently large, $J \cap J' \ne \emptyset$, implying that we can find our desired subinterval $I_j$.

Since $|I_j| = |w_k'|$, we will build code $\C_k'$ from code $\C_k$ by having $\C_k$ operate as if it were in the interval $I_j$.
In order to do that, we consider drawing two sequences of $(j-1)|w_k'|$ i.i.d. $\N(0,1)$ noise values and sharing them among all relays and the destination.
The relays will start operating at time $1$ in the interval $w_k'$ as if they were in time $(j-1)|w_k'| + 1$ in code $\C_k$ and had received, prior to that time, their corresponding shared noise sequence.
The destination will then use the relaying functions that the relays would have used in code $\C_k$ and apply them to the shared noise sequence of each relay, thus being able to simulate what the relays would have transmitted in code $\C_k$ prior to time $(j-1)|w_k'| + 1$.
This way the destination can simulate the signals it would have received prior to time $(j-1)|w_k'| + 1$, and start operating at time $1$ as if it were at time $(j-1)|w_k'| + 1$ in code $\C_k$.
Therefore, we see that, for this new code $\C_k'$,
\aln{
\Pr\left(\error(\C_k')\right) = \Pr\left(\left. \error(\C_k) \right| \nu_{B_k} \in I_j \right) \leq \frac{2(1+\eta)}{\eta} \Pr\left( \error(\C_k)   \right),
}
which tends to $0$ as $k \to \infty$, for any fixed $\eta > 0$.
Moreover, our new code $\C_k'$ will consume (in expectation) the same energy that code $\C_k$ would consume during $I_j$, conditioned on $\nu_{B_k} \in I_j$.
But this is clearly less than the total energy consumed by $\C_k$ conditioned on $\nu_{B_k} \in I_j$, and we have
\aln{
E \left[ \E_{\C_k'} \right] \leq E\left[\left. \E_{\C_k} \right| \nu_{B_k} \in I_j \right] \leq (1+\eta) E\left[ \E_{\C_k} \right].
}
By letting $d_{B_k}'=d_{B_k}$, we conclude that our new sequence of codes $\{\C_k'\}_{k=1}^{\infty}$ achieves an energy-per-bit $(1+\eta) e_b$ and communicates $B_k'$ bits on a channel with arrival distribution $\nu_{B_k'}$.
\end{proof}

\section{Proof of Lemma \ref{nollem}} \label{appnollem}

\begin{lemmarep}{\ref{nollem}}
Suppose we have a sequence of codes $\{\C_k\}_{k=1}^\infty$ achieving a finite energy-per-bit $e_b$ on the asynchronous diamond network in Figure \ref{netfig}.
Then we can build another sequence of codes $\{\C_k'\}$ with delay constraint $d_{B_k}'$ subexponential in $B_k' = B_k$, with non-overlapping transmission blocks of length $\ell_k$, for which 
\aln{
\liminf_{k \to \infty} \frac{E\left[\E_{\C_k'}[1:\nueff_{B_k}+\ell_k-1]\right]}{B_k} \leq (1+\eta)e_b,
}
for any arbitrarily small $\eta > 0$, and whose probability of error goes to $0$ as $k \to \infty$.
\end{lemmarep}

\begin{proof}
%
%
Our first step is to ``delay'' the entire coding scheme by $2 d_{B_k}$.
In order to do this, we shift all the encoding functions at the source, the relaying functions at the relays and the decoding functions at the destination by $2 d_{B_k}$ time steps.
More specifically, suppose the message $m$ arrives at time $\nu_{B_k}$, and let $\Enc_t(\nu_{B_k},m)$, for $t=\nu_{B_k},\nu_{B_k}+1,...,\nu_{B_k}+d_{B_k}-1$, be the signals transmitted by the source during $[\nu_{B_k}:\nu_{B_k}+d_{B_k}-1]$.
After the delaying operation, if the message arrives at time $\nu_{B_k}$, the source will wait until time $\nu_{B_k}+2d_{B_k}$ and transmit $\Enc_t(\nu_{B_k},m)$ at time $t+2d_{B_k}$ for $t=\nu_{B_k},\nu_{B_k}+1,...,\nu_{B_k}+d_{B_k}-1$.
Similarly, if according to code $\C_k$ relay $i$ transmitted, at time $t$, $X_i[t] = f_t (Y_i[1],Y_i[2],...,Y_i[t-1])$ (where $Y_{i}[t]$ is the received signal at relay $i$ at time $t$), then after the delaying operation relay $1$ will transmit, at time $t+2d_{B_k}$, $X_1[t+2d_{B_k}] = f_t(Y_i[2d_{B_k}+1],...,Y_i[2d_{B_k}+t-1])$.
Finally, if the destination, at time $t$, made its detection/decoding using a function $g_t(Y_d[1],...,Y_d[t-1])$ according to code $\C_k$, then it will, at time $t+2d_{B_k}$, make its detection/decoding by computing $g_t(Y_d[2d_{B_k}+1],Y_d[2d_{B_k}+2],...,Y_d[2d_{B_k}+t-1])$.
Clearly if we increase the delay constraint from $d_{B_k}$ to $3d_{B_k}$, this new code has the exact same error probability as $\C_k$.
We will refer to this delayed version of code $\C_k$ as $\C_k'$.

Next, consider partitioning the arrival interval $[1:A_k]$ into consecutive blocks of length $d_{B_k}$, and let $M_1, M_2 \subset [1:A_k]$ correspond to arrival times that belong to odd blocks and even blocks respectively.
If we assume for simplicity that $A_k = 2 q d_{B_k}$ for some $q \in \Z$, then the expected energy used by the delayed code $\C_k'$ can be written as
\aln{
E[\E_{\C_k'}] & = E[\E_{\C_k'}| \nu_{B_k} \in M_1] \Pr(\nu_{B_k} \in M_1) + E[\E_{\C_k'}| \nu_{B_k} \in M_2] \Pr(\nu_{B_k} \in M_2) \\
& = \frac12 E[\E_{\C_k'}| \nu_{B_k} \in M_1] + \frac12 E[\E_{\C_k'}| \nu_{B_k} \in M_2].
}
Therefore, we must have $E[\E_{\C_k'} | \nu_{B_k} \in M_1] \leq E[\E_{\C_k'} ]$ or $E[ \E_{\C_k'} | \nu_{B_k} \in M_2] \leq E[\E_{\C_k'}]$.
Let us suppose, without loss of generality, that the former is true. 
Then we will pick our set of transmission times $S \subset M_1 + 2d_{B_k}$.
More specifically, since we have a total of $q$ length $d_{B_k}$ blocks in $M_1$, we select our transmission times to be $t_i$, $i=1,..., q$, where $t_i \in [2 i d_{B_k}+1:(2i+1)d_{B_k}]$.
Notice that this guarantees that any two transmission times will be separated by at least $d_{B_k}$ time steps and at most $3 d_{B_k}$ time steps.
We will set $I_i = [2(i-1)d_{B_k}+1:2id_{B_k}]$, for $i=1,...,q$, and $\ell_k = d_{B_k}$.


For a given choice of transmission times $S = \{t_1,...,t_q\}$, the source will perform as follows. 
If the message arrives in the time window $I_i = [2(i-1)d_{B_k}+1:2id_{B_k}]$, then the source will wait until time $t_{i}$ (which occurs after the end of $I_i$) to transmit it.
This mapping operation performed by the source is depicted in Figure \ref{mapping}.
\begin{figure}[ht] 
     \centering
       \includegraphics[height=29mm]{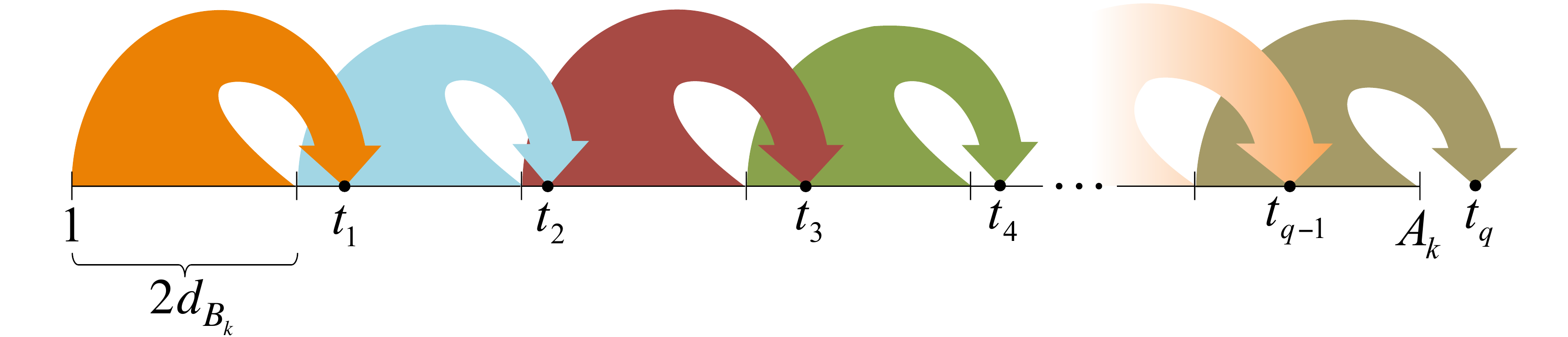} \caption{Mapping performed by the source from $\nu_{B_k}$ to $\nueff_{B_k}$. \label{mapping}}
\end{figure}
Clearly, for any choice of $S$, the decoding delay will be at most $4 d_{B_k}$, 
which is subexponential in $B_k$.

For each choice of $S$, we will let $\C_k^S$ be the resulting code with delay constraint $d_{B_k}^S = 4d_{B_k}$, and we consider choosing $S$ uniformly at random.
This is equivalent to picking each $t_i$ independently and uniformly at random from $[2 i d_{B_k}+1:(2i+1)d_{B_k}]$, for $i=1,...,q$.
For a given $S$, the expected energy of the new coding scheme is given by
\al{ \rescnt
E[\E_{\C_k^S}] & = \sum_{t=1}^{A_k} E[\E_{\C_k^S} | \nueff_{B_k} = t] \Pr(\nueff_{B_k} = t) \nonumber \\
& \eqnum \frac1q \sum_{t \in S} E[\E_{\C_k'} | \nu_{B_k} = t - 2d_{B_k}] \nonumber \\
& \eqnum \frac1q \sum_{t \in S} E[\E_{\C_k} | \nu_{B_k} = t-2d_{B_k}], \label{energys}
} \rescnt
where \cnt follows since, conditioned on $\nueff_{B_k} = t$, the new code $\C_k^S$ performs exactly as $\C_k'$ conditioned on $\nu_{B_k} = t-2d_{B_k}$ (due to the delay of $2d_{B_k}$), and \cnt follows because, for any $\nu_{B_k}$, $\C_k$ and $\C_k'$ spend the same amount of energy in expectation (although at different times). 
When we consider averaging over the ensemble of choices of $S$, we obtain
\al{\rescnt
\ol{E[\E_{\C_k^S}]} & =\frac1{ d_{B_k}^{q}} \sum_{S} E[\E_{\C_k^S}] = \frac1{q d_{B_k}^{q}} \sum_{S} \sum_{t \in S} E[\E_{\C_k}| \nu_{B_k} = t-2d_{B_k}] \nonumber \\
& = \frac{d_{B_k}^{q-1}}{q d_{B_k}^{q}} \sum_{t \in M_1} E[\E_{\C_k}| \nu_{B_k} = t] = \frac{2}{A_k} \sum_{t \in M_1} E[\E_{\C_k}| \nu_{B_k} = t] \nonumber \\
& \eqnum E[\E_{\C_k}| \nu_{B_k} \in M_1] \leq  E[\E_{\C_k}], \label{avgenergy}
} \rescnt
where \cnt follows from the fact that 
\aln{
E[\E_{\C_k}| \nu_{B_k} \in M_1] &= \sum_{t \in M_1} \Pr(\nu_{B_k}=t|\nu_{B_k} \in M_1) E[\E_{\C_k}| \nu_{B_k} = t] \\
& = \sum_{t \in M_1} \frac{\Pr(\nu_{B_k}=t)}{\Pr(\nu_{B_k} \in M_1)} E[\E_{\C_k} | \nu_{B_k} = t]  = \sum_{t \in M_1} \frac2{A_k} E[\E_{\C_k} | \nu_{B_k} = t].
}
Similar to (\ref{energys}), the probability of error of the new coding scheme for a fixed choice of $S$ satisfies
\al{
\Pr\left(\error(\C_k^S)\right) = \frac1q \sum_{t \in S} \Pr\left(\error(\C_k)|\nu_{B_k} = t-2d_{B_k}\right).
}
Similar to (\ref{avgenergy}), when we average over all choices of $S$, we obtain
\al{ \rescnt
\ol{\Pr(\error(\C_k^S))} & = \frac1{ d_{B_k}^{q}} \sum_{S} \Pr(\error(\C_k^S)) = \frac1{q d_{B_k}^{q}} \sum_{S} \sum_{t \in S} \Pr(\error(\C_k)| \nu_{B_k} = t-2d_{B_k})  \nonumber \\
& = \frac{d_{B_k}^{q-1}}{q d_{B_k}^{q}} \sum_{t \in M_1} \Pr(\error(\C_k)| \nu_{B_k} = t) \nonumber \\
& = \Pr(\error(\C_k)|\nu_{B_k} \in M_1)  \leqnum 2 \xi_k, \label{avgerror} \rescnt
}
where, 
in \cnt, we use the fact that
\aln{
\Pr\left(\error(\C_k)\right)  & = \Pr(\error(\C_k)|\nu_{B_k} \in M_1) \Pr(\nu_{B_k} \in M_1) +\Pr(\error(\C_k)|\nu_{B_k} \notin M_1) \Pr(\nu_{B_k} \notin M_1) \nonumber \\
& \geq  \frac12 \Pr(\error(\C_k)|\nu_{B_k} \in M_1),
}
and $\xi_k \defi \Pr\left(\error(\C_k)\right)$. 
Even though (\ref{avgenergy}) implies the existence of a choice of $S$ for which $E[\E_{\C_k^S}] \leq E[\E_{\C_k}]$ and (\ref{avgerror}) implies the existence of a choice of $S$ for which $\Pr\left(\error(\C_k^S)\right) \leq 2\xi_k$, where $\xi_k \to 0$ as $k \to \infty$, we have no guarantee that there exists an $S$ satisfying these two conditions simultaneously.
To fix this, we consider any small $\eta > 0$, and we notice that, when we choose $S$ uniformly at random over all possibilities, from Markov's inequality and inequalities (\ref{avgenergy}) and (\ref{avgerror}), we have
\al{
\Pr\left( E[\E_{\C_k^S}] \geq (1+\eta) E[\E_{\C_k}] \right) \leq \frac{\ol{E[\E_{\C_k^S}] }}{(1+\eta) E[\E_{\C_k}]} \leq \frac{1}{1+\eta}, \label{markov1}
}
\al{
\Pr\left( \Pr\left(\error(\C_k^S)\right)  \geq  \frac{4(1+\eta)\xi_k}{\eta}  \right) \leq \frac{\ol{\Pr\left(\error(\C_k^S)\right)} \eta}{4(1+\eta)\xi_k} \leq \frac{\eta}{2(1+\eta)}. \label{markov2}
}
We now use the union bound, (\ref{markov1}) and (\ref{markov2}) to conclude that 
\aln{
\Pr\left( E[\E_{\C_k^S}] \leq (1+\eta) E[\E_{\C_k}] \text{\; and \;} \Pr\left(\error(\C_k^S)\right) \leq  \frac{4(1+\eta)\xi_k}{\eta} \right) & \geq 1 - \frac{1}{1+\eta} - \frac{\eta}{2(1+\eta)} \\
& = \frac{\eta}{2(1+\eta)} > 0.
}
Therefore, for $\eta > 0$ arbitrarily small, there exists at least one set $S$, such that $E[\E_{\C_k^S}] \leq (1+\eta) E\left[\E_{\C_k}\right]$ and $\Pr\left(\error(\C_k^S)\right) \leq  \frac{4(1+\eta)\xi_k}{\eta}$.
Notice that $\frac{4(1+\eta)\xi_k}{\eta} \to 0$ for any fixed $\eta > 0$.
Thus, we will pick our set of transmission times to be one such $S$, which we will refer to as $S_k$.
Our new sequence of codes achieves an energy-per-bit at most $(1+\eta) e_b$, for any arbitrarily small $\eta > 0$.

Notice that, at a time $t \in [t_i:t_i+\ell_k-1]$, if there is a transmission happening, it must have started at $t_i$.
Therefore, any detection made by the destination at a time $\tau \in [t_i : t_i+\ell_k - 1]$ can wait and be outputted at time $\tau' = t_i+\ell_k - 1$.
This will not affect the error probability, since the probability of late decoding will not change.
The advantage is that the destination will now only make decisions at the end of a transmission block.
Thus, if we let $\ze_i \defi t_i + d_{B_k} -1$, for $i=1,...,|S_k|$, we can define the error event
\al{
L_i = \{ \tau = \ze_i, \nueff_{B_k} > t_i\} \cup \{ \tau = \ze_i, \nueff_{B_k} = t_i, \hat m \ne m \} \cup \{ \tau > \ze_i, \nueff_{B_k} = t_i \} \label{errorevents}
}
for $i=1,...,|S_k|$.
These three subevents can be read as false alarm at time $\ze_i$, wrong decoding at time $\ze_i$ and missed detection at time $\ze_i$, respectively. 
It is easy to see that any error event corresponds to $L_i$ for some $i \in \{1,...,|S_k|\}$, and $L_i \cap L_j = \emptyset$ if $i\ne j$.
Thus, the error probability of our code can be written as 
\al{ \rescnt
\Pr\left(\error(\C_k^S)\right) & = \sum_{i=1}^{|S_k|} \Pr(L_i) 
\eqnum \sum_{i=1}^{|S_k|} \Pr\left(\ol{L_{1:i-1}},  L_i,\nueff_{B_k} \geq t_i \right)  \nonumber  \\
& = \sum_{i=1}^{|S_k|} \Pr\left(  L_i \, \left|\, \ol{L_{1:i-1}}, \nueff_{B_k} \geq t_i \right. \right) \Pr\left(\ol{L_{1:i-1}},\nueff_{B_k} \geq t_i \right) \nonumber \\
& = \sum_{i=1}^{|S_k|} \Pr\left(  L_i \, \left|\, \ol{L_{1:i-1}}, \nueff_{B_k} \geq t_i \right. \right) \Pr(\nueff_{B_k} \geq t_i ) \prod_{j=1}^{i-1} \Pr\left(\ol{L_{j}}\, \left| \, \ol{L_{1:j-1}}, \nueff_{B_k} \geq t_i \right. \right), \label{error1}
} \rescnt
where 
\cnt follows since $ L_i$ implies $\nueff_{B_k} \geq t_i$ and $\ol{L_{1:i-1}} \defi \ol{L_{1}}\cap ... \cap \ol{L_{{i-1}}}$.
Then we notice that for each $i\in \{1,...,|S_k|\}$, conditioned on the fact that $\nueff_{B_k} \geq t_i$, the relays only received noise during time steps $1,..., t_i-1$.
Therefore, there is no actual information received up to time $t_i-1$. 
Thus, intuitively, the same performance should be achieved if, instead of using the actual noise received in $[1:t_i-1]$, the relays used a random noise sequence of size $t_i-1$ that is drawn before the communication session, and shared among relays and destination.
Notice that, in this case, the destination would not need to use its received signals during times $[1:t_i-1]$, since it can simulate the output of the relays, and simulate the AWGN channel between the relays and itself.

More formally, for each $t_i \in S_k$, we will draw two noise sequences of length $t_i-1$ (one for each relay) and share it among relays and destination.
Then, during the transmission block $[t_i:t_i+d_{B_k}-1]$, the relays compute their outputs assuming that the received signals during times $[1:t_i-1]$ were the corresponding noise sequence.
During the same transmission block, the destination simulates what its received signals would have been during times $[1:t_i-1]$ if the relays had in fact received the shared noise sequence.
This way, our resulting coding scheme will satisfy the third property of a coding scheme with non-overlapping transmission blocks.

However, we still need to specify, for each $t_i$, the distribution from which we draw the noise sequence for each relay.
Somewhat surprisingly, the natural choice of drawing the noise sequences i.i.d. $\N(0,1)$ does not work.
Instead, we will use the intuition provided by (\ref{error1}) to define how we draw the noise sequences.
For a given $t_i$, let $\vec N_{t_i-1}^1,\vec N_{t_i-1}^2 \in \R^{t_i-1}$ be the random vectors associated to the received signals at relays $1$ and $2$, conditioned on the fact that $\nueff_{B_k} \geq t_i$ (this guarantees that the relays actually received just noise in $[1:t_i-1]$).
To simplify the expressions, we define $\vec N_{t_i-1} = (\vec N_{t_i-1}^1,\vec N_{t_i-1}^2)$.
For each $t_i$, we will draw the pair of noise sequences according to the distribution of $\vec N_{t_i-1}$ conditioned on $\nueff_{B_k} \geq t_i$ and $\ol{L_{1:i-1}}$.
For the resulting code, which we call $\C_k^{S \prime}$, we define the error events $L_i'$ exactly as in (\ref{errorevents}).
Next we claim that, for our new scheme, for any $i \in \{1,...,|S_k|\}$ and $t \geq t_i$, we have
\al{
& \Pr\left(  L_i' \, \left|\, \ol{L_{1:i-1}'}, \nueff_{B_k} \geq t \right. \right) = \Pr\left(  L_i \, \left|\, \ol{L_{1:i-1}}, \nueff_{B_k} \geq t \right. \right). \label{probseq}
} 
To see this, notice that, conditioned on $\nueff_{B_k} \geq t$ and $\ol{L_{1:i-1}}$ in the case of $\C_k^S$, and conditioned on $\nueff_{B_k} \geq t$ and $\ol{L_{1:i-1}'}$ in the case of $\C_k^{S\prime}$, the distribution of the relays' received signals (or perceived received signals for $\C_k^{S \prime}$) during $[1:t_i-1]$ is the same, and, therefore, the distribution of their output signals during $[t_i:t_i+\ell_k - 1]$ will be the same.
Therefore, from equations (\ref{error1}) and (\ref{probseq}), we see that the probability of error of our new code $\C_k^{S \prime}$ is identical to the probability of error of our previous code, i.e., $\Pr\left(\error(\C_k^{S \prime})\right) = \Pr\left(\error(\C_k^S)\right)$.

Unfortunately, the same cannot be said about the average energy spent by the new code.
In particular, we notice that after the transmission block where the message was sent, the relays will keep operating based on the noise sequences that were drawn.
Therefore, in some sense, the relays are assuming that the message has not been sent yet, which may cause them to use more energy than in the previous coding scheme.
This is why in the statement of the Theorem we only require that the energy spent by the code up to time $\nueff_{B_k} + \ell_k - 1 = \nueff_{B_k} + d_{B_k} -1$ is at most $(1+\eta)e_b$.
In order to be able to bound the energy used by our code up to time $\nueff_{B_k}+d_{B_k}-1$, we consider making a slight modification to it.
If we let $\gamma_k = \Pr\left(\error(\C_k^{S \prime})\right)$, then we will have the source and the relays stay silent in the last $\sqrt{\gamma_k} |S_k|$ transmission blocks. 
We let $\C_k^{S \prime \prime}$ be the resulting code.
If $\nueff_{B_k} \geq t_{(1-\sqrt{\gamma_k})|S_k|}$, this modification will most likely cause an error, but since $\gamma_k \to 0$ as $k \to \infty$, this only occurs with probability $\sqrt{\gamma_k} \to 0$, as $k \to \infty$.
Thus it is clear that $\Pr\left(\error(\C_k^{S \prime \prime})\right) \to 0$ as $k \to \infty$.

Now let us consider the energy spent by our new code up to time $\nueff_{B_k} + d_{B_k} - 1$.
First we notice that, for $i < (1-\sqrt{\gamma_k})|S_k|$,
\al{ \rescnt
E\left[ \left. \E_{\C_k^{S \prime \prime}}{[1:\nueff_{B_k}+d_{B_k}-1]} \right| \nueff_{B_k} = t_i\right] & \eqnum \sum_{j=1}^{i} E\left[ \left. \E_{\C_k^{S \prime \prime}}{[t_j:t_j+d_{B_k}-1]} \right| \nueff_{B_k} = t_i \right] \nonumber \\
& \eqnum \sum_{j=1}^{i} E\left[ \left. \E_{\C_k^{S}}{[t_j:t_j+d_{B_k}-1]} \right| \nueff_{B_k} = t_i, \ol{L_{1:j-1}}  \right] \nonumber \\
& \leq \sum_{j=1}^{i} \frac{E\left[ \left. \E_{\C_k^{S}}{[t_j:t_j+d_{B_k}-1]} \right| \nueff_{B_k} = t_i  \right]}{\Pr(\ol{L_{1:j-1}} | \nueff_{B_k} = t_i)},  \label{bound1}
}\rescnt
where \cnt follows since energy is only spent during the transmission blocks, and \cnt follows because, from the way we drew our shared noise sequences, the expected energy spent in $[t_j:t_j+d_{B_k}-1]$ by code $\C_k^{S \prime \prime}$ for $j < (1-\sqrt{\gamma_k})|S_k|$ is the expected energy spent by code $\C_k^{S}$ in $[t_j:t_j+d_{B_k}-1]$, conditioned on $\ol{L_{1:j-1}}$.
For $i \geq (1-\sqrt{\gamma_k})|S_k|$, we have
\al{ \rescnt
E\left[ \left. \E_{\C_k^{S \prime \prime}}{[1:\nueff_{B_k}+d_{B_k}-1]} \right| \nueff_{B_k} = t_i\right] & = \sum_{j=1}^{(1-\sqrt{\gamma_k})|S_k|-1} E\left[ \left. \E_{\C_k^{S \prime \prime}}{[t_j:t_j+d_{B_k}-1]} \right| \nueff_{B_k} = t_i \right] \nonumber \\
& \leqnum \sum_{j=1}^{(1-\sqrt{\gamma_k})|S_k|-1} \frac{E\left[ \left. \E_{\C_k^{S}}{[t_j:t_j+d_{B_k}-1]} \right| \nueff_{B_k} = t_i  \right]}{\Pr(\ol{L_{1:j-1}} | \nueff_{B_k} = t_i)} \nonumber \\
& \eqnum \sum_{j=1}^{(1-\sqrt{\gamma_k})|S_k|-1} \frac{E\left[ \left. \E_{\C_k^{S}}{[t_j:t_j+d_{B_k}-1]} \right| \nueff_{B_k} = t_i  \right]}{\Pr(\ol{L_{1:j-1}} | \nueff_{B_k} = t_{(1-\sqrt{\gamma_k})|S_k|})},  \label{bound1b}
}\rescnt
where \cnt follows in the same way as the steps in (\ref{bound1}), and \cnt follows because, conditioned on $\nueff_{B_k} = t_i$ with $i \geq (1-\sqrt{\gamma_k})|S_k|$, or on $\nueff_{B_k} = t_{(1-\sqrt{\gamma_k})|S_k|}$, code $\C_k^S$ performs in the same way on any transmission block $[t_j:t_j+d_{B_k}-1]$ for $j < {(1-\sqrt{\gamma_k})|S_k|}$.
We also have that, for $i \geq j$,
\al{ \rescnt
\Pr(\ol{L_{1:j-1}} | \nueff_{B_k} = t_i) & \eqnum \Pr(\ol{L_{1:j-1}} | \nueff_{B_k} \geq t_i) \nonumber \\
& = 1 - \Pr\left( \left. \cup_{k=1}^{j-1} {L_{k}} \right| \nueff_{B_k} \geq t_i\right) \nonumber \\
& \geq 1 - \Pr\left(\error(\C_k^{S \prime}) | \nueff_{B_k} \geq t_i\right) \geq 1 - \frac{\Pr\left(\error(\C_k^{S \prime})\right)}{\Pr(\nueff_{B_k} \geq t_i)} = 1 - \frac{\gamma_k}{\Pr(\nueff_{B_k} \geq t_i)}, \label{bound2}
} \rescnt
where \cnt follows since the performance of the code up to transmission block $i-1$ is the same whether $\nueff_{B_k} = t_i$ or $\nueff_{B_k} > t_i$.
Finally, we obtain
\aln{ \rescnt
E\left[\E_{\C_k^{S \prime \prime}}{[1:\nueff_{B_k}+d_{B_k}-1]}\right] & = \sum_{i=1}^{|S_k|} |S_k|^{-1}  E\left[  \left. \E_{\C_k^{S \prime \prime}}{[1:\nueff_{B_k}+d_{B_k}-1]} \right| \nueff_{B_k} = t_i\right]\nonumber \\
& \leqnum  \sum_{i=1}^{(1-\sqrt{\gamma_k})|S_k|-1} |S_k|^{-1} \sum_{j=1}^{i} \frac{E\left[\left.\E_{\C_k^{S}}{[t_j:t_j+d_{B_k}-1]}\right| \nueff_{B_k} = t_i \right]}{\Pr(\ol{L_{1:j-1}} | \nueff_{B_k} = t_i)}   \nonumber \\
& \quad + \sum_{i=(1-\sqrt{\gamma_k})|S_k|}^{|S_k|} |S_k|^{-1} \sum_{j=1}^{i} \frac{E\left[\left.\E_{\C_k^{S}}{[t_j:t_j+d_{B_k}-1]}\right| \nueff_{B_k} = t_{i} \right]}{\Pr(\ol{L_{1:j-1}} | \nueff_{B_k} = t_{(1-\sqrt{\gamma_k})|S_k|})}   \nonumber \\
& \leqnum \sum_{i=1}^{(1-\sqrt{\gamma_k})|S_k|-1} |S_k|^{-1} \sum_{j=1}^{i} \frac{E\left[ \left. \E_{\C_k^{S}}{[t_j:t_j+d_{B_k}-1]}\right| \nueff_{B_k} = t_i  \right]}{1 - \frac{\gamma_k}{\Pr(\nueff_{B_k} \geq t_i)}}   \nonumber \\
& \quad + \sum_{i=(1-\sqrt{\gamma_k})|S_k|}^{|S_k|} |S_k|^{-1} \sum_{j=1}^{i} \frac{E\left[\left.\E_{\C_k^{S}}{[t_j:t_j+d_{B_k}-1]}\right| \nueff_{B_k} = t_{i} \right]}{1 - \frac{\gamma_k}{\Pr(\nueff_{B_k} \geq t_{(1-\sqrt{\gamma_k})|S_k|})}}   \nonumber \\
& \leqnum \sum_{i=1}^{|S_k|} |S_k|^{-1} \sum_{j=1}^{i} \frac{E\left[\E_{\C_k^{S}}{[t_j:t_j+d_{B_k}-1]}\left| \nueff_{B_k} = t_i \right. \right]}{1 - \sqrt{\gamma_k}}   \nonumber }
\al{
& = \frac{1}{1-\sqrt{\gamma_k}} \sum_{i=1}^{|S_k|} |S_k|^{-1} E\left[\left.\E_{\C_k^{S}}{[1:\nueff_{B_k}+d_{B_k}-1]} \right| \nueff_{B_k} = t_i\right]   \nonumber \\
& = \frac{1}{1-\sqrt{\gamma_k}}E\left[\E_{\C_k^S}{[1:\nueff_{B_k}+d_{B_k}-1]} \right]   \leq \frac{E[\E_{\C_k^{S}}]}{1-\sqrt{\gamma_k}},
}\rescnt
where 
\cnt follows from (\ref{bound1}) and (\ref{bound1b}), \cnt follows from (\ref{bound2}), and \cnt follows from the fact that for $i \leq (1-\sqrt{\gamma_k})|S_k|$, $\Pr(\nueff_{B_k} \geq t_i) \geq \sqrt{\gamma_k}$.
Thus, we conclude that
\aln{
\liminf_{k \to \infty} \frac{E\left[\E_{\C_k^{S \prime \prime}}[1:\nueff_{B_k}+d_{B_k}-1]\right]}{B_k} \leq (1+\eta)e_b,
}
since $\liminf_{k \to\infty} E\left[\E_{\C_k^S}\right]/B_k \leq (1+\eta)e_b$.
This concludes the proof.
\end{proof}

\newpage

\section{Proof of Lemma \ref{uniflemma}} \label{appuniflemma}

\begin{lemmarep}{\ref{uniflemma}}
Suppose we have a sequence of codes $\{\C_k\}_{k=1}^\infty$ with non-overlapping transmission blocks achieving a causal energy-per-bit $e_b$ on the asynchronous diamond network in Figure \ref{netfig}.
Then we can have  a sequence of codes $\{\C_k'\}$ that have non-overlapping transmission blocks, achieving a causal energy-per-bit $(1+\eta)e_b$ uniformly over the messages, for any $\eta > 0$.
\end{lemmarep}

\begin{proof}
Consider a sequence of codes $\{\C_k\}_{k=1}^\infty$ with non-overlapping transmission blocks achieving a causal energy-per-bit $e_b$, and fix some $\eta > 0$.
For a code $\C_k$ and each transmission time $t_i \in S_k$ 
we can define the set of messages
\aln{
\M_{t_i} = \left\{ m \in \{1,...,2^{B_k}\} \st E\left[\left. \E_{\C_k}[W(\nueff_{B_k})] \right | \nueff_{B_k} = t_i, \text{$m$ is sent } \right] \leq (1+\eta) E\left[\left. \E_{\C_k}[W(\nueff_{B_k})] \right | \nueff_{B_k} = t_i \right] \right\}.
}
To lower bound the size of $\M_{t_i}$ we notice that
\aln{
 E\left[\left. \E_{\C_k}[W(\nueff_{B_k})]  \right | \nueff_{B_k} = t_i \right]  & = \sum_{m=1}^{2^{B_k}} 2^{-{B_k}}
 E\left[\left. \E_{\C_k}[W(\nueff_{B_k})] \right | \nueff_{B_k} = t_i, \text{$m$ is sent } \right]  \nonumber \\
& \geq \sum_{m \notin \M_{t_i}}  2^{-{B_k}}
 E\left[\left. \E_{\C_k}[W(\nueff_{B_k})] \right | \nueff_{B_k} = t_i, \text{$m$ is sent } \right]  \nonumber \\
& \geq \sum_{m \notin \M_{t_i}}  2^{-{B_k}}
(1+\eta) E\left[\left. \E_{\C_k}[W(\nueff_{B_k})] \right | \nueff_{B_k} = t_i \right],
}
which implies that
\aln{
| \{1,...,2^{B_k}\} \setminus \M_{t_i} | \leq \frac{2^{B_k}}{1+\eta} \; \Rightarrow \; | \M_{t_i} | \geq \frac{\eta 2^{B_k}}{1+\eta}.
}
Therefore, for each transmission block $W(t_i)=[t_i:t_i+\ell_k-1]$, we will pick the $\frac{\eta 2^{B_k}}{1+\eta}$ messages $m$ for which \aln{E\left[\left. \E_{\C_k}[W(\nueff_{B_k})] \right | \nueff_{B_k} = t_i, \text{$m$ is sent } \right]} have the smallest values. 
Let $\psi = 1+1/\eta$.
For each transmission block $W(t_i)$, we fix any bijective mapping $\chi_{k,i}$ from $\{1,...,2^{B_k}/\psi\}$ to the $2^{B_k}/\psi$ messages chosen.
Then, for $m \in \{1,...,2^{B_k}/\psi\}$, we must have
\al{ \label{uniformexp}
E\left[\left. \E_{\C_k}[W(\nueff_{B_k})] \right | \nueff_{B_k} = t_i, \text{$\chi_{k,i}(m)$ is sent } \right] \leq (1+\eta) E\left[\left. \E_{\C_k}[W(\nueff_{B_k})] \right | \nueff_{B_k} = t_i \right].
}
From code $\C_k$ we can build code $\C_k'$, where $B_k'= {{B_k}-\log \psi}$, as the restriction of $\C_k$ according to $\chi_k$, i.e. $\C_{k}'=\C_k^{\chi}$.
Recall that, in order to show that $\{\C_k'\}$ achieves a causal energy-per-bit $(1+\eta)e_b$ uniformly over the messages, we need to show that $\{\C_k'\}$ achieves a causal energy-per-bit $(1+\eta)e_b$, and that, for any sequence of restrictions $\{\phi_k\}$, we have (\ref{uniformeq}).
Thus, we first consider a restriction of $\C_k'$ according to some arbitrary $\phi_k$, where $\phi_{k,i} : \{1,...,M_k\}\to \{1,...,2^{B_k}/\psi\} $, for some $M_k$.
Let $\W_{t_i} \subset \{1,...,2^{B_k}/\psi\}$ be the image of $\phi_{k,i}$.
Then, for any transmission time $t_i$, we have
\al{ \rescnt \label{stepenergy} 
E\left[\left.\E_{\C_k^{\prime \phi}}{[W(t_i)]}\right| \nueff_{B_k} = t_i \right] & = E\left[\left.\E_{\C_k^{\prime}}{[W(t_i)]}\right| \nueff_{B_k} = t_i, m \in \W_{t_i} \right] \nonumber \\
& = \sum_{m \in \W_{t_i}} M_k^{-1} E\left[\left.\E_{\C_k}{[W(t_i)]}\right| \nueff_{B_k} = t_i, \text{$\chi_{k,i}(m)$ is sent }\right] \nonumber \\
& \leqnum (1+\eta) \sum_{m \in \W_{t_i}} M_k^{-1} E\left[\left.\E_{\C_k}{[W(t_i)]}\right| \nueff_{B_k} = t_i \right] \nonumber \\
& = (1+\eta) E\left[\left.\E_{\C_k}{[W(t_i)]}\right| \nueff_{B_k} = t_i \right],
}\rescnt
where \cnt follows from (\ref{uniformexp}). 
Therefore, we have 
\al{ \rescnt \label{finalenergy}
E & \left[  \E_{\C_k^{\prime \phi}}{[1:\nueff_{B_k}+\ell_k-1]} \right]  = \sum_{i=1}^{|S_k|} |S_k|^{-1} E\left[\left.\E_{\C_k^{\prime \phi}}{[1:\nueff_{B_k}+\ell_k-1]}\right| \nueff_{B_k} = t_i \right]  \nonumber \\
& = \sum_{i =1}^{|S_k|} |S_k|^{-1} \sum_{j=1}^{i} E\left[\left.\E_{\C_k^{\prime \phi}}{[W(t_j)]}\right| \nueff_{B_k} = t_i \right]  \nonumber \\
& \leqnum \sum_{i =1}^{|S_k|} |S_k|^{-1} \left(\sum_{j=1}^{i-1} E\left[\left.\E_{\C_k}{[W(t_j)]}\right| \nueff_{B_k} = t_i \right] +(1+\eta) E\left[\left.\E_{\C_k}{[W(t_i)]}\right| \nueff_{B_k} = t_i \right] \right) \nonumber \\
& \leq (1+\eta) \sum_{i=1}^{|S_k|} |S_k|^{-1} E\left[\left.\E_{\C_k}{[1:\nueff_{B_k}+\ell_k-1]}\right| \nueff_{B_k} = t_i \right] \nonumber \\
& = (1+\eta) E\left[\E_{\C_k}{[1:\nueff_{B_k}+\ell_k-1]}\right],
} \rescnt
where 
\cnt follows from the fact that, prior to $\nueff_{B_k}$, $\C_k^{\prime \phi}$ performs exactly as $\C_k$ and from (\ref{stepenergy}).
%
Now, from (\ref{finalenergy}) we clearly have that 
\al{ \label{achievesuniformly}
\liminf_{k \to \infty} \frac{E\left[\E_{\C_k^{\prime \phi}}{[1:\nueff_{B_k}+\ell_k-1]}\right]}{B_k-\log \psi}
\leq (1+\eta) \liminf_{k \to \infty} \frac{E\left[\E_{\C_k}{[1:\nueff_{B_k}+\ell_k-1]}\right]}{B_k-\log \psi} = (1+\eta) e_b,
}
which means that, for any sequence of restrictions $\{\phi_k\}$, (\ref{uniformeq}) is satisfied.
Now, in order to see that $\{\C_k'\}$ achieves a causal energy-per-bit $(1+\eta) e_b$, we first notice that if we set $M_k=2^{B_k}/\psi$ and each $\phi_{k,i}$ to be the identity map, (\ref{achievesuniformly}) implies that
\aln{
\liminf_{k \to \infty} \frac{E\left[\E_{\C_k'}{[1:\nueff_{B_k}+\ell_k-1]}\right]}{B_k-\log \psi}
\leq  (1+\eta) e_b.
}
Moreover, the error probability of code $\C_k'$ satisfies
\aln{
\Pr\left(\error(\C_k')\right) & = \sum_{i=1}^{|S_k|} \sum_{m =1}^{2^{B_k}/\psi} |S_k|^{-1} \psi 2^{-B_k} \Pr\left(\error(\C_k)| \nueff_{B_k} =t_i, \text{$\chi_{k,i}(m)$ is sent } \right) \nonumber \\
& \leq \sum_{i=1}^{|S_k|} \sum_{m=1}^{2^{B_k}} |S_k|^{-1} \psi 2^{-B_k} \Pr\left(\error(\C_k)| \nueff_{B_k} =t_i, \text{$m$ is sent } \right) \nonumber \\
& = \psi \Pr\left(\error(\C_k)\right),
}
which tends to $0$ as $k \to \infty$, meaning that $\{\C_k'\}$ achieves a causal energy-per-bit $(1+\eta) e_b$.
Thus, we conclude that $\{\C_k'\}$ achieves a causal energy-per-bit $(1+\eta) e_b$ uniformly over the messages.
\end{proof}

\newpage 

\section{Proof of Lemma \ref{alphaeplemma}} \label{appalphaeplemma}

\begin{lemmarep}{\ref{alphaeplemma}}
There exists an $\alpha > 0$ and a non-negative sequence $\{\ep_k\}$ with $\ep_k \to 0$, such that
\aln{
\limsup_{k \to \infty}  \Pr \left( \nueff_{B_k} \in \T(\alpha,\C_k,\ep_k)  \right) = 1. 
}
\end{lemmarep}

\begin{proof}
We assume, by contradiction, that for all $\alpha > 0$ and all non-negative sequences $\{\ep_k\}$ with $\ep_k \to 0$, $\limsup_{k \to \infty} \Pr(\nueff_{B_k} \in \T(\alpha,\C_k,\ep_k)) < 1$. 
Notice that for any $\alpha > 0$ and any non-negative sequence $\{ \ep_k \}$ with $\ep_k \to 0$, if $t \in S_k \setminus \T(\alpha,\C_k,\ep_k)$, then 
\al{
\Pr \left(\left. \frac{\E_{\C_k}^{(r_1)}[W(\nueff_{B_k})]}{B_k} \leq \alpha \text{ and } \frac{\E_{\C_k}^{(r_2)}[W(\nueff_{B_k})]}{B_k} \leq \alpha \, \right| \nueff_{B_k} = t \right) > \ep_k. \label{probenergy0}
}
Consider some $\alpha$ and some non-negative sequence $\{\ep_k\}$ with $\ep_k \to 0$.
By assumption, we must have
\aln{
\limsup_{k \to \infty} \Pr(\nueff_{B_k} \in \T(\alpha,\C_k,\ep_k)) < 1- \delta
}
for some $\delta \in (0,1)$. 
Therefore, for some $k_0$ large enough, $\Pr(\nueff_{B_k} \notin \T(\alpha,\C_k,\ep_k)) \geq \delta/2$ if $k \geq k_0$, which implies that the set $S_k \setminus \T(\alpha,\C_k,\ep_k)$ is always non-empty for $k \geq k_0$.
In addition, we have that, for code $\C_k$,
\aln{
\Pr(\error(\C_k)) & \geq \Pr\left(\error(\C_k) | \nueff_{B_k} \notin \T(\alpha,\C_k,\ep_k)\right) \Pr\left(\nueff_{B_k} \notin \T(\alpha,\C_k,\ep_k)\right)\\
& \geq \Pr\left(\error(\C_k) | \nueff_{B_k} \notin \T(\alpha,\C_k,\ep_k)\right) \delta /2,
}
where $\Pr(\error(\C_k))\to 0$, as $k \to \infty$.
This implies that there exists at least one $t \in S_k \setminus \T(\alpha,\C_k,\ep_k)$, such that 
\al{
\Pr\left(\error(\C_k) | \nueff_{B_k} = t\right) \leq \frac{2 \Pr(\error(\C_k))}{\delta} \defi \xi_k, \label{proberror0}
}
for $k \geq k_0$.
Notice that $\xi_k \to 0$ as $k\to \infty$ as well.
To generate our contradiction, we will choose $\ep_k = \max(\xi_k^{1/4},1/B_k)$, which satisfies $\ep_k \to 0$ and $\ep_k>0$.

By noticing that the message $m$ is independent of $\nueff_{B_k}$, from (\ref{probenergy0}), we can write
\al{
\Pr & \left(\left. \frac{\E_{\C_k}^{(r_1)}[W(\nueff_{B_k})]}{B_k}  \leq \alpha \text{ and } \frac{\E_{\C_k}^{(r_2)}[W(\nueff_{B_k})]}{B_k} \leq \alpha \, \right| \nueff_{B_k} = t \right) \nonumber \\ 
& =  \sum_{m=1}^{2^{B_k}} 2^{-{B_k}} \Pr \left(\left. \frac{\E_{\C_k}^{(r_i)}[W(\nueff_{B_k})]}{B_k} \leq \alpha \text{ for } i=1,2  \, \right| \nueff_{B_k} = t, m \text{ is sent} \right) > \ep_k. \label{setsizestep}
}
Next we define the set of messages
\al{
\M_t = \left\{ m \st \Pr \left(\left. \frac{\E_{\C_k}^{(r_i)}[W(\nueff_{B_k})]}{B_k} \leq \alpha \text{ for } i=1,2\, \right| \nueff_{B_k} = t, m \text{ is sent} \right) > \frac{\ep_k}2 \right\}, \label{mtdef}
}
and, from (\ref{setsizestep}) we have
\aln{
\ep_k < & \sum_{m \in \M_t} 2^{-{B_k}} \Pr \left(\left. \frac{\E_{\C_k}^{(r_i)}[W(\nueff_{B_k})]}{B_k} \leq \alpha \text{ for } i=1,2  \, \right| \nueff_{B_k} = t, m \text{ is sent} \right) \\
& + \sum_{m \notin \M_t} 2^{-{B_k}} \Pr \left(\left. \frac{\E_{\C_k}^{(r_i)}[W(\nueff_{B_k})]}{B_k} \leq \alpha \text{ for } i=1,2  \, \right| \nueff_{B_k} = t, m \text{ is sent} \right) \\
& \leq \sum_{m \in \M_t} 2^{-{B_k}} + \sum_{m \notin \M_t} 2^{-{B_k}} \ep_k / 2 \\
& = 2^{-{B_k}}|\M_t| + (1-|\M_t|2^{-{B_k}}) \ep_k/2,
}
from which we conclude that
\al{
\frac{\ep_k}{2} < 2^{-{B_k}} |\M_t| \left(1-\frac{\ep_k}2\right) \; \Rightarrow \; |\M_t| > \frac{\ep_k 2^{B_k}}{2 - \ep_k}. \label{sizeofmt}
}
Now we can write
\aln{
\xi_k  & \geq \Pr (\error(\C_k) | \nueff_{B_k} = t) \\
& \geq \Pr \left( \left. \frac{\E_{\C_k}^{(r_j)}[W(\nueff_{B_k})]}{B_k} \leq \alpha, j=1,2 \, \right| \nueff_{B_k} = t \right) \\
& \quad \times \Pr \left(\error(\C_k) \left| \nueff_{B_k} = t, \frac{\E_{\C_k}^{(r_j)}[W(\nueff_{B_k})]}{B_k} \leq \alpha, j=1,2 \right.\right) \\
& > \ep_k \Pr \left(\error(\C_k) \left| \nueff_{B_k} = t, \frac{\E_{\C_k}^{(r_j)}[W(\nueff_{B_k})]}{B_k} \leq \alpha, j=1,2 \right.\right),
}
and we conclude that we have
\aln{
\Pr \left(\error(\C_k) \left|  \nueff_{B_k} = t, \frac{\E_{\C_k}^{(r_j)}[W(\nueff_{B_k})]}{B_k} \leq \alpha, j=1,2 \right.\right) \leq \frac{\xi_k}{\ep_k}
\leq \xi_k^{3/4}.
}
We can now write
\al{ 
\xi_k^{3/4}& \geq \Pr \left(\error(\C_k) \left|  \nueff_{B_k} = t, \frac{\E_{\C_k}^{(r_j)}[W(\nueff_{B_k})]}{B_k} \leq \alpha, j=1,2 \right.\right)  \nonumber  \\
& =  \sum_{m=1}^{2^B} \Pr \left( m \text{ is sent} \, \left| \nueff_{B_k} = t, \frac{\E_{\C_k}^{(r_j)}[W(\nueff_{B_k})]}{B_k} \leq \alpha, j=1,2 \right. \right) \nonumber \\
& \quad\quad \times \Pr \left(\error(\C_k) \left|  \nueff_{B_k} = t, \frac{\E_{\C_k}^{(r_j)}[W(\nueff_{B_k})]}{B_k} \leq \alpha, j=1,2,  m \text{ is sent} \right.\right). \label{proberror1}
}
Next we notice that, if $m \in \M_t$, we have
\aln{
\Pr & \left( m \text{ is sent} \, \left| \nueff_{B_k} = t, \frac{\E_{\C_k}^{(r_j)}[W(\nueff_{B_k})]}{B_k} \leq \alpha, j=1,2 \right. \right) = \\
 & \quad  = \frac{\Pr \left(  \left. \frac{\E_{\C_k}^{(r_j)}[W(\nueff_{B_k})]}{B_k} \leq \alpha, j=1,2  \, \right| \nueff_{B_k} = t, m \text{ is sent} \right) 
\Pr(m \text{ is sent }| \nueff_{B_k} = t)}{\Pr \left(  \left. \frac{\E_{\C_k}^{(r_j)}[W(\nueff_{B_k})]}{B_k} \leq \alpha, j=1,2  \, \right| \nueff_{B_k} = t \right)} \\
 & \quad  \geq \Pr \left(  \left. \frac{\E_{\C_k}^{(r_j)}[W(\nueff_{B_k})]}{B_k} \leq \alpha, j=1,2  \, \right| \nueff_{B_k} = t, m \text{ is sent} \right) 
\Pr(m \text{ is sent }| \nueff_{B_k} = t) \\
 & \quad  > \frac{\ep_k}2
\Pr(m \text{ is sent }| \nueff_{B_k} = t) = \frac{\ep_k}2 2^{-B_k} \geq \frac{\xi_k^{1/4}}2 2^{-B_k},
}
and, thus, from (\ref{proberror1}), we obtain
\al{
 & \xi_k^{3/4} \geq  \sum_{m \in \M_t} \frac{\xi_k^{1/4}}2 2^{-B_k} \Pr \left(\error(\C_k) \left|  \nueff_{B_k} = t,  \frac{\E_{\C_k}^{(r_j)}[W(\nueff_{B_k})]}{B_k} \leq \alpha, j=1,2,  m \text{ is sent} \right.\right)  \nonumber \\
\Rightarrow \;
 & 2 \, \xi_k^{2/4} 2^{B_k} \geq  \sum_{m \in \M_t} \Pr \left(\error(\C_k) \left|  \nueff_{B_k} = t, \frac{\E_{\C_k}^{(r_j)}[W(\nueff_{B_k})]}{B_k} \leq \alpha, j=1,2,  m \text{ is sent} \right.\right). \label{proberror2}
}
Now if we let 
\aln{
& \M_t' = \left\{ m \in \M_t \st \Pr \left(\error(\C_k) \left|  \nueff_{B_k} = t, \right. \right. \right. \nonumber \\ 
& \quad \quad \quad \left. \left. \left. \frac{\E_{\C_k}^{(r_j)}[W(\nueff_{B_k})]}{B_k} \leq \alpha, j=1,2,  m \text{ is sent} \right.\right) \leq \frac{4 (2-\ep_k)\xi_k^{2/4}}{\ep_k}  \right\},
}
we obtain 
\aln{
 2 \, \xi_k^{2/4} 2^{B_k} & \geq  \sum_{m \in \M_t \setminus \M_t'} \frac{4 (2-\ep_k)\xi_k^{2/4}}{\ep_k} 
\Rightarrow \; \frac{\ep_k 2^{B_k}}{2(2-\ep_k)} \geq |\M_t \setminus \M_t' | \\
\Rightarrow \;  |\M_t'| 
&\geq  \frac{\ep_k 2^{B_k}}{2(2-\ep_k)} \geq \frac{2^{B_k}}{4{B_k}},
}
where the last implication follows from (\ref{sizeofmt}).
Moreover, notice that, for  $m \in \M_t'$, we have
\al{
\Pr \left(\error(\C_k) \left|  \nueff_{B_k} = t, \frac{\E_{\C_k}^{(r_j)}[W(\nueff_{B_k})]}{B_k} \leq \alpha, j=1,2,  m \text{ is sent} \right.\right) \leq \frac{4 (2-\ep_k)\xi_k^{2/4}}{\ep_k} \leq 8 \, \xi_k^{1/4}, \label{proberror3}
}
which goes to $0$, as $k \to \infty$.

In order to generate our contradiction, we consider using this sequence of codes in the synchronous channel shown in Figure \ref{channel1}.
\begin{figure}[ht] 
     \centering
       \includegraphics[height=25mm]{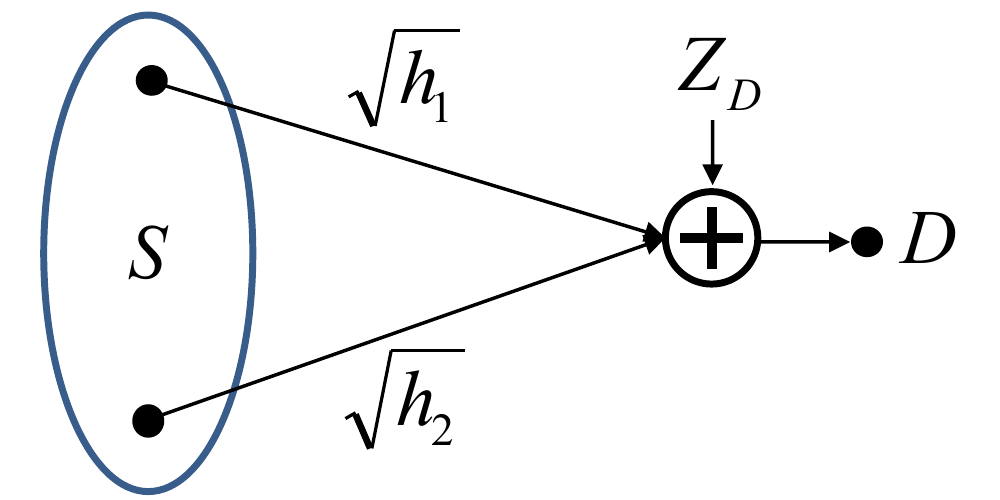} \caption{Synchronous channel considered.\label{channel1}}
\end{figure}
For each $k \geq k_0$, we can find an arrival time $t$, as described before, and a subset of messages $\M_t'$ containing at least $2^{B_k}/4{B_k}$ messages, each satisfying (\ref{mtdef}) and (\ref{proberror3}), when used over the original asynchronous channel.
In this synchronous channel, our source $S$ will receive a message chosen uniformly at random from $\M_t'$, and then play the role of both relays, since it possesses two separate antennas with channel gains $\sqrt h_1$ and $\sqrt h_2$ to the destination.
However, we change the scheme so that the source only needs to transmit what the relays would have transmitted during the transmission block $[t,t+d_{B_k}-1]$.

For a randomly selected message $m \in \M_t'$, the source proceeds as follows.
It draws two signal sequences of length $\tilde A_k = A_k + d_{B_k} -1$ using the joint distribution of the transmit signals $(X_1^{\tilde A_k},X_2^{\tilde A_k})$ of the relays when code $\C_k$ is used in the diamond network, conditioned on $\nueff_{B_k} = t$ and $m$ being sent.
If the resulting signals satisfy $\E_{\C_k}^{(r_1)}[W(\nueff_{B_k})] \leq \alpha B_k$ and $\E_{\C_k}^{(r_2)}[W(\nueff_{B_k})] \leq \alpha B_k$, then the source transmits the signals corresponding to the transmission block $[t,t+d_{B_k}-1]$ from each of them over their corresponding antennas.
Otherwise, the source repeats the process, until such transmit signals are found.
Notice that (\ref{mtdef}) guarantees that such a pair of transmit signals will eventually be found.

It is important to notice that the fact that our original sequence of codes $\{\C_k\}_{k=1}^\infty$ had non-overlapping transmission blocks guarantees that the destination applies its decoder for transmission block $[t:t+d_{B_k}-1]$ based only on the signals received during this interval (and not on signals received during $[1:t-1]$).
Therefore, the error probability only depends on the signals transmitted by the relays during $[t:t+d_{B_k}-1]$.
%
%
This allows us to conclude, from (\ref{proberror3}), that the error probability of this scheme, for any chosen message $m$, is upper bounded by $8 \, \xi_k^{1/4}$, and $8 \, \xi_k^{1/4} \to 0$, as $k \to \infty$.
The energy-per-bit achieved by this sequence of codes is given by
\aln{
\liminf_{k\to \infty} \frac{E[\E_{\C_k}]}{\log|\M_t'|} \leq \lim_{k \to \infty} \frac{2 \alpha B_k}{B_k - \log 4B_k} = 2\alpha.
}
However, since $\alpha > 0$ can be chosen arbitrarily small, this is a contradiction to the fact that the channel in Figure \ref{channel1} has a positive minimum energy-per-bit.
Therefore, we conclude that, for any sequence of asynchronous codes for the diamond network with error probability going to $0$, for some $\alpha > 0$ and some non-negative sequence $\{\ep_k\}$ such that $\ep_k \to 0$, we must have (\ref{limsup}) satisfied, which concludes the proof.
\end{proof}

\section{Proof of Lemma \ref{lim10lem}} \label{applim10lem}

\begin{lemmarep}{\ref{lim10lem}}
Suppose we have a sequence of codes $\{\C_k\}_{k=1}^\infty$ achieving a finite energy-per-bit $e_b$ on the asynchronous diamond network in Figure \ref{netfig}.
Consider any $\alpha > 0$ and any non-negative sequence $\{\ep_k\}$, with $\ep_k \to 0$.
Then, for any $\eta > 0$, we can have  a sequence of codes $\{\C_k'\}$ achieving a causal energy-per-bit $(1+\eta)e_b$ uniformly over the messages that have non-overlapping transmission blocks, 
and for which one of the following is true:
\begin{enumerate}
\item $\limsup_{k \to \infty} \Pr\left(\nueff_{B_k} \in \T_2(\alpha,\C_k,\ep_k)\right) = 1$,
\item $\liminf_{k \to \infty} \Pr\left(\nueff_{B_k} \in \T_2(\alpha,\C_k,\ep_k)\right) = 0$,
\end{enumerate}
where $\T_2(\alpha,\C_k,\ep_k)$ is defined in (\ref{two}).
\end{lemmarep}

\begin{proof}
Fix any small $\eta > 0$, any $\alpha>0$ and any non-negative sequence $\{\ep_k\}$ with $\ep_k \to 0$ as $k \to \infty$.
From Lemma \ref{nollem}, we know that the original sequence of codes can be converted into another sequence of codes with non-overlapping transmission blocks, achieving a causal energy-per-bit $(1+\eta)e_b$ uniformly over the messages.
Thus we will assume that our original sequence of codes $\{\C_k\}_{k=1}^\infty$ already satisfies these properties, and has a transmission block length $\ell_k$.
Notice that, if the set of transmission times is given by $S_k$, our delay for $\{\C_k\}_{k=1}^\infty$ is at most $2\frac{A_k}{|S_k|} + \ell_k$, which must be subexponential in $B_k$.
%
%
Now, suppose we have 
\aln{
\limsup_{k \to \infty}  \Pr \left( \nueff_{B_k} \in \T_2(\alpha,\C_k,\ep_k)  \right) = \ol\gamma \text{\quad and \quad} 
\liminf_{k \to \infty}  \Pr \left( \nueff_{B_k} \in \T_2(\alpha,\C_k,\ep_k)  \right) = \ul\gamma,
}
where $0 < \ul\gamma \leq \ol\gamma < 1$.
Let $\delta = \frac12 \min(\ul\gamma,1-\ol\gamma)$.
Then, for $k$ large enough, we must have
\al{
\Pr \left( \nueff_{B_k} \in \T_2(\alpha,\C_k,\ep_k)  \right) \geq \delta \text{\quad and \quad}
\Pr \left( \nueff_{B_k} \notin \T_2(\alpha,\C_k,\ep_k)  \right) \geq \delta, \label{deltabound}
}
for all $k$.
Now, for each $k$, notice that, since
\aln{
E\left[\E_{\C_k}{[1:\nueff_{B_k}+\ell_k-1]}\right] & = E\left[\left.\E_{\C_k}{[1:\nueff_{B_k}+\ell_k-1]}\right| \nueff_{B_k} \in \T_2(\alpha,\C_k,\ep_k)\right]\Pr \left( \nueff_{B_k} \in \T_2(\alpha,\C_k,\ep_k)  \right) \\
& + E\left[\left.\E_{\C_k}{[1:\nueff_{B_k}+\ell_k-1]}\right| \nueff_{B_k} \notin \T_2(\alpha,\C_k,\ep_k)\right]\Pr \left( \nueff_{B_k} \notin \T_2(\alpha,\C_k,\ep_k)\right),
} 
we must either have \aln{
& E\left[\left. \E_{\C_k}{[1:\nueff_{B_k}+\ell_k-1]}\right| \nueff_{B_k} \in \T_2(\alpha,\C_k,\ep_k)\right] \leq E\left[\E_{\C_k}{[1:\nueff_{B_k}+\ell_k-1]}\right] \text{\; or\;}   \nonumber \\
& E\left[\left.\E_{\C_k}{[1:\nueff_{B_k}+\ell_k-1]}\right| \nueff_{B_k} \notin \T_2(\alpha,\C_k,\ep_k)\right] \leq E\left[\E_{\C_k}{[1:\nueff_{B_k}+\ell_k-1]}\right].}
In the former case, we will define $\Tau_k$ to be the set of the $\delta |S_k|$ effective arrival times $t$ from $\T_2(\alpha,\C_k,\ep_k)$ with the smallest values of
\aln{
E\left[\left.\E_{\C_k}{[1:\nueff_{B_k}+\ell_k-1]}\right| \nueff_{B_k} = t \right].
}
In the latter case, we will define $\Tau_k$ to be the set of  the $\delta |S_k|$ effective arrival times $t \in \{t_1,...,t_{|S_k|}\}\setminus \T_2(\alpha,\C_k,\ep_k)$ with the smallest values of
\aln{
E\left[\left.\E_{\C_k}{[1:\nueff_{B_k}+\ell_k-1]} \right| \nueff_{B_k} = t \right].
}
Notice that the sequence $\Tau_k$ satisfies 
\al{
\Pr(\nueff_{B_k} \in \Tau_k) = \delta > 0,  \text{\, and \;} E\left[\left.\E_{\C_k}{[1:\nueff_{B_k}+\ell_k-1]}\right| \nueff_{B_k} \in \Tau_k\right] \leq E\left[\E_{\C_k}{[1:\nueff_{B_k}+\ell_k-1]}\right] \
\label{taubk}
}
for all $k$ large enough. 
Next, we use code $\C_k$ to build code $\C_k'$ in the following way.
Code $\C_k'$ will have $\delta |S_k|$ transmission times.
Notice that, from Definition \ref{nol}, $t_{i+1} - t_i \geq \ell_k+1$, which implies $|S_k| \leq A_k/\ell_k$.
Therefore, we have $\frac{A_k}{\delta |S_k|} > \frac{2 A_k}{|S_k|} > \ell_k$, and, if we choose our $\delta |S_k|$ transmission times to be $t_i'= \frac{i A_k}{\delta |S_k|}$, $i=1,...,\delta |S_k|$, there will be strictly more than $\ell_k$ time steps in between two consecutive transmission times.
We will perform a mapping from the $\delta |S_k|$ transmission times in $\Tau_k$ to the new transmission times $t_i' = \frac{i A_k}{\delta |S_k|}$, $i=1,...,\delta |S_k|$.
We will choose this mapping to preserve the order of the original transmission times in $\Tau_k$.
The source will now start the transmission of any message received in $\left[\frac{(i-1) A_k}{\delta |S_k|}+1:\frac{i A_k}{\delta |S_k|}\right]$ at time $t_i' = \frac{i A_k}{\delta |S_k|}$.
Moreover, if $t_j \in \Tau_k$ is mapped to $t_i'$, then
the encoding functions, relaying functions and decoding functions used at times $[t_i':t_i'+\ell_k - 1]$ will be the functions used in the original scheme during times $[t_j:t_j+\ell_k - 1]$.
At any time $t \notin \cup_{i=1}^{\delta |S_k|} [t_i':t_i'+\ell_k - 1]$, source, relays and destination will be inactive.
Notice that what allows us to perform this remapping of transmission blocks is property 3 in Definition \ref{nol}, which guarantees a sort of ``independence'' among the blocks.

For code $\C_k'$, the decoding delay will be at most $\frac{A_k}{\delta |S_k|} + \ell_k$, which is subexponential in $B_k$.
It is not difficult to see that code $\C_k'$ performs with an error probability not greater than the error probability of code $\C_k$ if the effective arrival distribution had been, instead of $\nueff_{B_k}$, a new effective distribution $\nueff_{B_k}'$, such that $\Pr(\nueff_{B_k}' = t) = \frac{1}{\delta |S_k|}$ if $t \in \Tau_k$ and $\Pr(\nueff_{B_k}' = t) = 0$ otherwise.
Thus, for code $\C_k'$, using (\ref{taubk}), we have
\aln{
\Pr\left(\error(\C_k')\right) \leq \Pr\left(\left.\error(\C_k) \right| \nueff_{B_k} \in \Tau_k\right) \leq \frac{\Pr\left(\error(\C_k)\right)}{\Pr(\nueff_{B_k}\in \Tau_k)} = \frac{\Pr\left(\error(\C_k)\right)}{\delta}, 
}
which goes to $0$, as $k \to \infty$, since $\delta$ is a positive constant.
Moreover, if $t_j \in \Tau_k$ is mapped to $t_i'$, then we have
\al{
\Pr \left(\left. \frac{\E_{\C_k}^{(r_2)}{[W(\nueff_{B_k})]}}{B_k} \leq \alpha \, \right| \nueff_{B_k} = t_j \right)  = \Pr \left(\left. \frac{\E_{\C_k'}^{(r_2)}{[W(\nueff_{B_k}')]}}{B_k} \leq \alpha \, \right| \nueff_{B_k}' = t_i' \right). \label{probeq}
}
This is the case since, conditioned on $\nueff_{B_k} = t_j$, the distribution of the transmit signals of relay 2 in $W(\nueff_{B_k})$ using code $\C_k$ is the same as the distribution, conditioned on $\nueff_{B_k}' = t_i'$, of the transmit signals of relay 2 in $W(\nueff_{B_k}')$ when using code $\C_k'$.
Furthermore, it is easy to see that $\C_k'$ also achieves a causal energy-per-bit $(1+\eta)e_b$ uniformly over the messages.
For our new code $\C_k'$, the set $\T_2(\alpha,\C_k,\ep_k)$ is defined in terms of the new effective arrival distribution $\nueff_{B_k}'$.
It is then not difficult to see that we will have, for each $B_k$, either
\aln{
\Pr \left( \nueff_{B_k}'\in \T_2(\alpha,\C_k,\ep_k)  \right) = 0  \text{\quad or \quad} \Pr \left( \nueff_{B_k}'\in \T_2(\alpha,\C_k,\ep_k)  \right) = 1,
}
depending on how we chose $\Tau_k$.
This clearly implies that
\al{
\liminf_{k \to \infty} \Pr \left( \nueff_{B_k}'\in \T_2(\alpha,\C_k,\ep_k)  \right) = 0  \text{\quad or \quad} \limsup_{k \to \infty} \Pr \left( \nueff_{B_k}'\in \T_2(\alpha,\C_k,\ep_k)  \right) = 1.
\label{0or1}
}
Moreover, from (\ref{taubk}), 
the expected energy used by $\C_k'$ up to $\nueff_{B_k}'+\ell_k-1$ satisfies
\aln{
E\left[\E_{\C_k'}{[1:\nueff_{B_k}'+\ell_k-1]}\right] \leq E\left[\left. \E_{\C_k}{[1:\nueff_{B_k}+\ell_k-1]} \right| \nueff_{B_k} \in \Tau_k\right] \leq E \left[ \E_{\C_k}{[1:\nueff_{B_k}+\ell_k-1]}\right],
}
which concludes the proof.
\end{proof}

\section{Proof of Lemma \ref{finallemma}} \label{appfinallemma}

\begin{lemmarep}{\ref{finallemma}}
Consider the network shown in Figure \ref{net2} in the asynchronous setting.
Suppose a sequence of codes $\{\C_k\}_{k=1}^\infty$ satisfies (\ref{extraconstraint}) and achieves a finite energy-per-bit.
Then we must have
\aln{
\liminf_{k \to\infty} \frac{E\left[\E_{\C_k}\right]}{B_k} \geq \gamma (1+\beta) \left(\frac{1}{g_1}+\frac{1}{h_1}\right) - f(\alpha),
}
where $f(\alpha)$ is a function satisfying $f(\alpha) \to 0$ as $\alpha \to 0$.
\end{lemmarep}

\begin{proof}
We start by considering the following network in the asynchronous setting, where we assume that there is a constraint of the form (\ref{extraconstraint}) on antenna $A_2$.
\begin{figure}[ht] 
     \centering
       \includegraphics[height=30mm]{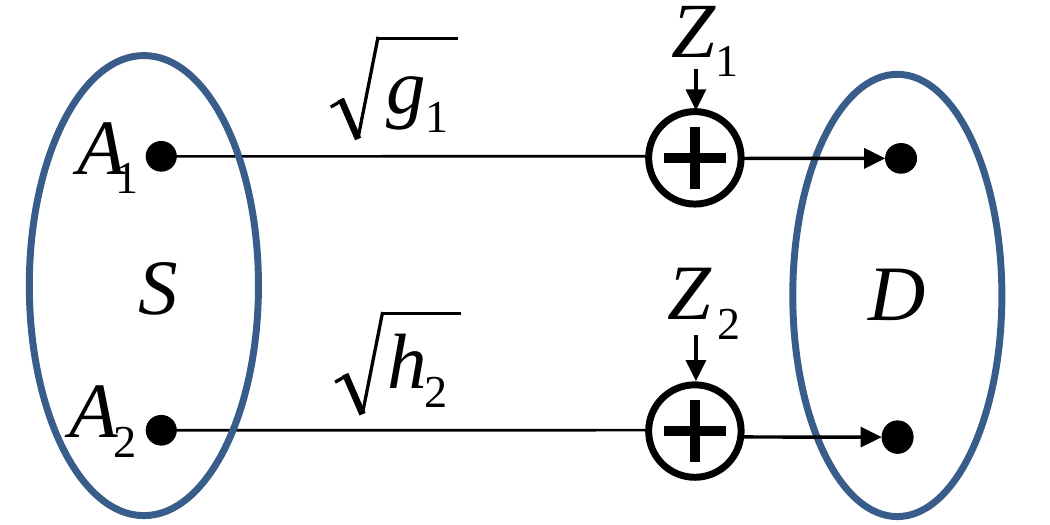} \caption{Network with parallel channels. \label{proofnet1}}
\end{figure}
Consider any sequence of codes $\{\C_k\}_{k=1}^\infty$ for this network achieving a finite energy-per-bit with delay $d_{B_k}$.
The expected energy used by code $\C_k$ can be written as
\aln{
E[\E_{\C_k}] & = E[\E_{\C_k}| \nu_{B_k} \leq A/2] \Pr(\nu_{B_k} \leq A/2) + E[\E_{\C_k}| \nu_{B_k} > A/2] \Pr(\nu_{B_k} > A/2) \\
& = \frac12 E[\E_{\C_k}| \nu_{B_k} \leq A/2] + \frac12 E[\E_{\C_k}| \nu_{B_k} > A/2].
}
Thus we must have either $E[\E_{\C_k}| \nu_{B_k} \leq A/2] \leq E[\E_{\C_k}]$ or  $E[\E_{\C_k}| \nu_{B_k} > A/2] \leq E[\E_{\C_k}]$.
Suppose the former case without much loss of generality.
Then we will modify code $\C_k$ to obtain a code $\C_k'$ that only uses $A_1$ in the following way.
An arrival time $\nu_{B_k} \in \{2t-1,2t\}$ will correspond to an arrival time $\nu_{B_k} = t$ in the original scheme, for $t=1,...,A/2$.
If a message arrives at time $\nu_{B_k} \in \{2t-1,tk\}$ the sequence of $d_{B_k}$ transmit signals on antenna $A_1$ will be sent at times $2t+1,2(t+1)+1,2(t+2)+1,...,2(t+d_{B_k})+1$.
The sequence of $d_{B_k}$ transmit signals that should be sent over antenna $A_2$ according to code $\C_k$ will be sent on antenna $A_1$ multiplied by a factor $\sqrt{\frac{h_2}{g_1}}$ at times $2t+2,2(t+1)+2,2(t+2)+2,...,2(t+d_{B_k})+2$.
Now the destination can simply interpret the signals received at times $2t+1$ and $2t+2$, for $t=1,...,A/2$ as the signals received on antennas $A_1$ and $A_2$ respectively.
With this interpretation of the received signals, the destination can apply the same decoder from code $\C_k$.
The delay of the new code is at most $d_{B_k}' = 2d_{B_k}+2$, and its error probability satisfies
\aln{
\Pr\left(\error(\C_k')\right) = \Pr\left(\error(\C_k)| \nu_{B_k} \leq A/2 \right) \leq 2 \Pr\left(\error(\C_k)\right),
}
and also tends to $0$ as $k \to \infty$.
The energy used by code $\C_k'$ satisfies
\al{ \label{energy1}
E\left[ \E_{\C_k'}\right] & = E\left[ \E_{\C_k}^{(A_1)}|\nu_{B_k} \leq A/2 \right] + \frac{h_2}{g_1}E\left[ \E_{\C_k}^{(A_2)}|\nu_{B_k} \leq A/2 \right].
}
In the case where $h_2 \leq g_1$, (\ref{energy1}) implies that 
\al{ 
E\left[ \E_{\C_k'}\right] & \leq E\left[ \E_{\C_k}^{(A_1)}|\nu_{B_k} \leq A/2 \right] + E\left[ \E_{\C_k}^{(A_2)}|\nu_{B_k} \leq A/2 \right] \nonumber \\
& = E\left[ \E_{\C_k}|\nu_{B_k} \leq A/2 \right] \leq E\left[ \E_{\C_k}\right]. \label{energy2}
}
In the case where $h_2 > g_1$, we notice that since code $\C_k$ must satisfy (\ref{extraconstraint}), we must have
\aln{
& 3\alpha B_k \geq E\left[ \E_{\C_k}^{(A_2)} \right] \geq \frac12 E\left[ \E_{\C_k}^{(A_2)}|\nu_{B_k} \leq A/2 \right],
}
and, thus,
\al{ \label{energy3}
6\alpha B_k \geq E\left[ \E_{\C_k}^{(A_2)}|\nu_{B_k} \leq A/2 \right].
}
Now by multiplying (\ref{energy3}) by $(1-h_2/g_1)$ (which is negative) and adding the resulting inequality to (\ref{energy1}), we obtain
\al{ 
E\left[ \E_{\C_k'}\right] - (h_2/g_1 - 1)6\alpha B_k & \leq E\left[ \E_{\C_k}^{(A_1)}|\nu_{B_k} \leq A/2 \right] + E\left[ \E_{\C_k}^{(A_2)}|\nu_{B_k} \leq A/2 \right] \nonumber \\
& = E\left[ \E_{\C_k}|\nu_{B_k} \leq A/2 \right] \leq E\left[ \E_{\C_k} \right].
\label{energy4}}
By combining (\ref{energy2}) and (\ref{energy4}) we can write, for all $h_2$ and $g_1$,
\al{ 
E\left[ \E_{\C_k'}\right] - \gamma \alpha B_k & \leq E\left[ \E_{\C_k} \right],
\label{energy5}}
where $\gamma = 6 \max[0,h_2/g_1 - 1]$.
Since $\{\C_k'\}$ is a sequence of codes for a point-to-point channel achieving a finite energy-per-bit, we know from Theorem \ref{pttopt} that it must satisfy
\aln{
\liminf_{k \to\infty} \frac{E\left[ \E_{\C_k'}\right]}{B_k} \geq \frac{\gamma (1+\beta)}{g_1},
}
and, thus,
\al{
\liminf_{k \to\infty} \frac{E\left[ \E_{\C_k}\right]}{B_k} \geq \liminf_{k \to\infty} \frac{E\left[ \E_{\C_k'}\right]}{B_k} - \gamma \alpha \geq \frac{\gamma (1+\beta)}{g_1} - \gamma \alpha.
\label{energy6}}
Next we consider the network in Figure \ref{proofnet2}, where we again assume that there is a constraint of the form (\ref{extraconstraint}) on antenna $A_2$.
\begin{figure}[ht] 
     \centering
       \includegraphics[height=30mm]{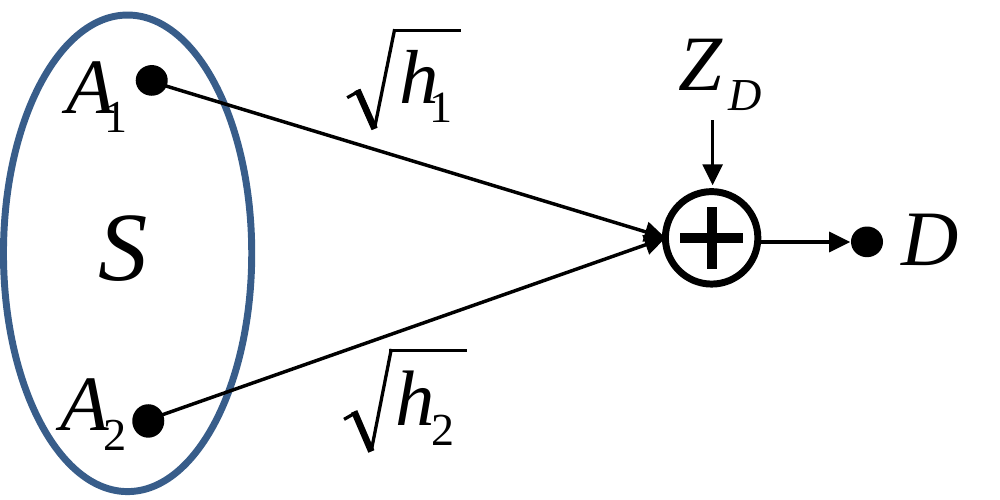} \caption{Two-input one-output network. \label{proofnet2}}
\end{figure}
Consider a sequence of codes $\{\C_k\}_{k=1}^\infty$ that achieves a finite energy-per-bit on this network.
In order for us to lower bound the energy-per-bit of this sequence of codes, we will notice that any sequence of codes of this network can be used on the network in Figure \ref{proofnet3}.
\begin{figure*}[ht] 
     \centering
     \subfigure[]{
       \includegraphics[height=34mm]{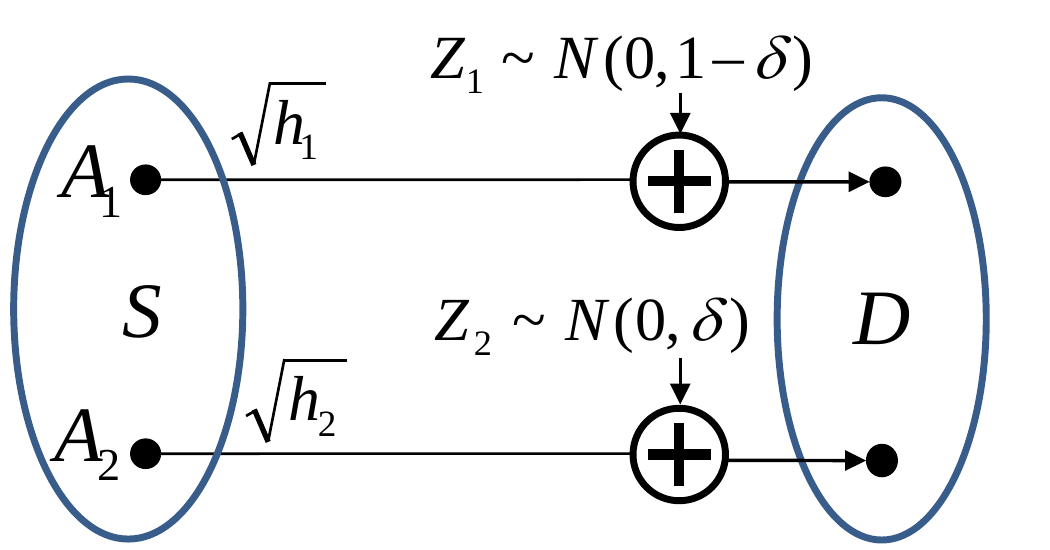} \label{proofnet3}} 
    \hspace{5mm}
    \subfigure[]{
       \includegraphics[height=34mm]{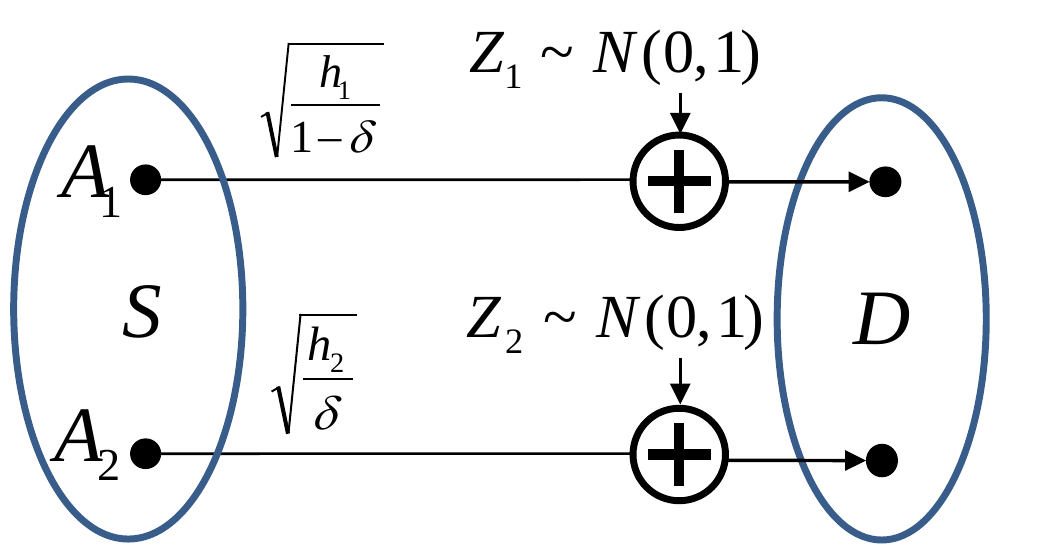} \label{proofnet4}}
    \hspace{0mm}
     \caption{Networks with parallel channels.}
\end{figure*}
The network in Figure \ref{proofnet3} is a network with two parallel channels just as the network in Figure \ref{proofnet1}, except that the additive Gaussian noise at each of the two receivers have variances $1-\delta$ and $\delta$, for some $\delta \in (0,1)$.
Any sequence of codes $\{\C_k\}_{k=1}^\infty$ for the network in Figure \ref{proofnet2}, can be directly used in the network in Figure \ref{proofnet3}.
The only modification that needs to be made is to have the destination add the signals received on each of its two receivers.
After this addition, the network effectively becomes the network from Figure \ref{proofnet2}.
Moreover, it is easy to see that the network from Figure \ref{proofnet3} is entirely equivalent to the network in Figure \ref{proofnet4}, since the SNR on each channel is the same.

Therefore, we can use our previous reasoning to lower bound the energy-per-bit achieved by the sequence of codes $\{\C_k\}_{k=1}^\infty$.
From (\ref{energy6}), we see that
\al{
\liminf_{k \to\infty} \frac{E\left[ \E_{\C_k}\right]}{B_k}  \geq \frac{\gamma (1+\beta)(1-\delta)}{h_1} - \gamma' \alpha,
\label{energy7}}
where $\gamma' = 6 \max\left[0,\frac{h_2 (1-\delta)}{h_1 \delta} - 1\right]$, and $\delta \in (0,1)$ is a free parameter that we can optimize over.
We will choose $\delta = \min[1/2, \sqrt{\alpha}]$.
Then we obtain
\al{
\liminf_{k \to\infty} \frac{E\left[ \E_{\C_k}\right]}{B_k} & \geq \frac{\gamma (1+\beta)}{h_1}-  \frac{\gamma (1+\beta)\delta}{h_1} - 6 \max\left[0,\frac{h_2 (1-\delta)}{h_1 \delta} - 1\right] \alpha \nonumber \\
& \geq \frac{\gamma (1+\beta)}{h_1}-  \frac{\gamma (1+\beta)\sqrt{\alpha}}{h_1} - 6 \frac{h_2 \alpha}{h_1 \delta} \nonumber \\
& \geq \frac{\gamma (1+\beta)}{h_1}-  \frac{\gamma (1+\beta)\sqrt{\alpha}}{h_1} - 6 \frac{h_2 (2\alpha + \sqrt{\alpha})}{h_1}.
\label{energy8}}
Now we are ready to prove the Lemma.
Suppose we have a sequence of codes $\{\C_k\}_{k=1}^\infty$ for the network in Figure \ref{net2} under the additional constraint (\ref{extraconstraint}). 
We first notice that these codes can be applied to the network in Figure \ref{proofnet1}.
In order to do that, we would have the destination considering its upper receiver to be relay $1$, and then simulating what relay $1$ would have transmitted and adding that to the received signals at the lower receiver.
The energy used when applying code $\C_k$ from the network in Figure \ref{net2} on the network in Figure \ref{proofnet1} is just the energy that $\C_k$ would consume on $A_1$ and $A_2$.
Therefore, from (\ref{energy6}), we have that
\al{
\liminf_{k \to\infty} \frac{E\left[ \E_{\C_k}^{(A_1)}\right]+E\left[ \E_{\C_k}^{(A_2)}\right]}{B_k} \geq \frac{\gamma (1+\beta)}{g_1} - \gamma \alpha.
\label{energy9}}
Similarly, we notice that the sequence of codes $\{\C_k\}_{k=1}^\infty$ for the network in Figure \ref{net2} can be applied to the network in Figure \ref{proofnet2}.
This time, the source from Figure \ref{proofnet2} computes what the source from Figure \ref{net2} would have transmitted over $A_1$ and simulates what relay $1$ would receive and transmit.
Then it transmits the simulated outputs of relay $1$ over $A_1$.
The signals transmitted on $A_2$ would be the same in both cases.
The energy consumed when using code $\C_k$ on the network in Figure \ref{proofnet2} is the energy that relay $1$ and source antenna $A_2$ would consume.
Therefore, from (\ref{energy8}), we obtain
\al{
\liminf_{k \to\infty} \frac{E\left[ \E_{\C_k}^{(r_1)}\right]+E\left[ \E_{\C_k}^{(A_2)}\right]}{B_k} \geq \frac{\gamma (1+\beta)}{h_1}-  \frac{\gamma (1+\beta)\sqrt{\alpha}}{h_1} - 6 \frac{h_2 (2\alpha+\sqrt{\alpha})}{h_1}.
\label{energy10}}
In order to lower bound the total energy-per-bit of the sequence of codes $\{\C_k\}_{k=1}^\infty$ we first compute
\al{ \rescnt
\liminf_{k \to\infty} & \frac{E\left[ \E_{\C_k}^{(r_1)}\right]+E\left[ \E_{\C_k}^{(A_1)}\right]+2 E\left[ \E_{\C_k}^{(A_2)}\right]}{B_k} \nonumber \\
& \geq \liminf_{k \to\infty} \frac{E\left[ \E_{\C_k}^{(A_1)}\right]+E\left[ \E_{\C_k}^{(A_2)}\right]}{B_k}  +\liminf_{k \to\infty} \frac{E\left[ \E_{\C_k}^{(r_1)}\right]+E\left[ \E_{\C_k}^{(A_2)}\right]}{B_k} \nonumber \\
& \geqnum \frac{\gamma (1+\beta)}{g_1} - \gamma \alpha + \frac{\gamma (1+\beta)}{h_1}-  \frac{\gamma (1+\beta)\sqrt{\alpha}}{h_1} - 6 \frac{h_2 (2\alpha+ \sqrt{\alpha})}{h_1} \nonumber \\
& = \gamma (1+\beta)\left(\frac{1}{g_1}+\frac{1}{h_1}\right) - \gamma \alpha -  \frac{\gamma (1+\beta)\sqrt{\alpha}}{h_1} - 6 \frac{h_2 (2\alpha+ \sqrt{\alpha})}{h_1},
\label{energy11}}\rescnt 
where \cnt follows from (\ref{energy9}) and (\ref{energy10}).
We also have that
\al{ \rescnt
\liminf_{k \to\infty} & \frac{E\left[ \E_{\C_k}^{(r_1)}\right]+E\left[ \E_{\C_k}^{(A_1)}\right]+2 E\left[ \E_{\C_k}^{(A_2)}\right]}{B_k} \nonumber \\
& \leq \liminf_{k \to\infty} \frac{E\left[ \E_{\C_k}^{(r_1)}\right]+E\left[ \E_{\C_k}^{(A_1)}\right]+ E\left[ \E_{\C_k}^{(A_2)}\right]}{B_k} + \limsup_{k \to\infty} \frac{E\left[ \E_{\C_k}^{(A_2)}\right]}{B_k} \nonumber \\
& = \liminf_{k \to\infty} \frac{E\left[ \E_{\C_k}\right]}{B_k} + \limsup_{k \to\infty} \frac{E\left[ \E_{\C_k}^{(A_2)}\right]}{B_k} \nonumber \\
& \leqnum  \liminf_{k \to\infty} \frac{E\left[ \E_{\C_k}\right]}{B_k} +3 \alpha, 
\label{energy12}}\rescnt
where \cnt follows from the constraint (\ref{extraconstraint}).
Finally, by combining (\ref{energy11}) and (\ref{energy12}) we obtain
\aln{
\liminf_{k \to\infty} \frac{E\left[ \E_{\C_k}\right]}{B_k} \geq \gamma (1+\beta)\left(\frac{1}{g_1}+\frac{1}{h_1}\right) - f(\alpha),
}
where
\aln{
f(\alpha) =  \gamma \alpha + \frac{\gamma (1+\beta)\sqrt{\alpha}}{h_1}+ 6 \frac{h_2 (2\alpha+ \sqrt{\alpha})}{h_1} + 3 \alpha,
}
which clearly satisfies $f(\alpha) \to 0$ as $\alpha \to 0$.\end{proof}

\bibliographystyle{unsrt}




\end{document}